\documentclass[letterpaper]{article} % DO NOT CHANGE THIS
\usepackage{aaai24}  % DO NOT CHANGE THIS
\usepackage{times}  % DO NOT CHANGE THIS
\usepackage{helvet}  % DO NOT CHANGE THIS
\usepackage{courier}  % DO NOT CHANGE THIS
\usepackage[hyphens]{url}  % DO NOT CHANGE THIS
\usepackage{graphicx} % DO NOT CHANGE THIS
\usepackage{natbib}  % DO NOT CHANGE THIS AND DO NOT ADD ANY OPTIONS TO IT
\usepackage{caption} % DO NOT CHANGE THIS AND DO NOT ADD ANY OPTIONS TO IT
\usepackage{algorithm}
\usepackage{algorithmic}
\usepackage{tabularx}
\usepackage{makecell}
\usepackage{multirow}
\usepackage{booktabs}
\usepackage{newfloat}
\usepackage{listings}
\DeclareCaptionStyle{ruled}{labelfont=normalfont,labelsep=colon,strut=off} % DO NOT CHANGE THIS
\floatstyle{ruled}
\newfloat{listing}{tb}{lst}{}
\floatname{listing}{Listing}
\usepackage{amsmath}
\usepackage{amssymb}
\usepackage[caption=false]{subfig}
\newcommand\norm[1]{\left\lVert#1\right\rVert}

\title{LDMVFI: Video Frame Interpolation with Latent Diffusion Models}
\author {
    Duolikun Danier,
    Fan Zhang,
    David Bull
}
\affiliations {
    University of Bristol\\
    \{duolikun.danier, fan.zhang, dave.bull\}@bristol.ac.uk
}
\usepackage{bibentry}
\begin{document}

\maketitle

\begin{abstract}
Existing works on video frame interpolation (VFI) mostly employ deep neural networks that are trained by minimizing the L1, L2, or deep feature space distance (e.g. VGG loss) between their outputs and ground-truth frames. However, recent works have shown that these metrics are poor indicators of perceptual VFI quality. Towards developing perceptually-oriented VFI methods, in this work we propose latent diffusion model-based VFI, LDMVFI. This approaches the VFI problem from a generative perspective by formulating it as a conditional generation problem. As the first effort to address VFI using latent diffusion models, we rigorously benchmark our method on common test sets used in the existing VFI literature. Our quantitative experiments and user study indicate that LDMVFI is able to interpolate video content with favorable perceptual quality compared to the state of the art, even in the high-resolution regime. Our code is available at \url{https://github.com/danier97/LDMVFI}.
\end{abstract}

\section{Introduction}
Video frame interpolation (VFI) aims to generate intermediate frames between two existing consecutive video frames. It is commonly used to synthetically increase frame rate, e.g., to produce jitter-free slow-motion content~\cite{jiang2018super}. VFI has also been used in video compression~\cite{wu2018video}, view synthesis~\cite{flynn2016deepstereo}, medical imaging~\cite{karargyris2010three} and animation production~\cite{siyao2021deep}.

Existing VFI methods~\cite{jiang2018super, xue2019video, lee2020adacof, niklaus2020softmax, kalluri2023flavr} are mostly based on deep neural networks. These deep models differ in architectural designs and motion modeling approaches, but are mostly trained to minimize the L1, L2, or VGG~\cite{Simonyan15vgg} feature distance between their outputs and the ground-truth intermediate frames. However, recent works~\cite{men2020visual, danier2022subjective} have shown that these optimization objectives are not indicative of the perceptual quality of interpolated videos, as they correlate poorly with human judgments. As a result, it has been reported~\cite{danier2022subjective} that existing methods, while achieving high PSNR values, tend to under-perform perceptually, especially under challenging scenarios involving dynamic textures with complex motion.

\begin{figure}[t]
\centering
\includegraphics[width=1\linewidth]{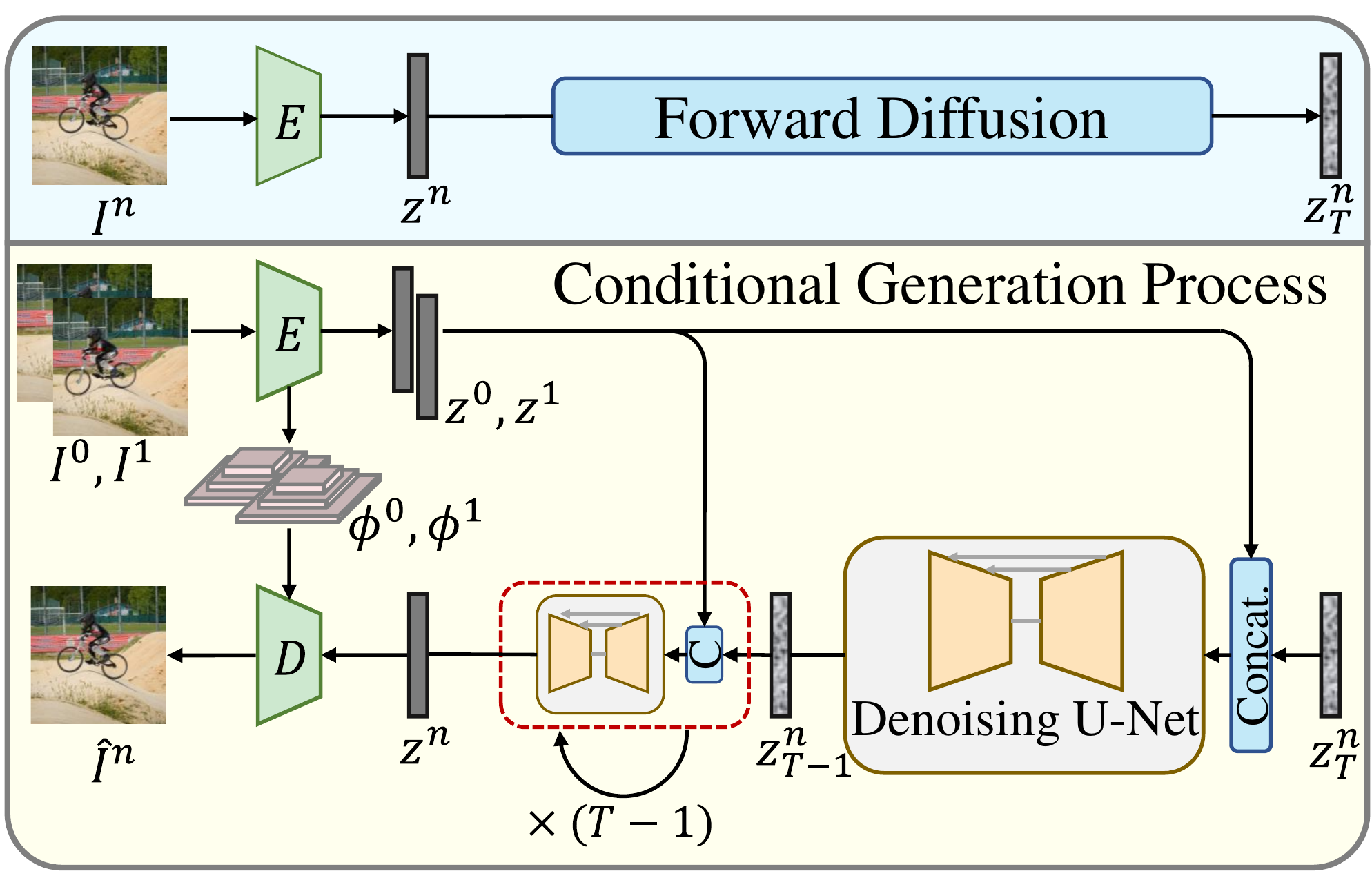}
\caption{Overview of the diffusion processes in LDMVFI. The encoder and decoder enable projection between image and latent spaces, and the diffusion processes take place in the latent space.}
\label{fig:overall}
\end{figure}

A potential approach to improve the perceptual performance of VFI methods is to develop more accurate perceptual metrics for training VFI models, and this is the focus of another research area, namely Video Quality Assessment~\cite{saha3perceptual}. In this work, instead of relying on particular metrics, we explore a new direction for perception-oriented VFI based on diffusion models~\cite{ho2020denoising, rombach2022high}. Diffusion models have recently exhibited remarkable performance in generating realistic, perceptually-optimized images and videos, reportedly outperforming other generative models including GANs and VAEs~\cite{dhariwal2021diffusion, ho2022cascaded}. However, despite their ability to synthesize high-fidelity visual content, the application of diffusion models for VFI has not been fully investigated.

In the above context, we propose a latent diffusion model for video frame interpolation (LDMVFI), where VFI is formulated as a conditional image generation problem. Specifically, we adopt the recently proposed \textit{latent diffusion models}~\cite{rombach2022high} (LDMs) within a framework comprising an autoencoding model that projects images into a latent space, and a denoising U-Net which performs reverse diffusion process in that latent space. To better adapt LDMs to VFI, we devise VFI-specific components, notably a novel vector quantization-based VFI-autoencoding model, VQ-FIGAN, with which our method shows superior performance over vanilla LDMs.

Despite the paradigmatic shift from the mainstream VFI methods, we adhere to the commonly adopted VFI benchmarking protocol and evaluate the proposed method on various VFI test sets, covering both low and high resolution content (up to 4K). Our results demonstrate that LDMVFI performs favorably against the state of the art in terms of three perceptual metrics~\cite{zhang2018unreasonable, danier2022flolpips, heusel2017gans}. We also conducted a user study to collect subjective judgments on the quality of full HD videos interpolated by our method benchmarked against several competitive counterparts; this further confirms the favorable perceptual performance of LDMVFI.

To the best of our knowledge, this work is the first to address VFI as a conditional generation problem using latent diffusion models, and the first to demonstrate the potential of the new paradigm for perception-oriented VFI. Our contributions are summarized below.

\begin{itemize}
    \item We present LDMVFI, a latent diffusion-based method that leverages the high-fidelity image synthesis ability of diffusion models to perform VFI. 
    \item We introduce novel VFI-specific components into LDMs, including a vector-quantized autoencoding model, VQ-FIGAN, which further enhances VFI performance.
    \item We demonstrate, through quantitative and qualitative experiments, that the proposed method performs favorably against the state of the art.
\end{itemize}

\section{Relate Work}
\paragraph{Video Frame Interpolation.} Existing VFI approaches are mostly based on deep learning, and can be generally categorized as flow-based or kernel-based. Flow-based methods rely on optical flow estimation to generate interpolated frames. To obtain the optical flow from input frames to the non-existent middle frame (or the other way around), some methods~\cite{jiang2018super, niklaus2018context, niklaus2020softmax, sim2021xvfi}  assume certain motion types to infer the intermediate optical flows using the flows between two input frames, while others~\cite{liu2017video,xue2019video,park2020bmbc,park2021asymmetric,lu2022video,kong2022ifrnet} directly estimate the intermediate flows. On the other hand, kernel-based methods argue that optical flows can be unreliable in dynamic texture scenes, so they predict locally adaptive convolution kernels to synthesize output pixels, allowing more flexible many-to-one mapping. While earlier methods~\cite{adaconv} in this class predict fixed-size kernels, more recent ones~\cite{lee2020adacof, ding2021cdfi, cheng2020video, cheng2021multiple, danier2022enhancing} tend to adopt deformable convolution~\cite{dai2017deformable} kernels. Other than these two classes, there are also attempts to combine flows and kernels~\cite{bao2019memc,danier2022st}, and to perform end-to-end frame synthesis~\cite{choi2020channel,kalluri2023flavr}. 

It is noted that the above methods are trained by optimizing PSNR-oriented loss functions, i.e., the L1/L2 distance between the model outputs and the ground-truth frames. To improve perceptual performance, some existing methods~\cite{niklaus2018context,niklaus2020softmax} use the VGG~\cite{Simonyan15vgg} feature-based loss in combination with the L1 loss. However, it has been previously reported~\cite{danier2022subjective} that these distances do not fully reflect the perceptual quality of interpolated videos, exhibiting poor correlation performance with subjective ground truth. As a result, it has been observed~\cite{danier2022subjective} that some state-of-the-art methods show unsatisfactory perceptual performance, especially on dynamic textures with complex motions.

\paragraph{Diffusion Models.} Recently, diffusion models (DMs)~\cite{ho2020denoising,rombach2022high} have demonstrated remarkable performance in synthesizing high-fidelity images and videos. In their original form~\cite{ho2020denoising}, DMs generate new images by progressively denoising a Gaussian noise image;  the process corresponds to the reverse of a Markov chain that gradually adds noise to a clean image. DMs have been reported to offer superior performance~\cite{dhariwal2021diffusion, ho2022cascaded} compared to GANs~\cite{goodfellow2020generative} and VAEs~\cite{Kingma2014vae} in image generation tasks. The only previous application of DMs on VFI is \cite{voleti2022mcvd}, but this work focused on low-resolution images and the model lacked any VFI-specific innovations, showing limited interpolation performance. The recently proposed latent diffusion models (LDMs) have demonstrated strong ability to synthesize high-resolution images by performing diffusion processes in latent space. However, we observe that LDMs have not previously been exploited for VFI.

\section{Proposed Method: LDMVFI}
Given two consecutive frames $I^0,I^1$ from a video, VFI aims to generate the non-existent intermediate frame $I^n$ where $n=0.5$ for $\times$2 upsampling. Approaching VFI from a generative perspective, our goal is to learn a parametric approximation of the conditional distribution $p(I^n|I^0,I^1)$ using a dataset $\mathcal{D}=\{I^0_s, I^n_s, I^1_s\}_{s=1}^S$. To achieve this, we adopt the latent diffusion models~\cite{rombach2022high} to perform conditional generation for VFI. The proposed LDMVFI contains two main components: (i) a VFI-specific \textbf{autoencoding model}, VQ-FIGAN, that projects frames into a latent space, and reconstructs the target frame from the latent encoding; (ii) a \textbf{denoising U-Net} that performs reverse diffusion process in the latent space for conditional image generation. Figure~\ref{fig:overall} shows the overview of the LDMVFI.

\begin{figure}[t]
\centering
\includegraphics[width=\linewidth]{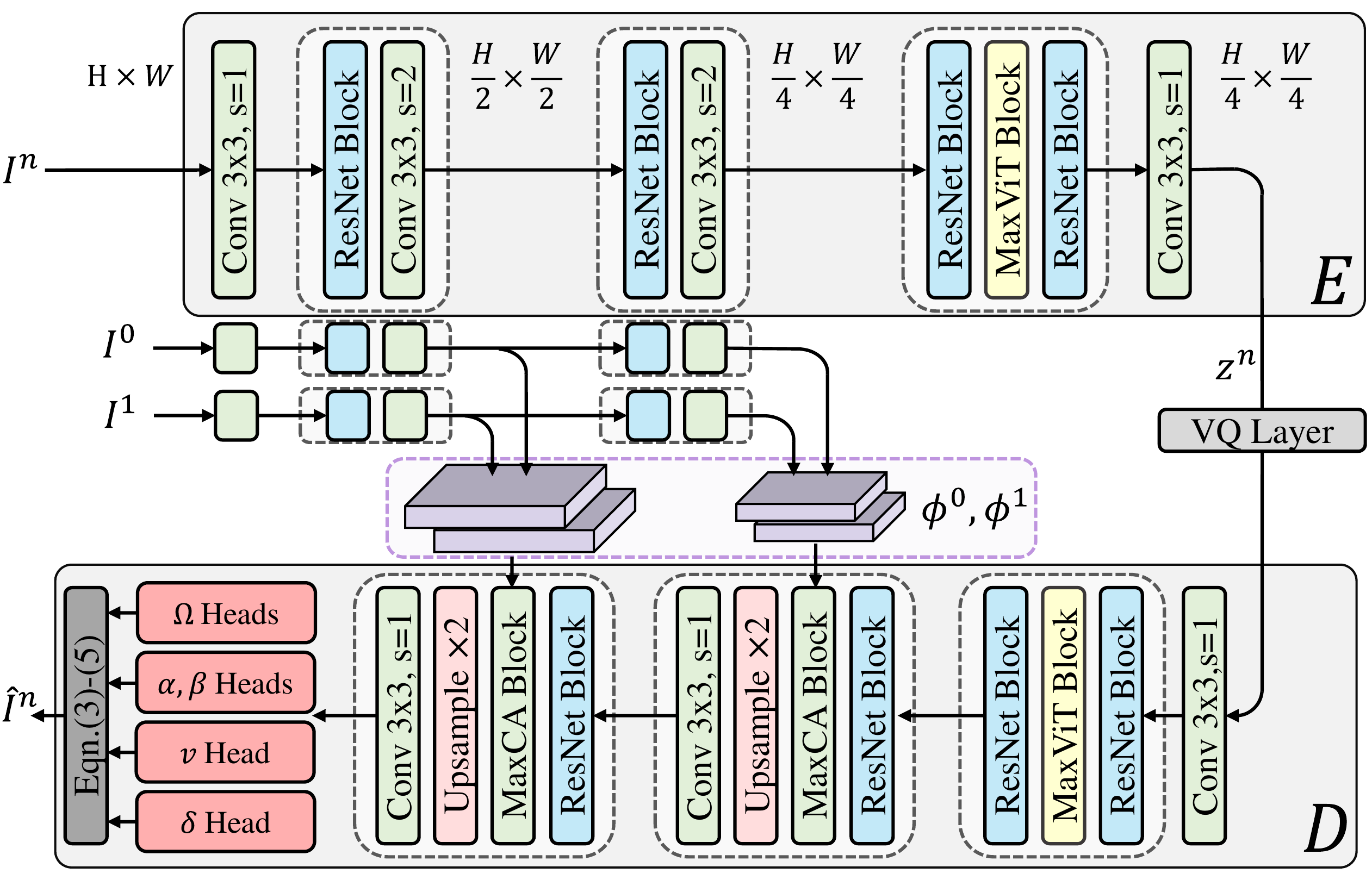}
\caption{The architecture of the VFI autoencoding model, VQ-FIGAN. It differs from the original VQGAN~\cite{esser2021taming} in three aspects: (i) the use of features extracted by the encoder from neighboring frames during the decoding via MaxViT-based cross attention; (ii) use of more efficient MaxViT block instead of the vanilla self-attention; (iii) frame synthesis via adaptive deformable convolution. The kernel ($\Omega$), offset ($\alpha,\beta$), visibility ($v$) and residual ($\delta$) heads contain 3$\times$\texttt{\{conv3x3, ReLU\}}.}
\label{fig:vqfinet}
\end{figure}

\subsection{Latent Diffusion Models}
Latent diffusion models (LDMs)~\cite{rombach2022high} are built upon the denoising diffusion probabilistic models~\cite{ho2020denoising} (referred to as diffusion models hereafter), which are a class of generative models that learns a data distribution $p(x)$ by learning the reverse process of a pre-defined Markov Chain of length $T$ (i.e. the forward process) that gradually adds Gaussian noise to the data. Specifically, a time-conditioned neural network $\epsilon_\theta(x_t,t)$ is trained to denoise the data at step $t=1,\dots,T$ with the objective
\begin{equation}
    \mathcal{L}_{\mathrm{DM}} = \mathbb{E}_{x_0,\epsilon\sim\mathcal{N}(\mathbf{0},\mathbf{I}),t\sim\mathcal{U}(1,T)} \big[ \norm{\epsilon - \epsilon_\theta(x_t,t)}^2 \big],
\end{equation}
where $x_t$ is sampled from the forward diffusion process and $\mathcal{U}(1,T)$ denotes uniform distribution over $\{1,\dots,T\}$. This corresponds to a reweighted version of the variational lower bound on $\log p(x)$. More details of diffusion processes and the loss are provided in Appendix~\ref{appendix:loss}. 

To sample new images, one can start from a Gaussian noise and gradually denoise it using the denoising network $\epsilon_\theta$.
Under the conditional generation setting, we can condition $\epsilon_\theta$ on an additional input $y$, which can denote the two input frames in the context of VFI.

Latent diffusion models contain an image encoder $E:x\mapsto z$ that encodes an image $x$ into a lower-dimensional latent representation $z$, and a decoder $D:z\mapsto x$ that reconstructs the image $x$. The forward and reverse diffusion processes then happen in the latent space, and the training objective for learning the reverse diffusion process becomes
\begin{equation}\small
    \mathcal{L}_{\mathrm{LDM}} = \mathbb{E}_{E(x_0),\epsilon\sim\mathcal{N}(\mathbf{0},\mathbf{I}),t\sim\mathcal{U}(1,T)} \big[ \norm{\epsilon - \epsilon_\theta(z_t,t)}^2 \big].
\end{equation}

By projecting images into a compact latent space, LDMs allow the diffusion process to concentrate on the semantically important portions of the data and enable more computationally efficient sampling process.

\subsection{Autoencoding with VQ-FIGAN}\label{sec:vqfigan}
In the original form of LDMs, the autoencoding model $\{E,D\}$ is considered as a perceptual image codec. Its design purpose is to project images into efficient latent representations where high-frequency details are removed during encoding and recovered in the decoding. However, in the context of VFI, such information is likely to affect the perceived quality of the interpolated videos, so the limited reconstruction ability of the decoder can negatively impact the VFI performance. To enhance high-frequency detail recovering, we propose a VFI-specific autoencoding model: VQ-FIGAN, which is illustrated in Figure~\ref{fig:vqfinet}. While the backbone model is similar to the original VQGAN~\cite{esser2021taming} used in \cite{rombach2022high}, there are three major differences.

Firstly, we take advantage of a property of the VFI task - that the neighboring frames are available during inference, in order to design a frame-aided decoder. Specifically, given the ground-truth target frame $I^n\in\mathbb{R}^{H\times W\times 3}$, the encoder $E$ produces the latent encoding $z^n = E(I^n)$, where $z^n\in\mathbb{R}^{\frac{H}{f}\times \frac{W}{f}\times 3}$, and $f$ is a hyper-parameter (Figure~\ref{fig:vqfinet} shows the case for $f=4$). Then, the decoder $D$ outputs a reconstructed frame, $\hat{I}^n$, taking input $z^n$, as well as the feature pyramids, $\phi^0, \phi^1$, extracted by $E$ from two neighboring frames, $I^0,I^1$. During decoding, these feature pyramids are fused with the decoded features from $z^n$ at multiple layers using the MaxCA blocks, which are newly designed MaxViT~\cite{tu2022maxvit}-based Cross Attention blocks, where the query embeddings for the attention mechanism are generated using decoded features from $z^n$, and the key and value embeddings are obtained from $\phi^0,\phi^1$.

Secondly, we notice the original VQGAN uses the full self-attention~\cite{vaswani2017attention} as in the vision transformer~\cite{dosovitskiy2021an}, which can be computationally heavy (quadratic complexity) and memory-intensive, especially when the image resolution is high. For more efficient inference on high-resolution (e.g. full HD) videos, we employ the recently proposed MaxViT block~\cite{tu2022maxvit} to perform self-attention. The multi-axis self-attention layer in the MaxViT block combines windowed attention and dilated grid attention to perform both local and global operations, while achieving linear complexity with respect to the input size.

Thirdly, instead of having the decoder directly output the reconstructed image $\hat{I}^n$, the proposed VQ-FIGAN outputs the deformable convolution-based interpolation kernels~\cite{dai2017deformable, lee2020adacof, cheng2020video} to enhance VFI performance. Specifically, given that $H,W$ are the frame height and width, the output of the decoder network contains parameters of the locally adaptive deformable convolution kernels of size $K\times K$: $\{\Omega^\tau, \alpha^\tau, \beta^\tau\}$ where $\tau=0,1$ indexes the input frames. Here $\Omega\in [0,1]^{H\times W\times K\times K}$ contains the weights of the kernels, and $\alpha,\beta\in\mathbb{R}^{H\times W\times K\times K}$ are their spatial offsets (horizontal and vertical respectively). The decoder also outputs a visibility map $v\in[0,1]^{H\times W}$ to account for occlusion~\cite{jiang2018super}, and a residual map $\delta\in\mathbb{R}^{H\times W}$ to further enhance VFI performance~\cite{cheng2021multiple}. To generate the interpolated frame, firstly locally adaptive deformable convolutions are performed for each input frame $\{I^\tau\}_{\tau=0,1}$:
\begin{gather}
\footnotesize
    I^{n\tau}(h,w) = \sum_{i=1}^K\sum_{j=1}^K \Omega^\tau_{h,w}(i,j)\cdot P^\tau_{h,w}(i,j), \label{eqn:defconv1} \\
    P^\tau_{h,w}(i,j) = I^\tau(h+\alpha^\tau_{h,w}(i,j), w+\beta^\tau_{h,w}(i,j)), \label{eqn:defconv2}
\end{gather}
in which $I^{n\tau}$ denotes the result obtained from $I^\tau$, and $P^\tau_{h,w}$ is the patch sampled from $I^\tau$ for output location $(h,w)$. These intermediate results are then combined using the visibility and residual maps:
\begin{gather}
\footnotesize
    \hat{I}^n = v\cdot I^{n0} + (1-v)\cdot I^{n1} + \delta. \label{eqn:defconv3}
\end{gather}
We adopt the separable deformable convolution implementation in \cite{cheng2020video} which exploits separability properties of kernels~\cite{rigamonti2013learning} to reduce memory requirements while maintaining VFI performance.

\paragraph{Training VQ-FIGAN.} We follow the original training settings of VQGAN in \cite{esser2021taming,rombach2022high}, where the loss function consists of an LPIPS-based~\cite{zhang2018unreasonable} perceptual loss, a patch-based adversarial loss~\cite{isola2017image} and a latent regularization term based on a vector quantization (VQ) layer~\cite{van2017neural}. We refer the readers to \cite{esser2021taming} for the details.

\subsection{Conditional Generation with LDM}
The trained VQ-FIGAN allows us to access a compact latent space in which we perform forward diffusion by gradually adding Gaussian noise to the latent $z^n$ of the target frame $I^n$ according to a pre-defined noise schedule, and learn the reverse (denoising) process to perform conditional generation. To this end, we adopt the noise-prediction parameterization~\cite{ho2020denoising} of DMs and train a denoising U-Net by minimizing the re-weighted variational lower bound on the conditional log-likelihood $\log p_\theta(z^n|z^0,z^1)$ where $z^0,z^1$ are the latent encodings of the two input frames. 

\paragraph{Training.} Specifically, the denoising U-Net $\epsilon_\theta$ takes as input the ``noisy'' latent encoding $z^n_t$ for the target frame $I^n$ (sampled from the $t$-th step in the forward diffusion process of length $T$), the diffusion step $t$, as well as the conditioning latents $z^0,z^1$ for the input frames $I^0,I^1$. It is trained to predict the noise added to $z^n$ at each time step $t$ by minimizing
\begin{equation}
    \mathcal{L} = \mathbb{E}_{z^n,z^0,z^1,\epsilon\sim\mathcal{N}(\mathbf{0},\mathbf{I}),t} \big[ \norm{\epsilon - \epsilon_\theta(z^n_t,t,z^0,z^1)}^2 \big],
\end{equation}
where $t\sim\mathcal{U}(1,T)$. The derivation and full details of the training procedure of $\epsilon_\theta$ are provide in Appendix~\ref{appendix:traininference}. Intuitively, the training is performed by alternately adding a random Gaussian noise to $z^n$ according to a pre-defined noising schedule, and having the network $\epsilon_\theta$ predict the noise added given the step $t$, conditioning on $z^0,z^1$.

\paragraph{Inference.} To interpolate $\hat{I}^n$ from $I^0, I^1$, we start by sampling a Gaussian noise $z^n_T$ in the latent space, and perform $T$ steps of denoising until we obtain $z^n_0$. Within each step, firstly the network $\epsilon_\theta$ predicts the noise $\hat{\epsilon}$. Then $z^n_{t-1}$ is calculated using $\hat{\epsilon}$ and relevant parameters of the pre-defined forward process. Finally, the decoder $D$ produces the interpolated frame from the denoised latent $z^n_0$, with the help of feature pyramids $\phi^0,\phi^1$ extracted by the encoder $E$ from $I^0,I^1$. Full details are provided in the Appendix~\ref{appendix:traininference}.

\paragraph{Network Architecture.} We employ the time-conditioned U-Net as in \cite{rombach2022high} for $\epsilon_\theta$, but with one modification: all the vanilla self-attention blocks~\cite{vaswani2017attention} are replaced with the aforementioned MaxViT blocks~\cite{tu2022maxvit} for computational efficiency. The conditioning mechanism for the U-Net is concatenation of $z^n_t$ and $z^0,z^1$ at the input. The architecture is detailed in Appendix~\ref{appendix:unet}.

\section{Experimental setup}

\paragraph{Implementation Details.} We set the downsampling factor of VQ-FIGAN to $f=32$, by repeating the \texttt{ResNetBlock+Conv3x3} layer in the encoder and the corresponding layer in the decoder three times (see Figure~\ref{fig:vqfinet}). The size of the kernels output by the decoder is $K=5$. Regarding the diffusion processes, following~\cite{rombach2022high}, we adopt a linear noise schedule and a codebook size of 8192 for vector quantization in VQ-FIGAN. We sample from all diffusion models with the DDIM~\cite{song2021denoising} sampler for 200 steps (details provided in Appendix~\ref{appendix:ddim}). We also follow \cite{rombach2022high} to train the VQ-FIGAN using the ADAM~\cite{kingma2014adam} optimizer and the denoising U-Net using the Adam-W optimizer~\cite{loshchilov2018decoupled}, with the initial learning rates set to $10^{-5}$ and $10^{-6}$ respectively. All models were trained until convergence, which corresponds to around 70 epochs for VQ-FIGAN, and around 60 epochs for the U-Net. NVIDIA RTX 3090 GPUs were used for all training and evaluation.

\begin{table*}[t]
\begin{center}
\resizebox{\linewidth}{!}{
\begin{tabular}{lcccccccccccccc}
    \toprule
    & \multicolumn{3}{c}{Middlebury} & \multicolumn{3}{c}{UCF-101}& \multicolumn{3}{c}{DAVIS}& \multicolumn{3}{c}{VFITex}& \multirow{2}[2]{*}{\makecell{RT \\ (sec)}} & \multirow{2}[2]{*}{\makecell{\#P \\ (M)}}\\
    \cmidrule(l{5pt}r{5pt}){2-4}\cmidrule(l{5pt}r{5pt}){5-7}\cmidrule(l{5pt}r{5pt}){8-10} \cmidrule(l{5pt}r{5pt}){11-13}
    &LPIPS$\downarrow$ & FloLPIPS$\downarrow$ & FID$\downarrow$ 
    &LPIPS$\downarrow$ & FloLPIPS$\downarrow$ & FID$\downarrow$ 
    &LPIPS$\downarrow$ & FloLPIPS$\downarrow$ & FID$\downarrow$
    &LPIPS$\downarrow$ & FloLPIPS$\downarrow$ & FID$\downarrow$ \\
    \midrule
    BMBC & 0.023 & \textbf{0.037} & 12.974 & 0.034 & 0.045 & 33.171 & 0.125 & 0.185 & 15.354 & 0.220 & 0.282 & 50.393 & 0.51 & 11.0 \\
    AdaCoF & 0.031 & 0.052 & 15.633 & 0.034 & 0.046 & 32.783 & 0.148 & 0.198 & 17.194 & 0.204 & 0.273 & 42.255 & 0.01 & 21.8 \\
    CDFI & 0.022 & 0.043 & 12.224 & 0.036 & 0.049 & 33.742 & 0.157 & 0.211 & 18.098 & 0.218 & 0.286 & 43.498 & 0.02 & 5.0 \\
    XVFI & 0.036 & 0.070 & 16.959 & 0.038 & 0.050 & 33.868 & 0.129 & 0.185 & 16.163 & 0.188 & 0.255 & 42.055 & 0.08 & 5.6 \\
    ABME & 0.027 & 0.040 & \textbf{11.393} & 0.058 & 0.069 & 37.066 & 0.151 & 0.209 & 16.931 & 0.254 & 0.341 & 53.317 & 0.27 & 18.1 \\
    IFRNet & 0.020 & 0.039 & 12.256 & 0.032 & 0.044 & 28.803 & 0.114 & 0.170 & 14.227 & 0.200 & 0.273 & 42.266 & 0.02 & 5.0 \\
    VFIformer & 0.031 & 0.065 & 15.634 & 0.039 & 0.051 & 34.112 & 0.191 & 0.242 & 21.702 & OOM & OOM & OOM & 1.74 & 5.0 \\
    ST-MFNet & N/A & N/A & N/A & 0.036 & 0.049 & 34.475 & 0.125 & 0.181 & 15.626 & 0.216 & 0.276 & 41.971 & 0.14 & 21.0 \\
    FLAVR & N/A & N/A & N/A & 0.035 & 0.046 & 31.449 & 0.209 & 0.248 & 22.663 & 0.234 & 0.295 & 56.690 & 0.02 & 42.1 \\
    MCVD & 0.123 & 0.138 & 41.053 & 0.155 & 0.169 & 102.054 & 0.247 & 0.293 & 28.002 & OOM & OOM & OOM & 52.55 & 27.3 \\
    \midrule
    LDMVFI & \textbf{0.019} & 0.044 & 16.167 & \textbf{0.026} & \textbf{0.035} & \textbf{26.301} & \textbf{0.107} & \textbf{0.153} & \textbf{12.554} & \textbf{0.150} & \textbf{0.207} & \textbf{32.316} & 8.48 & 439.0 \\
    \bottomrule
\end{tabular}
}
\caption{Quantitative comparison of LDMVFI ($f=32$) and 10 tested methods on Middlebury, UCF-101, DAVIS and VFITex. Note ST-MFNet and FLAVR require four input frames so cannot be evaluated on Middlebury dataset which contains frame triplets. For each column, the best result is \textbf{boldfaced}. The last two columns show the average run time (RT) needed to to interpolate one 480p frame, and the number of parameters (\#P) in each model.}
\label{tab:quant1}
\end{center}
\end{table*}

\begin{table*}[t]
\begin{center}
\resizebox{\linewidth}{!}{
\begin{tabular}{lcccccccccccc}
    \toprule
    & \multicolumn{3}{c}{SNU-FILM-Easy} & \multicolumn{3}{c}{SNU-FILM-Medium}& \multicolumn{3}{c}{SNU-FILM-Hard}& \multicolumn{3}{c}{SNU-FILM-Extreme}\\
    \cmidrule(l{5pt}r{5pt}){2-4}\cmidrule(l{5pt}r{5pt}){5-7}\cmidrule(l{5pt}r{5pt}){8-10}\cmidrule(l{5pt}r{5pt}){11-13}
    &LPIPS$\downarrow$ & FloLPIPS$\downarrow$ & FID$\downarrow$&LPIPS$\downarrow$ & FloLPIPS$\downarrow$ & FID$\downarrow$&LPIPS$\downarrow$ & FloLPIPS$\downarrow$ & FID$\downarrow$&LPIPS$\downarrow$ & FloLPIPS$\downarrow$ & FID$\downarrow$ \\
    \midrule
    BMBC & 0.020 & 0.031 & 6.162 & 0.034 & 0.059 & 12.272 & 0.068 & 0.118 & 25.773 & 0.145 & 0.237 & 49.519 \\
    AdaCoF & 0.021 & 0.033 & 6.587 & 0.039 & 0.066 & 14.173 & 0.080 & 0.131 & 27.982 & 0.152 & 0.234 & 52.848 \\
    CDFI & 0.019 & 0.031 & 6.133 & 0.036 & 0.066 & 12.906 & 0.081 & 0.141 & 29.087 & 0.163 & 0.255 & 53.916 \\
    XVFI & 0.022 & 0.037 & 7.401 & 0.039 & 0.072 & 16.000 & 0.075 & 0.138 & 29.483 & 0.142 & 0.233 & 54.449 \\
    ABME & 0.022 & 0.034 & 6.363 & 0.042 & 0.076 & 15.159 & 0.092 & 0.168 & 34.236 & 0.182 & 0.300 & 63.561 \\
    IFRNet & 0.019 & 0.030 & 5.939 & 0.033 & 0.058 & 12.084 & 0.065 & 0.122 & \textbf{25.436} & 0.136 & 0.229 & 50.047 \\
    ST-MFNet & 0.019 & 0.031 & 5.973 & 0.036 & 0.061 & \textbf{11.716} & 0.073 & 0.123 & 25.512 & 0.148 & 0.238 & 53.563 \\
    FLAVR & 0.022 & 0.034 & 6.320 & 0.049 & 0.077 & 15.006 & 0.112 & 0.169 & 34.746 & 0.217 & 0.303 & 72.673 \\
    MCVD & 0.199 & 0.230 & 32.246 & 0.213 & 0.243 & 37.474 & 0.250 & 0.292 & 51.529 & 0.320 & 0.385 & 83.156 \\
    \midrule
    LDMVFI & \textbf{0.014} & \textbf{0.024} & \textbf{5.752} & \textbf{0.028} & \textbf{0.053} & 12.485 & \textbf{0.060} & \textbf{0.114} & 26.520 & \textbf{0.123} & \textbf{0.204} & \textbf{47.042} \\
    \bottomrule
\end{tabular}
}
\caption{Quantitative comparison results on SNU-FILM (note VFIformer is not included because the GPU goes out of memory).}
\label{tab:quant2}
\end{center}
\end{table*}

\paragraph{Training Dataset.} We utilize the most commonly used training set in VFI, Vimeo90k~\cite{xue2019video}. However, previous works~\cite{sim2021xvfi} have discussed the limited range of motion magnitudes and the diversity of Vimeo90k. To better test the learning capability and performance of VFI methods on a wider range of scenarios, we follow \cite{danier2022st} to additionally incorporate samples from the BVI-DVC dataset~\cite{ma2020bvi}. The final training set thus comprises 64612 and 17600 frame triplets (using only the three frames in the center) from Vimeo90k-septuplets and BVI-DVC respectively. For data augmentation, we randomly crop $256\times 256$ patches and perform random flipping and temporal order reversing. It is noted that most existing works use only Vimeo90k-triplet for training, so for reference, we also provide evaluation results for LDMVFI trained on this dataset alone (see Appendix~\ref{appendix:vimeo}).

\begin{figure*}[t]
    \centering
    \includegraphics[width=\linewidth]{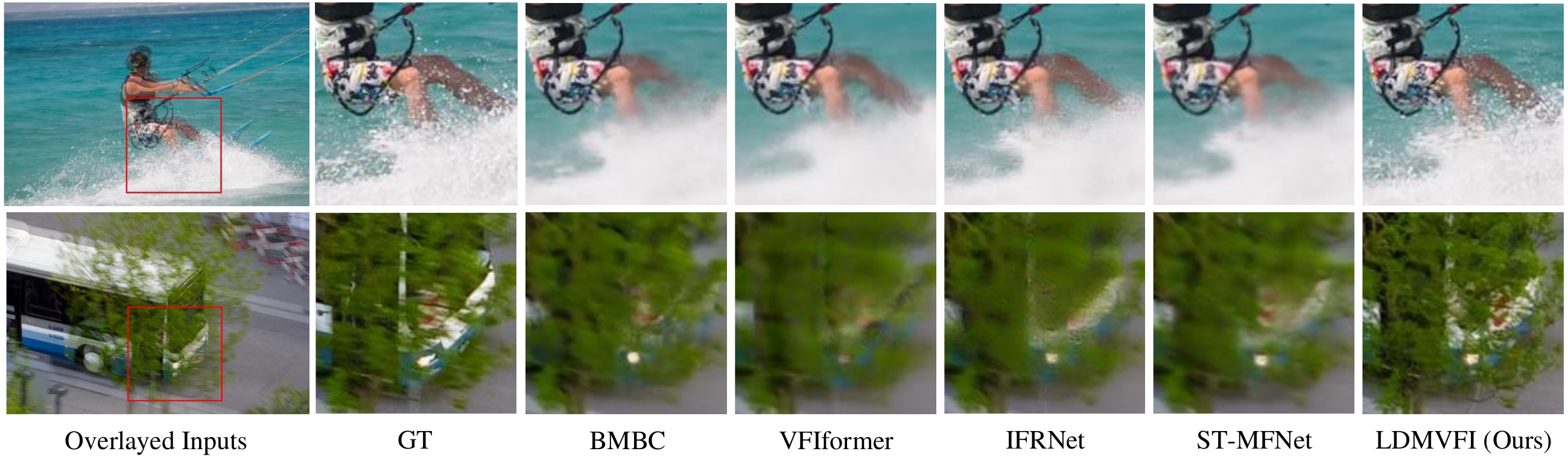}
    \caption{Visual examples of frames interpolated by the state-of-the-art methods and the proposed LDMVFI. Under large and complex motions, our method preserves the most high-frequency details, delivering superior perceptual quality.}
    \label{fig:visual}
\end{figure*}

\begin{figure}[t]
\centering
\includegraphics[width=\linewidth]{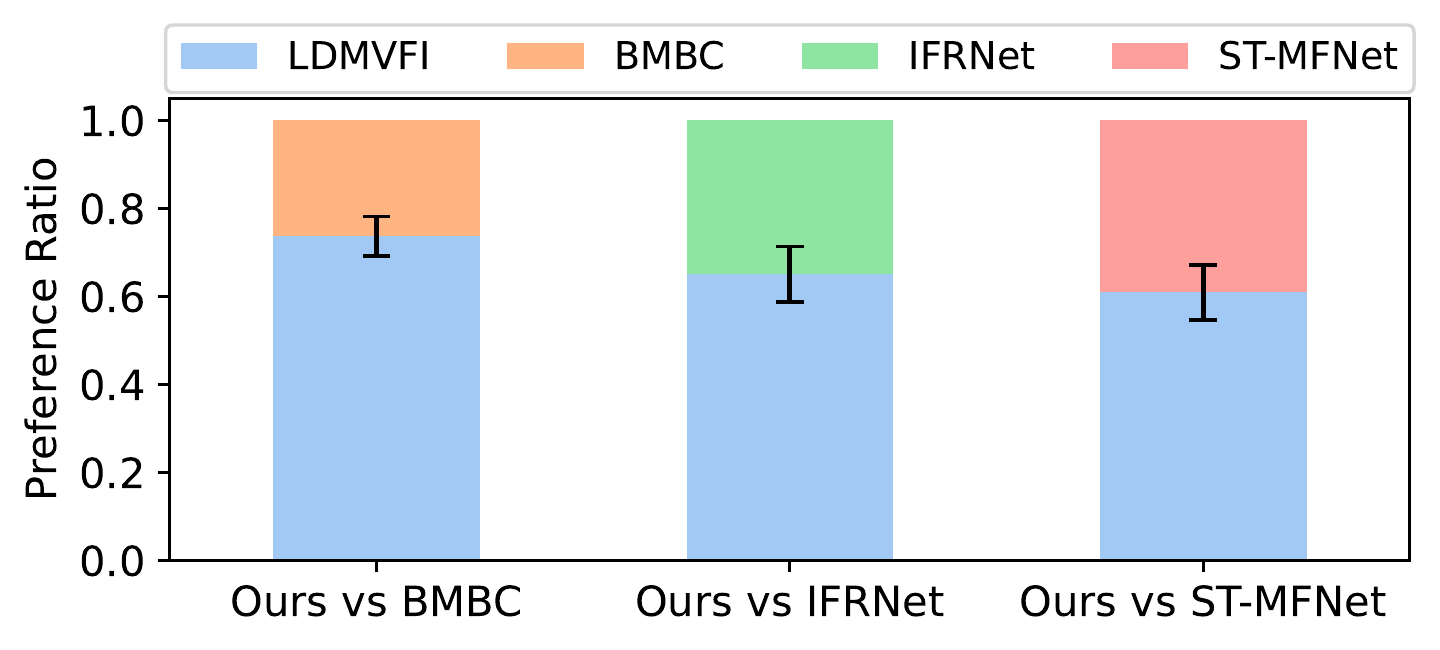}
\caption{Results of the user study in terms of preference ratio. Error bar reflects the standard error over test sequences.}
\label{fig:userstudy}
\end{figure}

\paragraph{Test Datasets.} We evaluate models on the most commonly used VFI benchmarks, including Middlebury~\cite{baker2011database}, UCF-101~\cite{soomro2012ucf101}, DAVIS~\cite{perazzi2016benchmark}, SNU-FILM~\cite{choi2020channel}, and VFITex~\cite{danier2022st}. These test sets cover resolutions from $225\times 225$ up to 4K, and various levels of VFI difficulties. To further assess the perceptual performance, we perform user study using the BVI-HFR~\cite{mackin2018study} dataset which covers a wide range of texture and motion types. 

\paragraph{Evaluation Methods.} As the main focus of this work is on improving the perceptual quality of interpolated content, we 
adopt a perceptual image quality metric LPIPS~\cite{zhang2018unreasonable}, and a bespoke VFI metric, FloLPIPS~\cite{danier2022flolpips} for performance evaluation. These metrics have shown superior correlation with human judgments of VFI quality compared to commonly used quality measurements, PSNR and SSIM~\cite{wang2004image}. We also evaluate FID~\cite{heusel2017gans} which measures the similarity between the distributions of interpolated and ground-truth frames; this was previously used as a perceptual metric for video compression~\cite{yang2022perceptual}, enhancement~\cite{yang2021ntire} and colorization~\cite{kang2023ntire}.

To measure the true perceptual performance of VFI methods, we also conducted a psychophysical experiment, in which the proposed method was compared against the state of the art (see Sec.~\ref{sec:userstudy}). For completeness, we also provide benchmark results based on PSNR and SSIM in Appendix~\ref{appendix:quantfull}, noting that these are limited in reflecting the perceptual quality of interpolated content~\cite{danier2022subjective} and are therefore not the focus of this paper.

\section{Experiments}
\subsection{Quantitative Evaluation}
The proposed LDMVFI was compared against 10 recent state-of-the-art VFI methods, including BMBC~\cite{park2020bmbc}, AdaCoF~\cite{lee2020adacof}, CDFI~\cite{ding2021cdfi}, XVFI~\cite{sim2021xvfi}, ABME~\cite{park2021asymmetric}, IFRNet~\cite{kong2022ifrnet}, VFIformer~\cite{lu2022video}, ST-MFNet~\cite{danier2022st}, FLAVR~\cite{kalluri2023flavr}, and MCVD~\cite{voleti2022mcvd}. It it noted that MCVD is the only existing diffusion-based VFI method. All these models were re-trained on our training dataset for fair comparison.

\paragraph{Performance.} Table~\ref{tab:quant1} shows the performance of the evaluated methods on the Middlebury, UCF-101, DAVIS, and VFITex test sets. It should be noted that the FID scores on Middlebury and UCF-101 might be unreliable because they contain very few test frames (12 and 100 respectively). It can be observed from the table that LDMVFI outperforms all the other VFI methods in most cases, and the performance gain against the second best method is most significant (approx. 20\%) on VFITex which contains mainly dynamic textures (e.g. fire, water, foliage) and exhibits complex motions (see further discussion on this in Appendix~\ref{appendix:ablation}). The model performance on the four splits of the SNU-FILM dataset are summarized in Table~\ref{tab:quant2}, which again demonstrates the favorable perceptual quality of videos interpolated by LDMVFI. It is also significant that the other diffusion-based VFI method, MCVD, does not perform satisfactorily overall, which implies that directly applying the original diffusion model formulation to VFI is not sufficient to enhance performance. This further shows the effectiveness of LDMVFI. Multi-frame (i.e. $\times$4) interpolation results are provided in Appendix~\ref{appendix:x4}.

\paragraph{Complexity.} The average time taken to interpolate a 480p frame on an RTX 3090 GPU and the number of parameters of each model are shown in Table~\ref{tab:quant1} (the last two columns). It is observed that the inference speed of LDMVFI is much lower compared to other methods, and this is mainly due to the iterative denoising operation performed during sampling. This is a common drawback of existing diffusion models. Various methods have been proposed to speed up sampling process of diffusion models~\cite{karras2022elucidating}, which can also be applied to LDMVFI. The number of parameters in LDMVFI is also large, and this is because we adopted (with some modifications, see Sec.~\ref{sec:vqfigan}) the existing denoising U-Net~\cite{rombach2022high} designed for generic image generation. We leave the design of a more efficient denoising U-Net and the improvement of LDMVFI sampling speed as future work. See more discussion on the limitations in Appendix~\ref{appendix:limitation}.

\subsection{Subjective Experiment}\label{sec:userstudy}

To further confirm the superior perceptual quality of the videos interpolated by LDMVFI compared to the state of the art, and also to measure its temporal consistency, we conducted a subjective experiment where human participants were hired to rate the quality of videos interpolated by ours and competing methods. 

\paragraph{Test Videos.} We use the 22 high-quality full HD 30fps videos from the BVI-HFR~\cite{mackin2018study} dataset as source content. These videos cover a wide range of video features related to motion and texture as shown by the analysis in the original paper~\cite{mackin2018study}, allowing for a more thorough benchmarking of VFI methods. To generate the test content, the 22 videos were first truncated to 5 seconds (150 frames) following \cite{moss2015optimal}. Then we used four different VFI methods to interpolate all videos to 60fps. Other than LDMVFI, the tested methods include ST-MFNet, IFRNet and BMBC, which showed the most competitive quantitative performance on the more challenging test sets (e.g. DAVIS, SNU-FILM-extreme). As a result, we obtain 88 test videos generated by the four VFI methods.

\begin{table*}[t]
\begin{center}
\resizebox{\linewidth}{!}{
\begin{tabular}{lcccccccccccc}
    \toprule
    & \multirow{2}[2]{*}{$f$} & \multirow{2}[2]{*}{\makecell{AE\\ Model}} & \multirow{2}[2]{*}{\makecell{Cond.\\ Mode}} & \multicolumn{3}{c}{Middlebury} & \multicolumn{3}{c}{UCF-101}& \multicolumn{3}{c}{DAVIS}\\
    \cmidrule(l{5pt}r{5pt}){5-7}\cmidrule(l{5pt}r{5pt}){8-10}\cmidrule(l{5pt}r{5pt}){11-13}
    & & & &LPIPS$\downarrow$ & FloLPIPS$\downarrow$ & FID$\downarrow$ &LPIPS$\downarrow$ & FloLPIPS$\downarrow$ & FID$\downarrow$ &LPIPS$\downarrow$ & FloLPIPS$\downarrow$ & FID$\downarrow$ \\
    \midrule
    V1 & 32 & frame & concat & 0.077 & 0.085 & 40.399 & 0.063 & 0.067 & 60.742 & 0.168 & 0.200 & 21.812 \\
    V2 & 32 & MaxCA+frame & concat & 0.028 & 0.045 & 19.485 & 0.032 & 0.041 & 29.578 & 0.135 & 0.176 & 19.980 \\
    \midrule
    V3 & 8 & MaxCA+kernel & concat & 0.018 & 0.046 & 19.552 & 0.028 & 0.036 & 27.278 & 0.125 & 0.168 & 15.535 \\
    V4 & 16 & MaxCA+kernel & concat & 0.017 & 0.037 & 12.539 & 0.026 & 0.035 & 26.805 & 0.107 & 0.154 & 12.720 \\
    V5 & 64 & MaxCA+kernel & concat & 0.022 & 0.044 & 15.041 & 0.026 & 0.036 & 26.055 & 0.114 & 0.157 & 12.330 \\
    \midrule
    V6 & 32 & MaxCA+kernel & N/A & 0.100 & 0.108 & 38.538 & 0.042 & 0.053 & 32.189 & 0.193 & 0.214 & 8.639 \\
    \midrule
    Ours & 32 & MaxCA+kernel & concat & 0.019 & 0.044 & 16.167 & 0.026 & 0.035 & 26.301 & 0.107 & 0.153 & 12.554 \\
    \bottomrule
\end{tabular}
}
\caption{Ablation experiment results, showing performance of variants of the proposed LDMVFI.}
\label{tab:ablation}
\end{center}
\end{table*}

\paragraph{Test Methodology.} Following previous works~\cite{liu2017video,niklaus2018context,kalluri2023flavr}, the 2AFC approach is adopted for the subjective experiment, where the participant is asked to choose the one with better perceived quality from a pair of videos. Specifically, in each test session, the participant was displayed 66 pairs of videos where one video in each pair is interpolated by LDMVFI and the other is interpolated by ST-MFNet, IFRNet or BMBC. The display order of the 66 pairs, and the order of test videos within each pair are both randomized. The user is unaware of the methods used for generating the videos. Each pair is presented twice to the participant, who is then asked to provide an answer to the question ``which of the two videos is of higher quality?''. Twenty participants were hired in total. See Appendix~\ref{appendix:userstudy} for more details.

\paragraph{Results.} After collecting all the user data, for each of the 22 source sequences, the ratio of users that preferred LDMVFI is calculated. Figure~\ref{fig:userstudy} reports the average preference ratios for LDMVFI and the standard error over the sequences. It can be seen that in all comparisons, LDMVFI achieved higher preference ratios. T-test analysis (see Appendix~\ref{appendix:userstudy}) on the sequence-wise preference ratios shows that the advantage of LDMVFI over the other three tested methods is statistically significant at 95\% confidence level. These results further confirm the superior perceptual performance of LDMVFI.

\paragraph{Visual Examples.} Figure~\ref{fig:visual} shows example frames interpolated by LDMVFI and the competing methods, clearing demonstrating that the LDMVFI results have the best visual quality. See more visual examples in Appendix~\ref{appendix:visual}.

\subsection{Ablation Study}\label{sec:ablation}
In this section we experimentally validate and study different components and hyper-parameters in LDMVFI. The ablation study results are summarized in Table~\ref{tab:ablation}, which shows evaluation results of 6 variants of the proposed model on three test sets. See Appendix~\ref{appendix:ablation} for full ablation study results, where more aspects of the model are analyzed, including discussions on the variability of LDMVFI results.

\paragraph{Effectiveness of VQ-FIGAN.} To validate the effectiveness of the proposed VQ-FIGAN design, we tested two variants of the model: V1 and V2. V2 outputs the frame $\hat{I}^n$ directly instead of predicting the deformable kernels, i.e. the convolutional heads after the last \texttt{Conv3x3} layer are removed (see Figure~\ref{fig:vqfinet}). For V1, a further change is made by removing the feature-aided reconstruction process that involves $\phi^0,\phi^1$ and replacing \texttt{MaxCABlock}s in the decoder with \texttt{ResNetBlock}s. As such, there is no information from the neighbor frames during decoding. Note that V1 is similar to the original VQGAN~\cite{esser2021taming}. Table~\ref{tab:ablation} shows that without the deformable convolution-based synthesis, the performance of V2 sees an evident decrease. Furthermore, V1 shows a more severe drop in performance indicating the effectiveness of using features of neighbor frames during reconstruction.

\paragraph{Downsampling Factor $f$.} Here we study how the dimension of the latent space affects the VFI performance. Specifically, to obtain V4 ($f=16$), we remove one \texttt{ResNetBlock+Conv3x3} layer and one \texttt{ResNetBlock+MaxCABlock+Upsample+Conv3x3} layer from the encoder and decoder of VQ-FIGAN respectively (see Figure~\ref{fig:vqfinet}). We repeat this process once to obtain V3 ($f=8$). To create V5 ($f=64$) we add these layers instead of removing them. Table~\ref{tab:ablation} shows that as $f$ increases from $8$ to $32$, there is generally an increasing trend in model performance (except that V4 outperformed LDMVFI on Middlebury). Such improvement is more obvious on DAVIS which mainly contains challenging large motion content. However, looking at $f=64$, the general performance deteriorates (in terms of LPIPS and FloLPIPS). The reason for this can be that to a reasonable extent, increasing $f$ allows the VQ-FIGAN decoder to make use of more information from the neighboring frames which can benefit frame interpolation, while preserving sufficient information for conditional generation in the latent space. However, if the downsampling is too aggressive, the information needed to perform reverse latent diffusion can be insufficient, resulting in degraded quality of latent generation. A similar trade-off between downsampling factor and model performance was also observed in \cite{rombach2022high}.

\paragraph{Effectiveness of Diffusion Process.} Here (V6) the denoising U-Net, hence the reverse diffusion process, is removed, and the decoder directly takes as input the concatenation of $z^0$ and $z^1$. Table~\ref{tab:ablation} shows that without the diffusion process, the performance drops significantly, implying the effectiveness of the latent diffusion.

\section{Conclusion}
In this work we propose LDMVFI, the first approach that addresses video frame interpolation as a conditional generation problem using latent diffusion models. It contains two major components: an autoencoding model that provides access to a compact latent space, and a denoising U-Net that performs reverse diffusion on latent representations. To leverage latent diffusion models for VFI, we present several innovative designs including a VFI-specific autoencoding network, VQ-FIGAN, which employs efficient self-attention modules and deformable kernel-based frame synthesis techniques. LDMVFI was comprehensively evaluated on a wide range of test sets (including 4K content) using both quantitative metrics and subjective experiments. The results demonstrate its favorable perceptual performance over the state of the art.

\paragraph{Acknowledgements.}
This work was supported in part by the China Scholarship Council, in part by the University of Bristol, and in part by the UK Research and Innovation (UKRI) MyWorld Strength in Places Program. 

\bibliography{aaai24}

\appendix
\onecolumn

\section{Details of LDMVFI Loss Function (Diffusion Part)}\label{appendix:loss}

In this section, we derive the training objective of the denoising U-Net in LDMVFI which is responsible for performing conditional generation. As mentioned in the main paper, diffusion models consist of a forward diffusion and a reverse denoising process. The forward diffusion process is defined by a Markov chain that gradually adds noise to a ``clean'' image $x_0$ using a pre-defined noise schedule $\{\beta_t\}_{t=1}^T$ in $T$ steps, with conditional probabilities
\begin{gather}
    q(x_t|x_{t-1}) = \mathcal{N}(x_t; \sqrt{1-\beta_t}x_{t-1}, \beta_t \mathbf{I}), \\
    \Rightarrow \quad q(x_{1:T}|x_0)=\prod_{t=1}^T q(x_t|x_{t-1}).
\end{gather}
Let $\alpha_t=1-\beta_t$ and $\bar{\alpha}_t = \prod_{i=1}^t\alpha_i$, according to the conditional independence property of Markov chain, one can sample from the forward diffusion process at an arbitrary time step $t$ with
\begin{equation}
    q(x_t|x_0) = \mathcal{N}(x_t; \sqrt{\bar{\alpha}_t}x_0, (1-\bar{\alpha}_t)\mathbf{I}).
\end{equation}
In order to sample $x_t$ from this distribution in practice, we can use a reparameterization
\begin{equation}
    x_t = \sqrt{\bar{\alpha}_t}x_0 + \sqrt{1-\bar{\alpha}_t}\epsilon \quad \text{where} \quad \epsilon \sim \mathcal{N}(\mathbf{0}, \mathbf{I}). \label{eqn:xt}
\end{equation}
Here the noise schedule $\{\beta_t\}_{t=1}^T$ is designed such that $\bar{\alpha}_T \approx 0$ and $q(x_T|x_0)\approx \mathcal{N}(x_T; \mathbf{0}, \mathbf{I})$. That is, as the forward diffusion process comes to an end, the last state of the image becomes close to a pure Gaussian noise.

Given the forward diffusion process, one could generate new samples by starting from pure Gaussian noise and sampling from the reverse conditionals $q(x_{t-1}|x_t)$. However, $q(x_{t-1}|x_t)$ is intractable, so one can use a $\theta$-parameterized Gaussian distribution 
\begin{gather}
    p_\theta(x_{t-1}|x_t) = \mathcal{N}(x_{t-1}; \mu_\theta(x_t, t), \sigma_t^2\mathbf{I}), \\
    \Rightarrow \quad p_\theta(x_{0:T}) = p(x_T)\prod_{t=1}^T p_\theta(x_{t-1}|x_t),
\end{gather} 
to approximate the reverse Markov chain (provided that $\beta_t$ is sufficiently small in each forward step~\cite{sohl2015deep}). Here $\mu_\theta$ corresponds to a neural network, and $\sigma_t^2$ can be set to a value based on $\beta_t$~\cite{ho2020denoising,li2022srdiff,rombach2022high}. One can then derive~\cite{yang2022diffusion,ho2020denoising} the variational lower bound on the data log-likelihood:
\begin{equation}
    \mathbb{E}_{q(x_0)}[-\log p_\theta(x_0)] \leq \mathbb{E}_{q(x_0)q(x_{1:T}|x_0)} \Big[ -\log \frac{p_\theta(x_{0:T})}{q(x_{1:T}|x_0)} \Big] =: L,
\end{equation}
and training can be done by minimizing $L$. It can be shown \cite{sohl2015deep,ho2020denoising} that $L$ can be decomposed as
\begin{equation}
    L=\mathbb{E}_{q} \Big[ D_\mathrm{KL}\big(q(x_T|x_0)||p_\theta(x_T)\big) + \sum_{t=2}^T D_\mathrm{KL}\big(q(x_{t-1}|x_t,x_0)||p_\theta(x_{t-1}|x_t)\big) - \log p_\theta(x_0|x_1) \Big]. \label{eqn:vlb}
\end{equation}
The first term in (\ref{eqn:vlb}) can be ignored because the prior $p_\theta(x_T)$ can be set to a standard normal distribution, and the last term was handled by a separate decoder in \cite{ho2020denoising}, leaving the second term as the main focus for learning the reverse diffusion process. The $q(x_{t-1}|x_t,x_0)$ in the second term is tractable and it can be derived as
\begin{equation}
    q(x_{t-1}|x_t,x_0) = \mathcal{N}(x_{t-1}; \tilde{\mu}_t(x_t,x_0), \tilde{\beta}_t\mathbf{I}),
\end{equation}
where 
\begin{gather}
    \tilde{\mu}_t(x_t,x_0) = \frac{\sqrt{\bar{\alpha}_{t-1}}\beta_t}{1-\bar{\alpha}_t} x_0 +\frac{\sqrt{\alpha_t}(1-\bar{\alpha}_{t-1})}{1-\bar{\alpha}_t}x_t,\\
    \tilde{\beta}_t=\frac{1-\bar{\alpha}_{t-1}}{1-\bar{\alpha}_t}\beta_t.
\end{gather}
Since both $q(x_{t-1}|x_t,x_0)$ and $p_\theta(x_{t-1}|x_t)$ are Gaussian distributions, the KL-divergence in the second term in (\ref{eqn:vlb}) takes the form
\begin{equation}
    L_{t-1}=:D_\mathrm{KL}\big( q(x_{t-1}|x_t,x_0)||p_\theta(x_{t-1}|x_t) \big) = \mathbb{E}_q \Big[ \frac{1}{2\sigma_t^2}\norm{\tilde{\mu}(x_t,x_0) - \mu_\theta(x_t,t)}^2 \Big]. \label{eqn:lt-1}
\end{equation}
In \cite{ho2020denoising}, it is noted that plugging (\ref{eqn:xt}) into $\tilde{\mu}(x_t,x_0)$, the latter can be written as
\begin{equation}
    \tilde{\mu}(x_t,x_0) = \frac{1}{\sqrt{\alpha_t}}\big( x_t - \frac{1-\alpha_t}{\sqrt{1-\bar{\alpha}_t}}\epsilon \big) \label{eqn:mutilde}.
\end{equation}
Therefore, it is proposed that instead of predicting $\mu_\theta(x_t,t)$ directly, we can predict the noise $\epsilon_\theta(x_t,t)$ then infer $\mu_\theta(x_t,t)$ with
\begin{equation}
    \mu_\theta(x_t,t) = \frac{1}{\sqrt{\alpha_t}}\big( x_t - \frac{1-\alpha_t}{\sqrt{1-\bar{\alpha}_t}}\epsilon_\theta(x_t,t) \big). \label{eqn:mutheta}
\end{equation}
Plugging in (\ref{eqn:mutilde}) and (\ref{eqn:mutheta}) into (\ref{eqn:lt-1}), the loss then reads
\begin{align}
        L_{t-1}  & = \mathbb{E}_{x_0\sim q(x_0),\epsilon\sim \mathcal{N}(\mathbf{0},\mathbf{I})} \Big[ \frac{\beta_t^2}{2\sigma_t^2\alpha_t(1-\bar{\alpha}_t)} \norm{\epsilon - \epsilon_\theta(x_t,t)}^2\Big] + C \\
         & = \mathbb{E}_{x_0\sim q(x_0),\epsilon\sim \mathcal{N}(\mathbf{0},\mathbf{I})} \Big[ \frac{\beta_t^2}{2\sigma_t^2\alpha_t(1-\bar{\alpha}_t)} \norm{\epsilon - \epsilon_\theta(\sqrt{\bar{\alpha}_t}x_0 + \sqrt{1-\bar{\alpha}_t}\epsilon,t)}^2\Big] + C, \label{eqn:ddpmloss}
\end{align}
where $C$ absorbs terms independent of $\theta$. It was observed in \cite{ho2020denoising} that setting the multiplicative term before the norm in (\ref{eqn:ddpmloss}) to 1 provides improved performance, i.e.
\begin{equation}
    L_\mathrm{DM} = \mathbb{E}_{x_0,\epsilon\sim\mathcal{N}(\mathbf{0},\mathbf{I}),t\sim\mathcal{U}(1,T)} \big[ \norm{\epsilon - \epsilon_\theta(x_t,t)}^2 \big],
\end{equation}
which corresponds to a re-weighted version of the variational lower bound. Latent diffusion models (LDMs) adopt the similar overall framework derived above, but performs the diffusion processes in a lower-dimensional latent space provided by an autoencoding model that contains an encoder $E:x\mapsto z$ and a decoder $D:z\mapsto x$. Accordingly, the noise predictor $\epsilon_\theta$ in LDMs are trained to optimize
\begin{equation}
    L_\mathrm{LDM} = \mathbb{E}_{E(x_0),\epsilon\sim\mathcal{N}(\mathbf{0},\mathbf{I}),t\sim\mathcal{U}(1,T)} \big[ \norm{\epsilon - \epsilon_\theta(z_t,t)}^2 \big].
\end{equation}
In LDMVFI, the noise prediction also conditions on the latent encodings $z^0=E(I^0),z^1=E(I^1)$ of the two input frames $I^0,I^1$, and the loss we use becomes
\begin{align}
    \mathcal{L}  & = \mathbb{E}_{z^n,z^0,z^1,\epsilon\sim\mathcal{N}(\mathbf{0},\mathbf{I}),t\sim\mathcal{U}(1,T)} \big[ \norm{\epsilon - \epsilon_\theta(z^n_t,t,z^0,z^1)}^2 \big] \\
     & = \mathbb{E}_{z^n,z^0,z^1,\epsilon\sim\mathcal{N}(\mathbf{0},\mathbf{I}),t\sim\mathcal{U}(1,T)} \big[ \norm{\epsilon - \epsilon_\theta(\sqrt{\bar{\alpha}_t}z^n_0 + \sqrt{1-\bar{\alpha}_t}\epsilon,t,z^0,z^1)}^2 \big].
\end{align}

\section{Details of LDMVFI Training and Inference}\label{appendix:traininference}
The full training and inference algorithms are summarized in Algorithm~\ref{alg:trainfull} and \ref{alg:inferencefull}. Here the symbols and equations refer to those derived in Section~\ref{appendix:loss} above.

\section{Details of DDIM Sampling Process}\label{appendix:ddim}
As stated in the main paper, in order  to sample from LDMVFI and other diffusion models, we use the DDIM~\cite{song2021denoising} sampler, which has been shown to achieve sampling quality on par with the full original sampling method (Algorithm~\ref{alg:inferencefull}), but with fewer steps. We refer the reader to the original paper for details on the derivation and design of DDIM, and present the DDIM sampling procedure for LDMVFI in Algorithm~\ref{alg:ddim}.

\begin{figure}[t]
\begin{minipage}[t]{\textwidth}
\begin{algorithm}[H]
\caption{Training} \label{alg:trainfull}
\small
\begin{algorithmic}[1]
    \STATE\textbf{Input}: dataset $\mathcal{D}=\{I^0_s, I^n_s, I^1_s\}_{s=1}^S$ of consecutive frame triplets, maximum diffusion step $T$, noise schedule $\{\beta_t\}_{t=1}^T$
    \STATE\textbf{Load}: pre-trained VQ-FIGAN encoder $E$
    \STATE\textbf{Initialize}: denoising U-Net $\epsilon_\theta$
    \STATE Compute $\{\bar{\alpha}_t\}_{t=1}^T$ from $\{\beta_t\}_{t=1}^T$
    \REPEAT
      \STATE Sample $(I^0, I^n, I^1) \sim \mathcal{D}$
      \STATE Encode $z^0=E(I^0), z^n=E(I^n), z^1=E(I^1)$
      \STATE Sample $t\sim\mathcal{U}(1,T)$
      \STATE Sample $\epsilon\sim\mathcal{N}(\mathbf{0}, \mathbf{I})$
      \STATE $z^n_t = \sqrt{\bar{\alpha}_t}z^n + \sqrt{1-\bar{\alpha}_t}\epsilon$ 
      \STATE Take a gradient descent step on \\ \quad\quad\quad\quad\quad\quad $\nabla_\theta\norm{\epsilon - \epsilon_\theta(z^n_t,t,z^0,z^1)}^2$
    \UNTIL{converged}
\end{algorithmic}
\end{algorithm}
\end{minipage}
\end{figure}

\begin{figure}[t]
\centering
\begin{minipage}[t]{\textwidth}
\begin{algorithm}[H]
  \caption{Inference} \label{alg:inferencefull}
  \small
  \begin{algorithmic}[1]
    \STATE\textbf{Input}: original frames $I^0,I^1$, noise schedule $\{\beta_t\}_{t=1}^T$, maximum diffusion step $T$
    \STATE\textbf{Load}: pre-trained denoising U-Net $\epsilon_\theta$, VQ-FIGAN encoder $E$ and decoder $D$
    \STATE Compute $\{\bar{\alpha}_t\}_{t=1}^T$ from $\{\beta_t\}_{t=1}^T$
    \STATE Sample $z^n_T \sim \mathcal{N}(\mathbf{0}, \mathbf{I})$
    \STATE Encode $z^0=E(I^0), z^1=E(I^1)$ and store features $\phi^0, \phi^1$ extracted by $E$
    \FOR{$t=T,\dots, 1$}
      \STATE Predict noise $\hat{\epsilon} = \epsilon_\theta(z^n_t,t,z^0,z^1)$
      \STATE $\mu_\theta = \frac{1}{\sqrt{\alpha_t}}\big( z^n_t - \frac{1-\alpha_t}{\sqrt{1-\bar{\alpha}_t}}\hat{\epsilon} \big)$
      \STATE $\sigma^2_t = \tilde{\beta}_t$
      \STATE Sample $\zeta \sim \mathcal{N}(\mathbf{0}, \mathbf{I})$
      \STATE $z^n_{t-1}=\mu_\theta + \sigma_t \zeta$
    \ENDFOR
    \STATE \textbf{return} $\hat{I}^n=D(z^n_0,\phi^0,\phi^1)$ as the interpolated frame
  \end{algorithmic}
\end{algorithm}
\end{minipage}
\end{figure}

\begin{figure}[t]
\centering
\begin{minipage}[t]{\textwidth}
\begin{algorithm}[H]
  \caption{DDIM Sampling for LDMVFI} \label{alg:ddim}
  \small
  \begin{algorithmic}[1]
    \STATE\textbf{Input}: original frames $I^0,I^1$, noise schedule $\{\beta_t\}_{t=1}^T$, maximum DDIM step $\mathcal{T}$
    \STATE\textbf{Load}: pre-trained denoising U-Net $\epsilon_\theta$, VQ-FIGAN encoder $E$ and decoder $D$
    \STATE Compute $\{\bar{\alpha}_t\}_{t=1}^T$ from $\{\beta_t\}_{t=1}^T$
    \STATE Sample $z^n_\mathcal{T} \sim \mathcal{N}(\mathbf{0}, \mathbf{I})$
    \STATE Encode $z^0=E(I^0), z^1=E(I^1)$ and store features $\phi^0, \phi^1$ extracted by $E$
    \FOR{$t=\mathcal{T},\dots, 1$}
      \STATE Predict noise $\hat{\epsilon} = \epsilon_\theta(z^n_t,t,z^0,z^1)$
      \STATE $\hat{z}^n_0 = \frac{1}{\sqrt{\bar{\alpha}_t}}(z^n_t - \sqrt{1-\bar{\alpha}_t}\hat{\epsilon})$
      \STATE $ z^n_{t-1} = \sqrt{\bar{\alpha}_{t-1}}\hat{z}^n_0 + \sqrt{1 - \bar{\alpha}_{t-1}} \hat{\epsilon}$
    \ENDFOR
    \STATE \textbf{return} $\hat{I}^n=D(z^n_0,\phi^0,\phi^1)$ as the interpolated frame
  \end{algorithmic}
\end{algorithm}
\end{minipage}
\end{figure}

\clearpage
\section{Denoising U-Net Architecture}\label{appendix:unet}
\begin{figure}[t]
\centering
\includegraphics[width=0.5\linewidth]{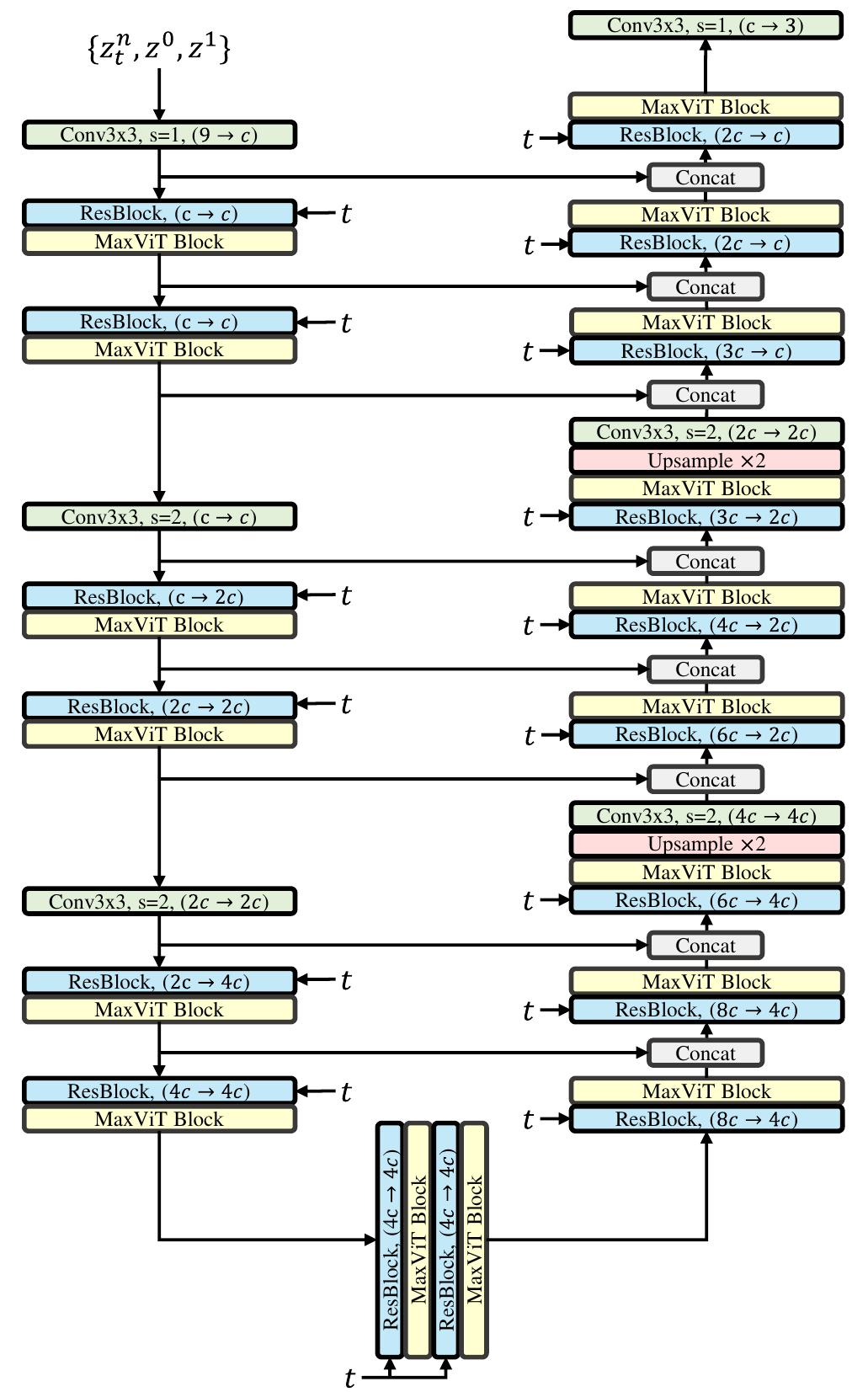}
\caption{The architecture of the denoising U-Net. The hyper-parameter $c$ is a base channel size, which is set to 256 in LDMVFI. In each block, the $(\cdot \rightarrow \cdot)$ indicates the input and output channels of the block.}
\label{fig:unet}
\end{figure}

The high-level architecture of the denoising U-Net used in LDMVFI is shown in Figure~\ref{fig:unet}. As stated in the main paper, one modification from its original version in \cite{rombach2022high} is to replace all vanilla self-attention layers with the MaxViT blocks~\cite{tu2022maxvit}. Figure~\ref{fig:unet} demonstrates how the hyper-parameter $c$ (the base channel size) affects the overall model size. We refer the reader to the original papers~\cite{ho2020denoising,dhariwal2021diffusion,rombach2022high} and to our code for full details of the U-Net. 

\section{Full Quantitative Evaluation Results}\label{appendix:quantfull}
The full evaluation results of LDMVFI and the compared VFI methods on all test sets (Middlebury~\cite{baker2011database}, UCF-101~\cite{soomro2012ucf101}, DAVIS~\cite{perazzi2016benchmark}, VFITex and SNU-FILM~\cite{choi2020channel}) in terms of all metrics (PSNR, SSIM~\cite{wang2004image}, LPIPS~\cite{zhang2018unreasonable}, FloLPIPS~\cite{danier2022flolpips} and FID~\cite{heusel2017gans}) are summarized in Table~\ref{tab:quantfull}.

\begin{table}[H]
\begin{center}\footnotesize
\begin{tabular}{lcccccccccc}
    \toprule
    & \multicolumn{5}{c}{Middlebury} & \multicolumn{5}{c}{UCF-101}\\
    \cmidrule(l{5pt}r{5pt}){2-6}\cmidrule(l{5pt}r{5pt}){7-11}
     & PSNR$\uparrow$ & SSIM$\uparrow$ & LPIPS$\downarrow$ & FloLPIPS$\downarrow$ & FID$\downarrow$ & PSNR$\uparrow$ & SSIM$\uparrow$ & LPIPS$\downarrow$ & FloLPIPS$\downarrow$ & FID$\downarrow$  \\
    \midrule
    BMBC & 36.368 & 0.982 & 0.023 & 0.037 & 12.974 & 32.576 & 0.968 & 0.034 & 0.045 & 33.171 \\
    AdaCoF & 35.256 & 0.975 & 0.031 & 0.052 & 15.633 & 32.488 & 0.968 & 0.034 & 0.046 & 32.783 \\
    CDFI & 36.205 & 0.981 & 0.022 & 0.043 & 12.224 & 32.541 & 0.968 & 0.036 & 0.049 & 33.742 \\
    XVFI & 34.724 & 0.975 & 0.036 & 0.070 & 16.959 & 32.224 & 0.966 & 0.038 & 0.050 & 33.868 \\
    ABME & 37.639 & 0.986 & 0.027 & 0.040 & 11.393 & 32.055 & 0.967 & 0.058 & 0.069 & 37.066 \\
    IFRNet & 36.368 & 0.983 & 0.020 & 0.039 & 12.256 & 32.716 & 0.969 & 0.032 & 0.044 & 28.803 \\
    VFIformer & 35.566 & 0.977 & 0.031 & 0.065 & 15.634 & 32.745 & 0.968 & 0.039 & 0.051 & 34.112 \\
    ST-MFNet & N/A & N/A & N/A & N/A & N/A & 33.383 & 0.970 & 0.036 & 0.049 & 34.475 \\
    FLAVR & N/A & N/A & N/A & N/A & N/A & 33.224 & 0.969 & 0.035 & 0.046 & 31.449 \\
    MCVD & 20.539 & 0.820 & 0.123 & 0.138 & 41.053 & 18.775 & 0.710 & 0.155 & 0.169 & 102.054 \\
    \midrule
    LDMVFI & 34.033 & 0.971 & 0.019 & 0.044 & 16.167 & 32.186 & 0.963 & 0.026 & 0.035 & 26.301 \\
    \bottomrule
\end{tabular}
\begin{tabular}{lcccccccccc}
    \toprule
    & \multicolumn{5}{c}{DAVIS} & \multicolumn{5}{c}{VFITex}\\
    \cmidrule(l{5pt}r{5pt}){2-6}\cmidrule(l{5pt}r{5pt}){7-11}
     & PSNR$\uparrow$ & SSIM$\uparrow$ & LPIPS$\downarrow$ & FloLPIPS$\downarrow$ & FID$\downarrow$ & PSNR$\uparrow$ & SSIM$\uparrow$ & LPIPS$\downarrow$ & FloLPIPS$\downarrow$ & FID$\downarrow$  \\
    \midrule
    BMBC & 26.835 & 0.869 & 0.125 & 0.185 & 15.354 & 27.337 & 0.904 & 0.220 & 0.282 & 50.393 \\
    AdaCoF & 26.234 & 0.850 & 0.148 & 0.198 & 17.194 & 27.394 & 0.904 & 0.204 & 0.273 & 42.255 \\
    CDFI & 26.471 & 0.857 & 0.157 & 0.211 & 18.098 & 27.577 & 0.906 & 0.218 & 0.286 & 43.498 \\
    XVFI & 26.475 & 0.861 & 0.129 & 0.185 & 16.163 & 27.625 & 0.907 & 0.188 & 0.255 & 42.055 \\
    ABME & 26.861 & 0.865 & 0.151 & 0.209 & 16.931 & 26.765 & 0.901 & 0.254 & 0.341 & 53.317 \\
    IFRNet & 27.313 & 0.877 & 0.114 & 0.170 & 14.227 & 27.770 & 0.909 & 0.200 & 0.273 & 42.266 \\
    VFIformer & 26.241 & 0.850 & 0.191 & 0.242 & 21.702 & OOM & OOM & OOM & OOM & OOM \\
    ST-MFNet & 28.287 & 0.895 & 0.125 & 0.181 & 15.626 & 29.175 & 0.929 & 0.216 & 0.276 & 41.971 \\
    FLAVR & 27.104 & 0.862 & 0.209 & 0.248 & 22.663 & 28.471 & 0.915 & 0.234 & 0.295 & 56.690 \\
    MCVD & 18.946 & 0.705 & 0.247 & 0.293 & 28.002 & OOM & OOM & OOM & OOM & OOM \\
    \midrule
    LDMVFI & 25.541 & 0.833 & 0.107 & 0.153 & 12.554 & 27.001 & 0.891 & 0.150 & 0.207 & 32.316 \\
    \bottomrule
\end{tabular}
% snufilm
\begin{tabular}{lcccccccccc}
    \toprule
    & \multicolumn{5}{c}{SNU-FILM-Easy} & \multicolumn{5}{c}{SNU-FILM-Medium}\\
    \cmidrule(l{5pt}r{5pt}){2-6}\cmidrule(l{5pt}r{5pt}){7-11}
     & PSNR$\uparrow$ & SSIM$\uparrow$ & LPIPS$\downarrow$ & FloLPIPS$\downarrow$ & FID$\downarrow$ & PSNR$\uparrow$ & SSIM$\uparrow$ & LPIPS$\downarrow$ & FloLPIPS$\downarrow$ & FID$\downarrow$  \\
    \midrule
    BMBC & 39.809 & 0.990 & 0.020 & 0.031 & 6.162 & 35.437 & 0.978 & 0.034 & 0.059 & 12.272 \\
    AdaCoF & 39.632 & 0.990 & 0.021 & 0.033 & 6.587 & 34.919 & 0.975 & 0.039 & 0.066 & 14.173 \\
    CDFI & 39.881 & 0.990 & 0.019 & 0.031 & 6.133 & 35.224 & 0.977 & 0.036 & 0.066 & 12.906 \\
    XVFI & 38.903 & 0.989 & 0.022 & 0.037 & 7.401 & 34.552 & 0.975 & 0.039 & 0.072 & 16.000 \\
    ABME & 39.697 & 0.990 & 0.022 & 0.034 & 6.363 & 35.280 & 0.977 & 0.042 & 0.076 & 15.159 \\
    IFRNet & 39.881 & 0.990 & 0.019 & 0.030 & 5.939 & 35.668 & 0.979 & 0.033 & 0.058 & 12.084 \\
    ST-MFNet & 40.775 & 0.992 & 0.019 & 0.031 & 5.973 & 37.111 & 0.984 & 0.036 & 0.061 & 11.716 \\
    FLAVR & 40.161 & 0.990 & 0.022 & 0.034 & 6.320 & 36.020 & 0.979 & 0.049 & 0.077 & 15.006 \\
    MCVD & 22.201 & 0.828 & 0.199 & 0.230 & 32.246 & 21.488 & 0.812 & 0.213 & 0.243 & 37.474 \\
    \midrule
    LDMVFI & 38.674 & 0.987 & 0.014 & 0.024 & 5.752 & 33.996 & 0.970 & 0.028 & 0.053 & 12.485 \\
    \bottomrule
\end{tabular}
% \vspace{0.6em}
% \caption{Quantitative comparison results on SNU-FILM (note VFIformer is not included because the GPU goes OOM).}
% \vspace{-1em}
% \label{tab:quant2full1}
% \end{center}
% \end{table}

% \begin{table}[t]
% \begin{center}\footnotesize
\begin{tabular}{lcccccccccc}
    \toprule
    & \multicolumn{5}{c}{SNU-FILM-Hard} & \multicolumn{5}{c}{SNU-FILM-Extreme}\\
    \cmidrule(l{5pt}r{5pt}){2-6}\cmidrule(l{5pt}r{5pt}){7-11}
     & PSNR$\uparrow$ & SSIM$\uparrow$ & LPIPS$\downarrow$ & FloLPIPS$\downarrow$ & FID$\downarrow$ & PSNR$\uparrow$ & SSIM$\uparrow$ & LPIPS$\downarrow$ & FloLPIPS$\downarrow$ & FID$\downarrow$  \\
    \midrule
    BMBC & 29.942 & 0.933 & 0.068 & 0.118 & 25.773 & 24.715 & 0.856 & 0.145 & 0.237 & 49.519 \\
    AdaCoF & 29.477 & 0.925 & 0.080 & 0.131 & 27.982 & 24.650 & 0.851 & 0.152 & 0.234 & 52.848 \\
    CDFI & 29.660 & 0.929 & 0.081 & 0.141 & 29.087 & 24.645 & 0.854 & 0.163 & 0.255 & 53.916 \\
    XVFI & 29.364 & 0.928 & 0.075 & 0.138 & 29.483 & 24.545 & 0.853 & 0.142 & 0.233 & 54.449 \\
    ABME & 29.643 & 0.929 & 0.092 & 0.168 & 34.236 & 24.541 & 0.853 & 0.182 & 0.300 & 63.561 \\
    IFRNet & 30.143 & 0.935 & 0.065 & 0.122 & 25.436 & 24.954 & 0.859 & 0.136 & 0.229 & 50.047 \\
    ST-MFNet & 31.698 & 0.951 & 0.073 & 0.123 & 25.512 & 25.810 & 0.874 & 0.148 & 0.238 & 53.563 \\
    FLAVR & 30.577 & 0.938 & 0.112 & 0.169 & 34.746 & 25.206 & 0.861 & 0.217 & 0.303 & 72.673 \\
    MCVD & 20.314 & 0.766 & 0.250 & 0.292 & 51.529 & 18.464 & 0.694 & 0.320 & 0.385 & 83.156 \\
    \midrule
    LDMVFI & 28.547 & 0.917 & 0.060 & 0.114 & 26.520 & 23.934 & 0.837 & 0.123 & 0.204 & 47.042 \\
    \bottomrule
\end{tabular}
% \vspace{0.6em}
\caption{Quantitative comparison results on Middlebury, UCF-101, DAVIS, VFITex, and four subsets of SNU-FILM. Note ST-MFNet and FLAVR require four input frames so cannot be evaluated on Middlebury that contains frame triplets. VFIformer evaluation on SNU-FILM causes the GPU to go out of memory (OOM).}
% \vspace{-1em}
\label{tab:quantfull}
\end{center}
\end{table}

\begin{table}[H]
\begin{center}\footnotesize
\begin{tabular}{lcccccccccc}
    \toprule
    & \multicolumn{5}{c}{Middlebury} & \multicolumn{5}{c}{UCF-101}\\
    \cmidrule(l{5pt}r{5pt}){2-6}\cmidrule(l{5pt}r{5pt}){7-11}
     & PSNR$\uparrow$ & SSIM$\uparrow$ & LPIPS$\downarrow$ & FloLPIPS$\downarrow$ & FID$\downarrow$ & PSNR$\uparrow$ & SSIM$\uparrow$ & LPIPS$\downarrow$ & FloLPIPS$\downarrow$ & FID$\downarrow$  \\
    \midrule
    BMBC & 36.787 & 0.984 & 0.021 & 0.036 & 14.968 & 32.729 & 0.969 & 0.032 & 0.042 & 33.024 \\
    AdaCoF & 35.715 & 0.978 & 0.029 & 0.052 & 15.634 & 32.610 & 0.968 & 0.033 & 0.044 & 31.787 \\
    CDFI & 37.140 & 0.985 & 0.011 & 0.022 & 6.645 & 32.653 & 0.968 & 0.024 & 0.033 & 23.856 \\
    XVFI & 36.103 & 0.981 & 0.022 & 0.048 & 13.609 & 32.650 & 0.968 & 0.033 & 0.044 & 28.753 \\
    ABME & 37.639 & 0.986 & 0.027 & 0.040 & 11.393 & 32.055 & 0.967 & 0.058 & 0.069 & 37.066 \\
    IFRNet & 37.356 & 0.986 & 0.015 & 0.030 & 10.029 & 32.843 & 0.969 & 0.031 & 0.042 & 27.925 \\
    VFIformer & 38.404 & 0.987 & 0.015 & 0.024 & 9.439 & 32.959 & 0.970 & 0.034 & 0.047 & 32.734 \\
    \midrule
    LDMVFI & 34.230 & 0.974 & 0.019 & 0.041 & 35.745 & 32.160 & 0.964 & 0.026 & 0.034 & 25.792 \\
    \bottomrule
\end{tabular}
\begin{tabular}{lcccccccccc}
    \toprule
    & \multicolumn{5}{c}{DAVIS} & \multicolumn{5}{c}{VFITex}\\
    \cmidrule(l{5pt}r{5pt}){2-6}\cmidrule(l{5pt}r{5pt}){7-11}
     & PSNR$\uparrow$ & SSIM$\uparrow$ & LPIPS$\downarrow$ & FloLPIPS$\downarrow$ & FID$\downarrow$ & PSNR$\uparrow$ & SSIM$\uparrow$ & LPIPS$\downarrow$ & FloLPIPS$\downarrow$ & FID$\downarrow$  \\
    \midrule
    BMBC & 26.293 & 0.855 & 0.131 & 0.187 & 15.136 & 26.784 & 0.899 & 0.201 & 0.266 & 35.493 \\
    AdaCoF & 25.992 & 0.840 & 0.146 & 0.197 & 16.102 & 27.003 & 0.894 & 0.194 & 0.264 & 31.371 \\
    CDFI & 26.403 & 0.849 & 0.106 & 0.149 & 11.418 & 27.278 & 0.899 & 0.160 & 0.228 & 32.765 \\
    XVFI & 26.222 & 0.850 & 0.141 & 0.200 & 14.856 & 23.352 & 0.871 & 0.247 & 0.316 & 45.873 \\
    ABME & 26.861 & 0.865 & 0.151 & 0.209 & 16.931 & 26.765 & 0.901 & 0.254 & 0.341 & 53.317 \\
    IFRNet & 27.080 & 0.868 & 0.106 & 0.156 & 12.422 & 27.790 & 0.910 & 0.180 & 0.249 & 36.431 \\
    VFIformer & 27.204 & 0.869 & 0.127 & 0.184 & 14.407 & OOM & OOM & OOM & OOM & OOM \\
    \midrule
    LDMVFI & 25.073 & 0.819 & 0.125 & 0.172 & 14.093 & 26.355 & 0.887 & 0.161 & 0.224 & 36.831 \\
    \bottomrule
\end{tabular}
% snufilm
\begin{tabular}{lcccccccccc}
    \toprule
    & \multicolumn{5}{c}{SNU-FILM-Easy} & \multicolumn{5}{c}{SNU-FILM-Medium}\\
    \cmidrule(l{5pt}r{5pt}){2-6}\cmidrule(l{5pt}r{5pt}){7-11}
     & PSNR$\uparrow$ & SSIM$\uparrow$ & LPIPS$\downarrow$ & FloLPIPS$\downarrow$ & FID$\downarrow$ & PSNR$\uparrow$ & SSIM$\uparrow$ & LPIPS$\downarrow$ & FloLPIPS$\downarrow$ & FID$\downarrow$  \\
    \midrule
    BMBC & 39.897 & 0.990 & 0.018 & 0.029 & 5.661 & 35.310 & 0.977 & 0.034 & 0.060 & 12.122 \\
    AdaCoF & 39.801 & 0.990 & 0.019 & 0.030 & 6.007 & 35.050 & 0.975 & 0.036 & 0.065 & 13.287 \\
    CDFI & 40.116 & 0.990 & 0.013 & 0.021 & 4.641 & 35.501 & 0.978 & 0.024 & 0.045 & 10.035 \\
    XVFI & 39.554 & 0.989 & 0.020 & 0.033 & 6.629 & 35.062 & 0.976 & 0.037 & 0.069 & 14.021 \\
    ABME & 39.697 & 0.990 & 0.022 & 0.034 & 6.363 & 35.280 & 0.977 & 0.042 & 0.076 & 15.159 \\
    IFRNet & 40.107 & 0.991 & 0.017 & 0.027 & 5.429 & 35.851 & 0.979 & 0.029 & 0.050 & 10.745 \\
    VFIformer & 40.204 & 0.991 & 0.018 & 0.028 & 5.682 & 36.028 & 0.980 & 0.033 & 0.056 & 11.188 \\
    \midrule
    LDMVFI & 38.890 & 0.988 & 0.013 & 0.023 & 5.430 & 33.975 & 0.971 & 0.027 & 0.056 & 12.315 \\
    \bottomrule
\end{tabular}
% \vspace{0.6em}
% \caption{Quantitative comparison results on SNU-FILM (note VFIformer is not included because the GPU goes OOM).}
% \vspace{-1em}
% \label{tab:quant2full1}
% \end{center}
% \end{table}

% \begin{table}[t]
% \begin{center}\footnotesize
\begin{tabular}{lcccccccccc}
    \toprule
    & \multicolumn{5}{c}{SNU-FILM-Hard} & \multicolumn{5}{c}{SNU-FILM-Extreme}\\
    \cmidrule(l{5pt}r{5pt}){2-6}\cmidrule(l{5pt}r{5pt}){7-11}
     & PSNR$\uparrow$ & SSIM$\uparrow$ & LPIPS$\downarrow$ & FloLPIPS$\downarrow$ & FID$\downarrow$ & PSNR$\uparrow$ & SSIM$\uparrow$ & LPIPS$\downarrow$ & FloLPIPS$\downarrow$ & FID$\downarrow$  \\
    \midrule
    BMBC & 29.328 & 0.927 & 0.075 & 0.133 & 26.373 & 23.924 & 0.843 & 0.152 & 0.246 & 51.292 \\
    AdaCoF & 29.463 & 0.924 & 0.075 & 0.127 & 26.707 & 24.307 & 0.844 & 0.148 & 0.233 & 50.278 \\
    CDFI & 29.745 & 0.928 & 0.056 & 0.099 & 20.829 & 24.542 & 0.847 & 0.121 & 0.198 & 43.418 \\
    XVFI & 29.510 & 0.927 & 0.075 & 0.132 & 28.167 & 24.435 & 0.848 & 0.143 & 0.231 & 50.851 \\
    ABME & 29.643 & 0.929 & 0.092 & 0.168 & 34.236 & 24.541 & 0.853 & 0.182 & 0.300 & 63.561 \\
    IFRNet & 30.150 & 0.934 & 0.058 & 0.109 & 22.512 & 24.795 & 0.856 & 0.128 & 0.216 & 45.458 \\
    VFIformer & 30.256 & 0.935 & 0.069 & 0.124 & 24.268 & 24.921 & 0.858 & 0.146 & 0.235 & 47.378 \\
    \midrule
    LDMVFI & 28.144 & 0.911 & 0.068 & 0.124 & 28.726 & 23.349 & 0.827 & 0.139 & 0.226 & 51.022 \\
    \bottomrule
\end{tabular}
% \vspace{0.6em}
\caption{Quantitative comparison results on Middlebury, UCF-101, DAVIS, VFITex, and four subsets of SNU-FILM for models trained with Vimeo90k-triplet only.}
\label{tab:vimeoquantfull}
\vspace{-2em}
\end{center}
\end{table}

\section{Additional Quantitative Evaluation with Vimeo90k-triplet Training}\label{appendix:vimeo}
In the main paper and Section~\ref{appendix:quantfull} the evaluation results for models trained using the combination of Vimeo90k-septuplet and BVI-DVC are presented. Here we additionally include evaluation results for the models trained only with the Vimeo90k-triplet dataset. Note that here we only include models whose pre-trained versions (on the Vimeo90k-triplet dataset) are publicly available. These results are summerized in Table~\ref{tab:vimeoquantfull}. Comparing Table~\ref{tab:vimeoquantfull} and Table~\ref{tab:quantfull}, it is noticed that for some methods, the re-trained version on Vimeo90k+BVI-DVC show some extent of PSNR performance drop on the Middlebury and UCF-101 testsets. This can be due to the fact that Middlebury and UCF-101 show higher distribution overlap with the pre-training dataset Vimeo90k-triplet, while re-training on BVI-DVC+Vimeo90k shifts the training data distribution. Such shift towards content with larger and more complex motion (i.e. content from BVI-DVC) indeed shows benefits for the more challenging test sets such as DAVIS and VFITex, which are reflected in the increase in overall PSNR performance after re-training.

\begin{table}[t]
\begin{center}
\resizebox{\linewidth}{!}{
\begin{tabular}{lccccccccccccc}
    \toprule
    & \multirow{2}[2]{*}{$f$} & \multirow{2}[2]{*}{$c$} & \multirow{2}[2]{*}{\makecell{AE\\ Model}} & \multirow{2}[2]{*}{\makecell{Cond.\\ Mode}} & \multicolumn{3}{c}{Middlebury} & \multicolumn{3}{c}{UCF-101}& \multicolumn{3}{c}{DAVIS}\\
    \cmidrule(l{5pt}r{5pt}){6-8}\cmidrule(l{5pt}r{5pt}){9-11}\cmidrule(l{5pt}r{5pt}){12-14}
    & & & & &LPIPS$\downarrow$ & FloLPIPS$\downarrow$ & FID$\downarrow$ &LPIPS$\downarrow$ & FloLPIPS$\downarrow$ & FID$\downarrow$ &LPIPS$\downarrow$ & FloLPIPS$\downarrow$ & FID$\downarrow$ \\
    \midrule
    V1 & 32 & 256 & frame & concat & 0.077 & 0.085 & 40.399 & 0.063 & 0.067 & 60.742 & 0.168 & 0.200 & 21.812 \\
    V2 & 32 & 256 & MaxCA+frame & concat & 0.028 & 0.045 & 19.485 & 0.032 & 0.041 & 29.578 & 0.135 & 0.176 & 19.980 \\
    \midrule
    V3 & 8 & 256 & MaxCA+kernel & concat & 0.018 & 0.046 & 19.552 & 0.028 & 0.036 & 27.278 & 0.125 & 0.168 & 15.535 \\
    V4 & 16 & 256 & MaxCA+kernel & concat & 0.017 & 0.037 & 12.539 & 0.026 & 0.035 & 26.805 & 0.107 & 0.154 & 12.720 \\
    V5 & 64 & 256 & MaxCA+kernel & concat & 0.022 & 0.044 & 15.041 & 0.026 & 0.036 & 26.055 & 0.114 & 0.157 & 12.330 \\
    \midrule
    V6 & 32 & 256 & MaxCA+kernel & N/A & 0.100 & 0.108 & 38.538 & 0.042 & 0.053 & 32.189 & 0.193 & 0.214 & 8.639 \\
    \midrule
    V7 & 32 & 64 & MaxCA+kernel & concat & 0.019 & 0.044 & 16.101 & 0.026 & 0.035 & 26.376 & 0.107 & 0.153 & 12.627 \\
    V8 & 32 & 128 & MaxCA+kernel & concat & 0.019 & 0.044 & 16.026 & 0.026 & 0.035 & 26.340 & 0.107 & 0.153 & 12.611 \\
    \midrule
    V9 & 32 & 256 & MaxCA+kernel & MaxCA & 0.019 & 0.044 & 16.005 & 0.026 & 0.035 & 26.334 & 0.106 & 0.153 & 12.571 \\
    \midrule
    Ours & 32 & 256 & MaxCA+kernel & concat & 0.019 & 0.044 & 16.167 & 0.026 & 0.035 & 26.301 & 0.107 & 0.153 & 12.554 \\
    \bottomrule
\end{tabular}
}
\caption{Ablation experiment results, showing performance of variants of the proposed LDMVFI.}
\label{tab:ablationfull1}
\end{center}
\end{table}

\begin{table}[t]
\begin{center}
\resizebox{\linewidth}{!}{
\begin{tabular}{lcccccccccccc}
    \toprule
    & \multicolumn{3}{c}{SNU-FILM-Easy} & \multicolumn{3}{c}{SNU-FILM-Medium} & \multicolumn{3}{c}{SNU-FILM-Hard}& \multicolumn{3}{c}{SNU-FILM-Extreme}\\ \cmidrule(l{5pt}r{5pt}){2-4}\cmidrule(l{5pt}r{5pt}){5-7}\cmidrule(l{5pt}r{5pt}){8-10}\cmidrule(l{5pt}r{5pt}){11-13}
    &LPIPS$\downarrow$ & FloLPIPS$\downarrow$ & FID$\downarrow$ &LPIPS$\downarrow$ & FloLPIPS$\downarrow$ & FID$\downarrow$ &LPIPS$\downarrow$ & FloLPIPS$\downarrow$ & FID$\downarrow$ &LPIPS$\downarrow$ & FloLPIPS$\downarrow$ & FID$\downarrow$ \\
    \midrule
    V1 & 0.054 & 0.056 & 14.010 & 0.068 & 0.082 & 20.572 & 0.104 & 0.142 & 35.429 & 0.172 & 0.243 & 72.617 \\
    V2 & 0.020 & 0.028 & 7.779 & 0.035 & 0.058 & 15.138 & 0.076 & 0.126 & 32.653 & 0.150 & 0.233 & 69.730 \\
    \midrule
    V3 & 0.016 & 0.026 & 6.374 & 0.030 & 0.052 & 11.701 & 0.068 & 0.119 & 27.183 & 0.144 & 0.227 & 55.596 \\
    V4 & 0.014 & 0.024 & 5.982 & 0.027 & 0.052 & 11.987 & 0.059 & 0.112 & 25.088 & 0.122 & 0.204 & 47.623 \\
    V5 & 0.016 & 0.027 & 6.191 & 0.031 & 0.058 & 13.060 & 0.064 & 0.116 & 26.439 & 0.129 & 0.211 & 48.336\\
    \midrule
    V6 & 0.083 & 0.086 & 20.134 & 0.103 & 0.111 & 24.124 & 0.141 & 0.160 & 29.157 & 0.201 & 0.237 & 35.021 \\
    \midrule
    V7 & 0.014 & 0.024 & 5.758 & 0.028 & 0.053 & 12.483 & 0.060 & 0.114 & 26.401 & 0.123 & 0.204 & 47.022\\
    V8 & 0.014 & 0.024 & 5.754 & 0.028 & 0.053 & 12.483 & 0.060 & 0.114 & 26.380 & 0.123 & 0.204 & 46.882\\
    \midrule
    V9 & 0.014 & 0.024 & 5.757 & 0.028 & 0.053 & 12.477 & 0.060 & 0.114 & 26.525 & 0.122 & 0.204 & 46.906 \\
    \midrule
    Ours & 0.014 & 0.024 & 5.752 & 0.028 & 0.053 & 12.485 & 0.060 & 0.114 & 26.520 & 0.123 & 0.204 & 47.042 \\
    \bottomrule
\end{tabular}
}
% \vspace{1mm}
\caption{Ablation experiment results on SNU-FILM. Full details of the variants (V1-8) can be found in the main paper.}
\label{tab:ablationfull2}
\vspace{-1em}
\end{center}
\end{table}

\section{Full Ablation Study Results}\label{appendix:ablation}
In the main paper, due to space limitations, we presented selected ablation experiments on only three test sets. Here we provide the full ablation study, including additional experiments and evaluation on additional test sets.

\paragraph{Effect of Model Size.} We investigate how the model size of the denoising U-Net affects the performance. The denoising U-Net contains multiple layers of ResNet and MaxViT blocks, where the number of feature channels is a multiple of a base channel number $c$ (detailed in Figure~\ref{fig:unet}). In the default LDMVFI, $c=256$. This corresponds to a total of 439.0M parameters. We experiment with $c=128,64$ (V8,V7), which reduce the model parameters to 126.8M and 48.7M respectively. It is observed from Table~\ref{tab:ablationfull1} and \ref{tab:ablationfull2} that as $c$ is decreased, there is no significant change in performance. This may be because when $f=32$, the latent dimension is low enough ($8\times 8$ for training data) that the model capacity is still sufficient even when $c=64$. To verify this, we additionally tested $c=64,128,256$ for the case of $f=8$ (latent dimension is $32\times 32$ for training data). The results are shown in Table~\ref{tab:ablationfull3}, where it is observed that as $c$ decreases, there is a clearer decreasing trend in overall model performance.

\paragraph{Conditioning Mechanism.} In LDMVFI, the mechanism for conditioning the denoising U-Net on the latents $z^0,z^1$ is concatenation, i.e., we concatenate $z^n_t, z^0, z^1$ to form the U-Net input. In \cite{rombach2022high}, an alternative cross attention-based conditioning mechanism is introduced. Based on this, We create a variant of LDMVFI (V9) where the conditioning is done using MaxCA blocks at different layers of the U-Net. As shown in Table~\ref{tab:ablationfull1} and \ref{tab:ablationfull2}, V9 shows highly similar performance to LDMVFI. Given the limited improvement, we adopted the simpler concatenation-based conditioning for LDMVFI.

\begin{table}[t]
\begin{center}
\resizebox{\linewidth}{!}{
\begin{tabular}{lccccccccccccc}
    \toprule
    & \multirow{2}[2]{*}{$f$} & \multirow{2}[2]{*}{$c$} & \multirow{2}[2]{*}{\makecell{AE\\ Model}} & \multirow{2}[2]{*}{\makecell{Cond.\\ Mode}} & \multicolumn{3}{c}{Middlebury} & \multicolumn{3}{c}{UCF-101}& \multicolumn{3}{c}{DAVIS}\\
    \cmidrule(l{5pt}r{5pt}){6-8}\cmidrule(l{5pt}r{5pt}){9-11}\cmidrule(l{5pt}r{5pt}){12-14}
    & & & & &LPIPS$\downarrow$ & FloLPIPS$\downarrow$ & FID$\downarrow$ &LPIPS$\downarrow$ & FloLPIPS$\downarrow$ & FID$\downarrow$ &LPIPS$\downarrow$ & FloLPIPS$\downarrow$ & FID$\downarrow$ \\
    \midrule
    V10 & 8 & 64 & MaxCA+kernel & concat & 0.020 & 0.040 & 13.800 & 0.029 & 0.037 & 28.831 & 0.133 & 0.179 & 17.937 \\
    V11 & 8 & 128 & MaxCA+kernel & concat & 0.019 & 0.035 & 13.528 & 0.028 & 0.036 & 26.917 & 0.131 & 0.175 & 17.746 \\
    V3 & 8 & 256 & MaxCA+kernel & concat & 0.018 & 0.046 & 19.552 & 0.028 & 0.036 & 27.278 & 0.125 & 0.168 & 15.535 \\
    \bottomrule
\end{tabular}
}
\resizebox{\linewidth}{!}{
\begin{tabular}{lcccccccccccc}
    \toprule
    & \multicolumn{3}{c}{SNU-FILM-Easy} & \multicolumn{3}{c}{SNU-FILM-Medium} & \multicolumn{3}{c}{SNU-FILM-Hard}& \multicolumn{3}{c}{SNU-FILM-Extreme}\\ \cmidrule(l{5pt}r{5pt}){2-4}\cmidrule(l{5pt}r{5pt}){5-7}\cmidrule(l{5pt}r{5pt}){8-10}\cmidrule(l{5pt}r{5pt}){11-13}
    &LPIPS$\downarrow$ & FloLPIPS$\downarrow$ & FID$\downarrow$ &LPIPS$\downarrow$ & FloLPIPS$\downarrow$ & FID$\downarrow$ &LPIPS$\downarrow$ & FloLPIPS$\downarrow$ & FID$\downarrow$ &LPIPS$\downarrow$ & FloLPIPS$\downarrow$ & FID$\downarrow$ \\
    \midrule
    V10 & 0.016 & 0.026 & 6.255 & 0.032 & 0.060 & 13.312 & 0.076 & 0.132 & 30.786 & 0.157 & 0.243 & 62.075 \\
    V11 & 0.016 & 0.024 & 6.256 & 0.031 & 0.054 & 13.329 & 0.073 & 0.127 & 30.343 & 0.155 & 0.241 & 66.383 \\
    V3 & 0.016 & 0.026 & 6.374 & 0.030 & 0.052 & 11.701 & 0.068 & 0.119 & 27.183 & 0.144 & 0.227 & 55.596 \\
    \bottomrule
\end{tabular}
}
\caption{Ablation experiment results on $c$ when $f=8$.}
\label{tab:ablationfull3}
\end{center}
\end{table}

\begin{figure}
    \centering
    \subfloat{\includegraphics[width=0.264\linewidth]{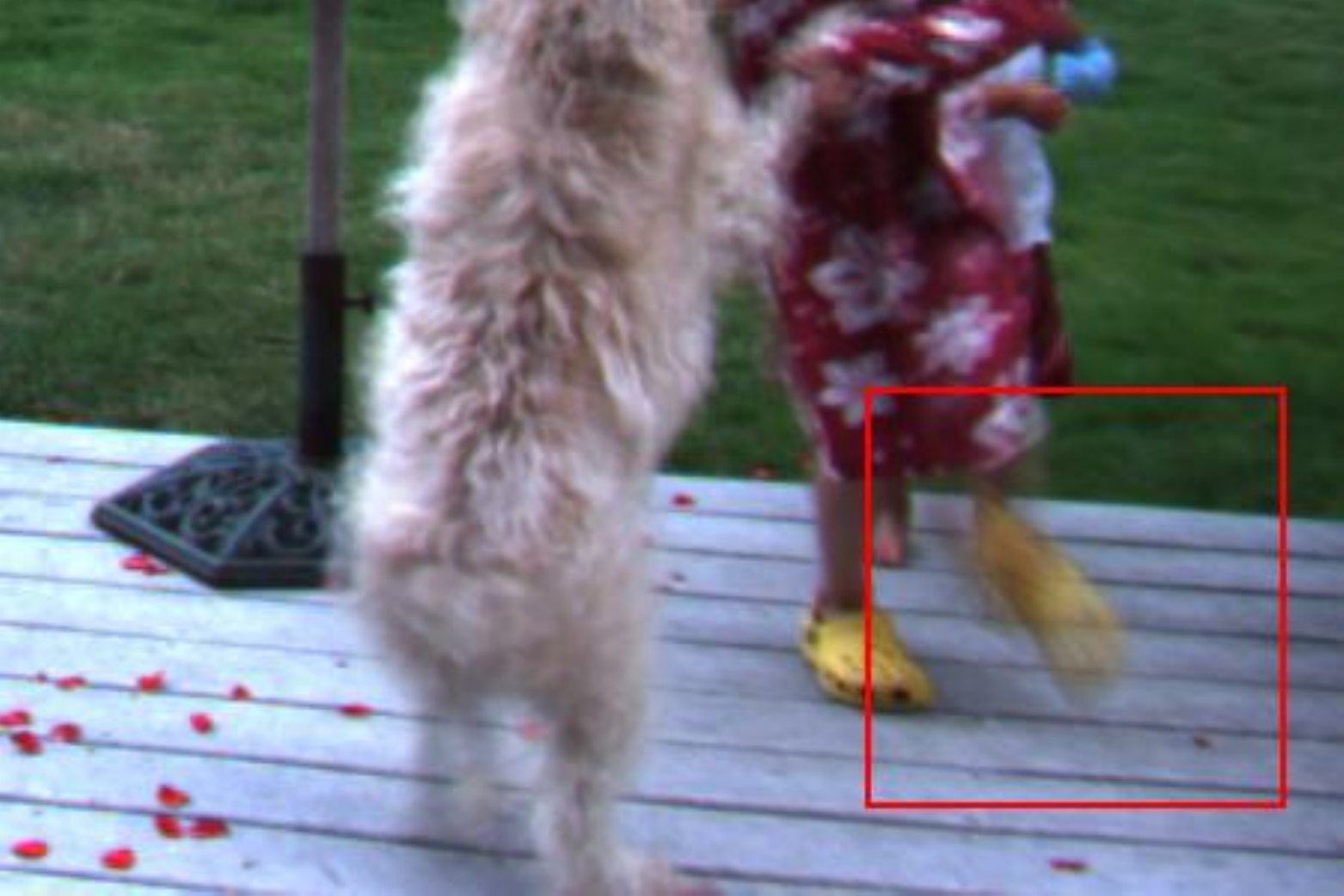}}\;\!\!
	\subfloat{\includegraphics[width=0.176\linewidth]{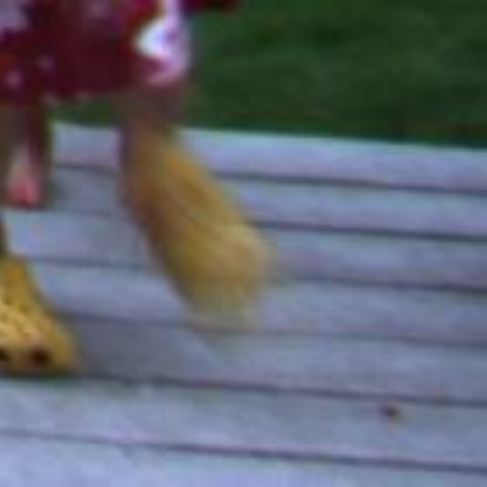}}\;\!\!
	\subfloat{\includegraphics[width=0.176\linewidth]{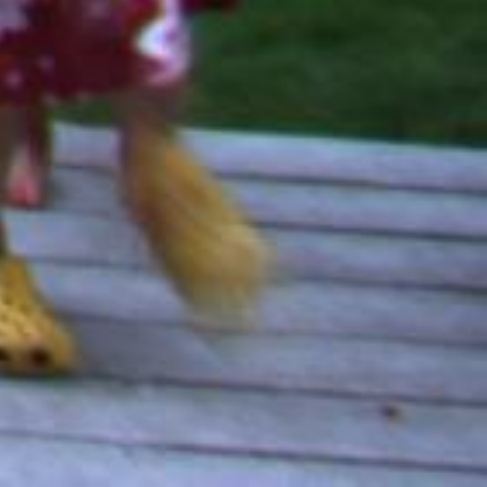}}\;\!\!
    \subfloat{\includegraphics[width=0.176\linewidth]{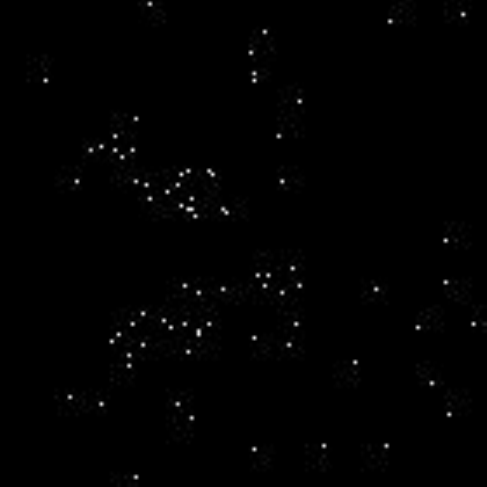}}\;\!\!
    \subfloat{\includegraphics[width=0.176\linewidth]{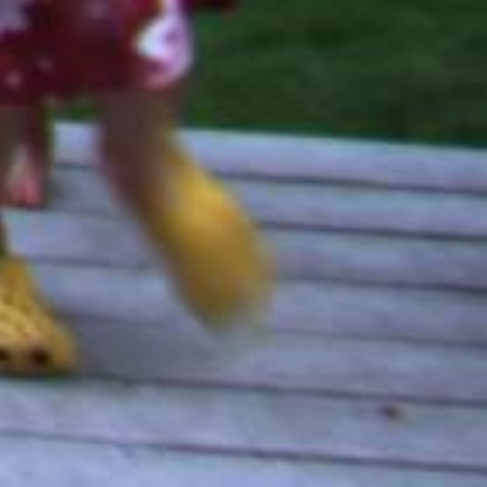}}
    \setcounter{subfigure}{0}
    \subfloat[Overlaid Inputs] {\includegraphics[width=0.264\linewidth]{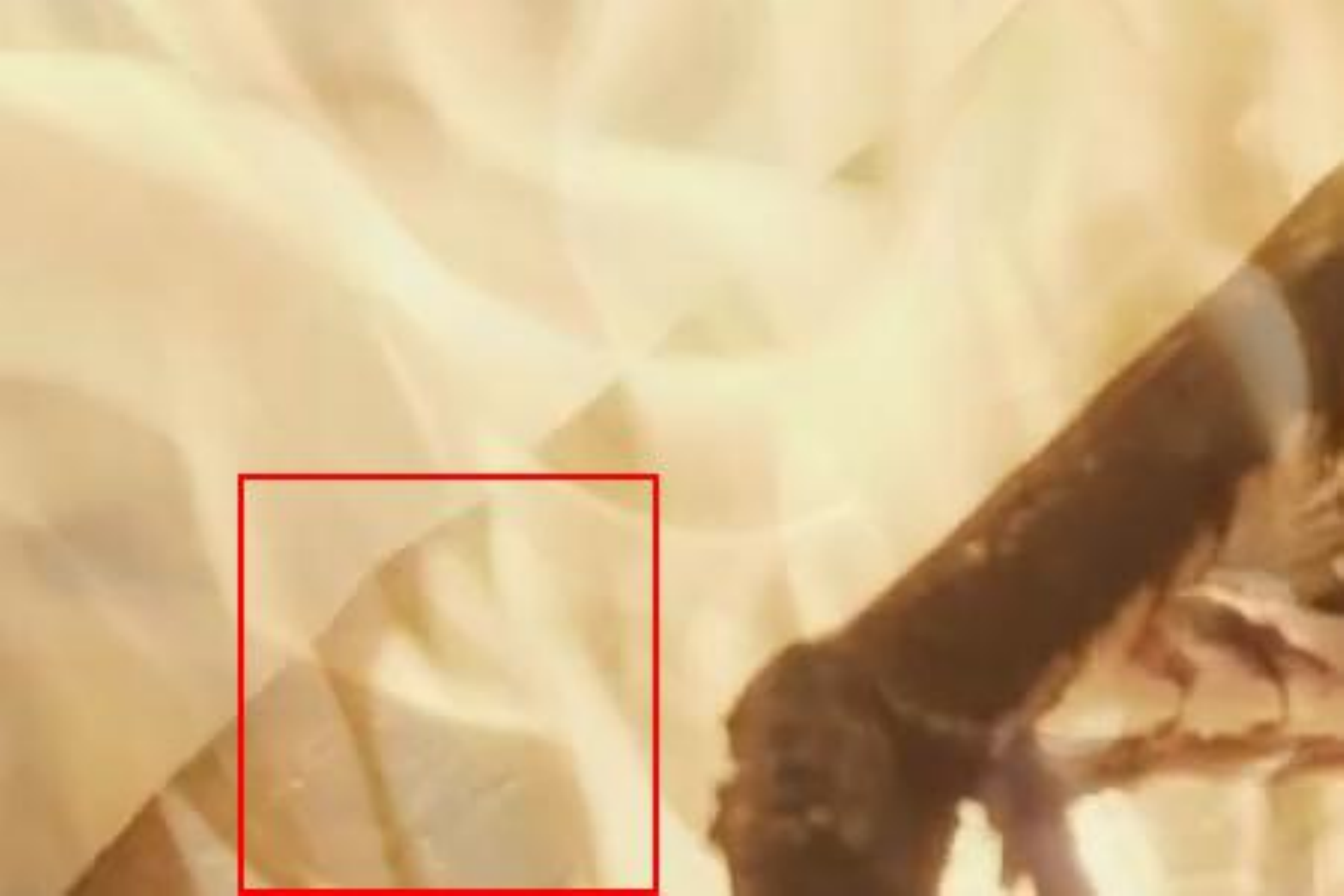}}\;\!\!
	\subfloat[Sample 1] {\includegraphics[width=0.176\linewidth]{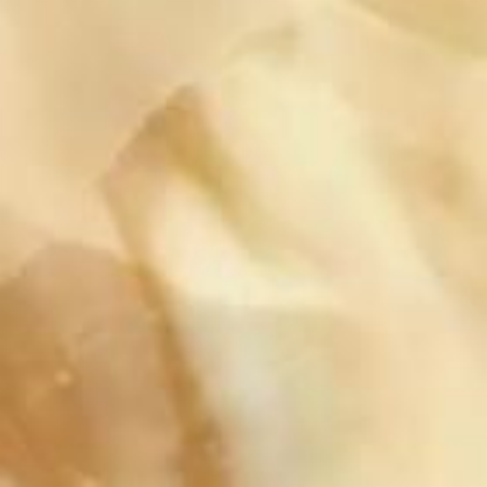}}\;\!\!
	\subfloat[Sample 2] {\includegraphics[width=0.176\linewidth]{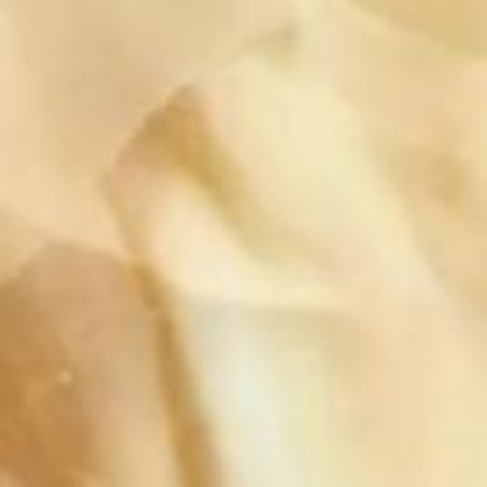}}\;\!\!
    \subfloat[Residual] {\includegraphics[width=0.176\linewidth]{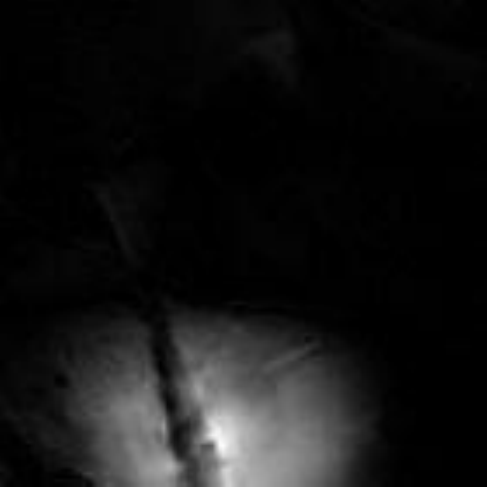}}\;\!\!
    \subfloat[IFRNet] {\includegraphics[width=0.176\linewidth]{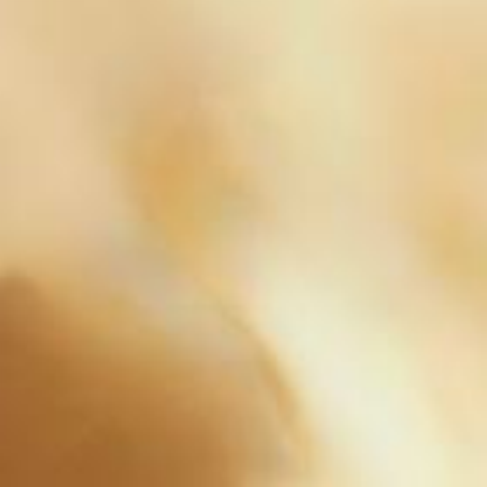}}\\
\caption{Visual examples on variability of LDMVFI results. Residual is the difference between Sample 1 and Sample 2.}
\label{fig:variability}
\end{figure}

\paragraph{Variability of VFI results.} When non-deterministic samplers are used (e.g. DDPM~\cite{ho2020denoising}), diffusion models can produce varied results given the same input. We investigate such variability in VFI by sampling multiple times from LDMVFI using DDPM given the same frames. As illustrated in Figure~\ref{fig:variability}, under relatively regular motions (e.g. that of rigid objects), there is little variation in the result, as shown by the residual image (the fourth column) between Sample 1 and Sample 2. This is desirable because in such cases there might only be one ground-truth frame and variation in the interpolation result can lead to temporal inconsistency. On the other hand, for dynamic textures (e.g. water) where the underlying motion can be quite irregular and random, LDMVFI does produce varied results. It is also noted that these results are much sharper than those of IFRNet, and this can be due to the fact that as a generative model, LDMVFI draws single samples from a distribution, while existing methods like IFRNet that are trained on L1 losses tend to learn an average of many possible results (i.e. the over-smoothing issue discussed in \cite{li2022srdiff}), which manifest as blur and causes flickering (see the supplementary video). This is further supported by the large performance gain of LDMVFI over existing methods on the VFITex dataset (Table~\ref{tab:quantfull}).

\section{Details of Subjective Experiment}\label{appendix:userstudy}
The user study and the use of human data have undergone an internal ethics review and have been approved by the Institutional Review Board.

The subjective experiment was conducted in a lab-based environment. The monitor used to display videos was a BENQ XL2720Z (59.8$\times$33.6cm screen size). The spatial resolution of the display was set to $1920\times 1080$ and the frame rate was set to $60$Hz. The viewing distance of the participants was 1 meter, approximately three times screen height~\cite{itu2002500}.

In the main paper, we state that t-tests are performed between the sequence-wise preference ratios of the proposed method and the three tested baselines. Here we report the $p$ values for the t-tests in Table~\ref{tab:ttest}.

\begin{table}[H]
\begin{center}
\begin{tabular}{l|c}
    \toprule
    Comparison & $p$ value \\
    \midrule
    Ours vs BMBC & $3.8\times10^{-9}$ \\
    Ours vs IFRNet & $1.6\times10^{-3}$ \\
    Ours vs ST-MFNet & $1.7\times10^{-2}$\\
    \bottomrule
\end{tabular}
\caption{The results of t-test analysis on the user study results.}
\label{tab:ttest}
\end{center}
\end{table}

\begin{table}[H]
\centering
% \small
\begin{tabular}{l|ccccc}
    \toprule
     &PSNR$\uparrow$ &SSIM$\uparrow$ & LPIPS$\downarrow$ &FloLPIPS$\downarrow$ & FID$\downarrow$ \\
    \midrule
    ST-MFNet & 21.709 & 0.780 & 0.226 & 0.297 & 28.939 \\
    IFRNet & 22.745 & 0.813 & 0.186 & 0.270 & 23.122 \\
    \midrule
    LDMVFI & 22.194 & 0.753 & 0.167 & 0.204 & 14.493\\
    \bottomrule
\end{tabular}
\caption{Multi-frame ($\times$4) evaluation results on DAVIS.}
\label{tab:davisx4}
\end{table}

\section{Multi-Frame ($\times$4) Interpolation}\label{appendix:x4}
The $\times$4 up-sampling performance of LDMVFI is evaluated on the DAVIS~\cite{perazzi2016benchmark} dataset. Specifically, in each sequence of DAVIS, every 2nd, 3rd and 4th frames are dropped and considered as the ground truth. The 3rd frame is firstly interpolated from the 1st and the 5th frame, then the 2nd frame is interpolated using the 1st and the 3rd frames (previously interpolated), and the 4th frame is interpolated in a similar way. For this experiment, we compared LDMVFI with the two most competitive approaches according to Table~\ref{tab:quantfull}: IFRNet~\cite{kong2022ifrnet} and ST-MFNet~\cite{danier2022st}. Table~\ref{tab:davisx4} shows that LDMVFI achieves the best scores in terms of the perceptual metrics (i.e. LPIPS, FloLPIPS and FID).

\section{Limitations and Societal Impact}\label{appendix:limitation}
\paragraph{Limitations.} Firstly, as discussed in the main paper, the proposed LDMVFI shows a much slower inference speed compared to the other state-of-the-art methods. This is a common drawback of diffusion models~\cite{ho2020denoising,rombach2022high} mainly due to the iterative reverse process during generation. Various  techniques~\cite{salimans2022progressive,karras2022elucidating,lu2022dpm} have been proposed to speed up the sampling process and these can also be applied to LDMVFI. Secondly, the number of model parameters in LDMVFI is also larger than other methods. The main component of LDMVFI that accounts for the large model size is the denoising U-Net, which was developed in previous work~\cite{ho2020denoising,dhariwal2021diffusion,rombach2022high} and modified in here by replacing the self-attention layers with MaxViT blocks~\cite{tu2022maxvit}. To reduce the model size, techniques such as knowledge distillation~\cite{hinton2015distilling, morris2023st} and model compression~\cite{modelcompression} can be used. The exploration of these techniques, as well as the accelerated diffusion sampling methods in the context of VFI, remain for future work. Finally, we noticed that LDMVFI can generate unsatisfactory results under extremely large motion, which is a common challenge faced by many state-of-the-art VFI methods.

\paragraph{Potential Negative Societal Impact.} Firstly, in general, generative models for images and videos can be used to generate inappropriately manipulated content or in unethical ways (e.g. ``deep fake'' generation)~\cite{denton2021ethical}. Secondly, the two-stage training strategy, large model size and slow inference speed of LDMVFI mean that large-scale training and evaluation processes can consume significant amounts of energy, leading to increased carbon footprint~\cite{lacoste2019quantifying}. We refer the readers to \cite{denton2021ethical} and \cite{lacoste2019quantifying} for more detailed discussions on these matters.

\section{Attribution of Assets: Code and Data}
In this section, we summarize the sources and licenses of all the datasets and code used in the work. The attribution of datasets and code are shown in Table~\ref{tab:license1} and \ref{tab:license2}. For MCVD~\cite{voleti2022mcvd}, we used the \texttt{smmnist\_DDPM\_big5.yml} configuration\footnote{\url{https://github.com/voletiv/mcvd-pytorch/blob/master/configs/smmnist_DDPM_big5.yml}}, setting both the numbers of previous and future frames to 1, and the number of interpolated frames also to 1.
\vfill

% ===============================================================
\begin{table}[t]
\centering
% \resizebox{\linewidth}{!}{
% \scriptsize
\begin{tabular}{p{5cm}|p{7cm}|p{4cm}}
\toprule
\textbf{Dataset}  & \textbf{Dataset URL} &\textbf{License / Terms of Use} \\ 
\midrule
Vimeo-90k~\cite{xue2019video} & \url{http://toflow.csail.mit.edu} &MIT license. \\
\midrule
BVI-DVC~\cite{ma2020bvi} & \url{https://fan-aaron-zhang.github.io/BVI-DVC/} (original videos); \url{https://github.com/danielism97/ST-MFNet} (quintuplets) & All sequences are allowed for academic research.\\
\midrule
UCF101~\cite{soomro2012ucf101} & \url{https://www.crcv.ucf.edu/research/data-sets/ucf101/} & No explicit license terms, but compiled and made available for research use by the University of Central Florida. \\
\midrule
DAVIS~\cite{perazzi2016benchmark} & \url{https://davischallenge.org} & BSD license. \\
\midrule
SNU-FILM~\cite{choi2020channel} & \url{https://myungsub.github.io/CAIN/} & MIT license .\\
\midrule
Middlebury~\cite{baker2011database} & \url{https://vision.middlebury.edu/flow/data/} & All sequences are available for research use.\\
\midrule
VFITex~\cite{danier2022st} & \url{https://github.com/danier97/ST-MFNet} & All sequences are available for research use.\\
\midrule
BVI-HFR~\cite{mackin2018study} & \url{https://fan-aaron-zhang.github.io/BVI-HFR/} & Non-Commercial Government Licence for public sector information. \\
\bottomrule
\end{tabular}
% }
\caption{License information for the datasets used in this work.}
\label{tab:license1}
\end{table}
% ==============================================================

% ===============================================================
\begin{table}[H]
\centering
\begin{tabular}{p{5cm}|p{7cm}|p{4cm}}
\toprule
\textbf{Method}  & \textbf{Source code URL} & \textbf{License / Teams of Use} \\ 
\midrule
BMBC~\cite{park2020bmbc} & \url{https://github.com/JunHeum/BMBC} & MIT license. \\
\midrule
AdaCoF~\cite{lee2020adacof} & \url{https://github.com/HyeongminLEE/AdaCoF-pytorch} & MIT license. \\
\midrule
CDFI~\cite{ding2021cdfi} & \url{https://github.com/tding1/CDFI} & Available for research use. \\
\midrule
XVFI~\cite{sim2021xvfi} & \url{https://github.com/JihyongOh/XVFI} & Research and education only. \\
\midrule
ABME~\cite{park2021asymmetric} & \url{https://github.com/JunHeum/ABME} & MIT license. \\
\midrule
IFRNet~\cite{kong2022ifrnet} & \url{https://github.com/ltkong218/IFRNet} & MIT license. \\
\midrule
VFIformer~\cite{lu2022video} & \url{https://github.com/dvlab-research/VFIformer} & Available for research use. \\
\midrule
ST-MFNet~\cite{danier2022st} & \url{https://github.com/danielism97/ST-MFNet} & MIT license. \\
\midrule
FLAVR~\cite{kalluri2023flavr} & \url{https://github.com/tarun005/FLAVR} & Apache-2.0 license. \\
\midrule
MCVD~\cite{voleti2022mcvd} & \url{https://github.com/voletiv/mcvd-pytorch} & MIT license. \\
\bottomrule
\end{tabular}
\caption{License information for the code assets used in this work.}
\label{tab:license2}
\end{table}

% ==============================================================

\section{Additional Qualitative Examples}\label{appendix:visual}

\begin{figure}[H]
    \centering
    \subfloat {\includegraphics[width=0.195\linewidth]{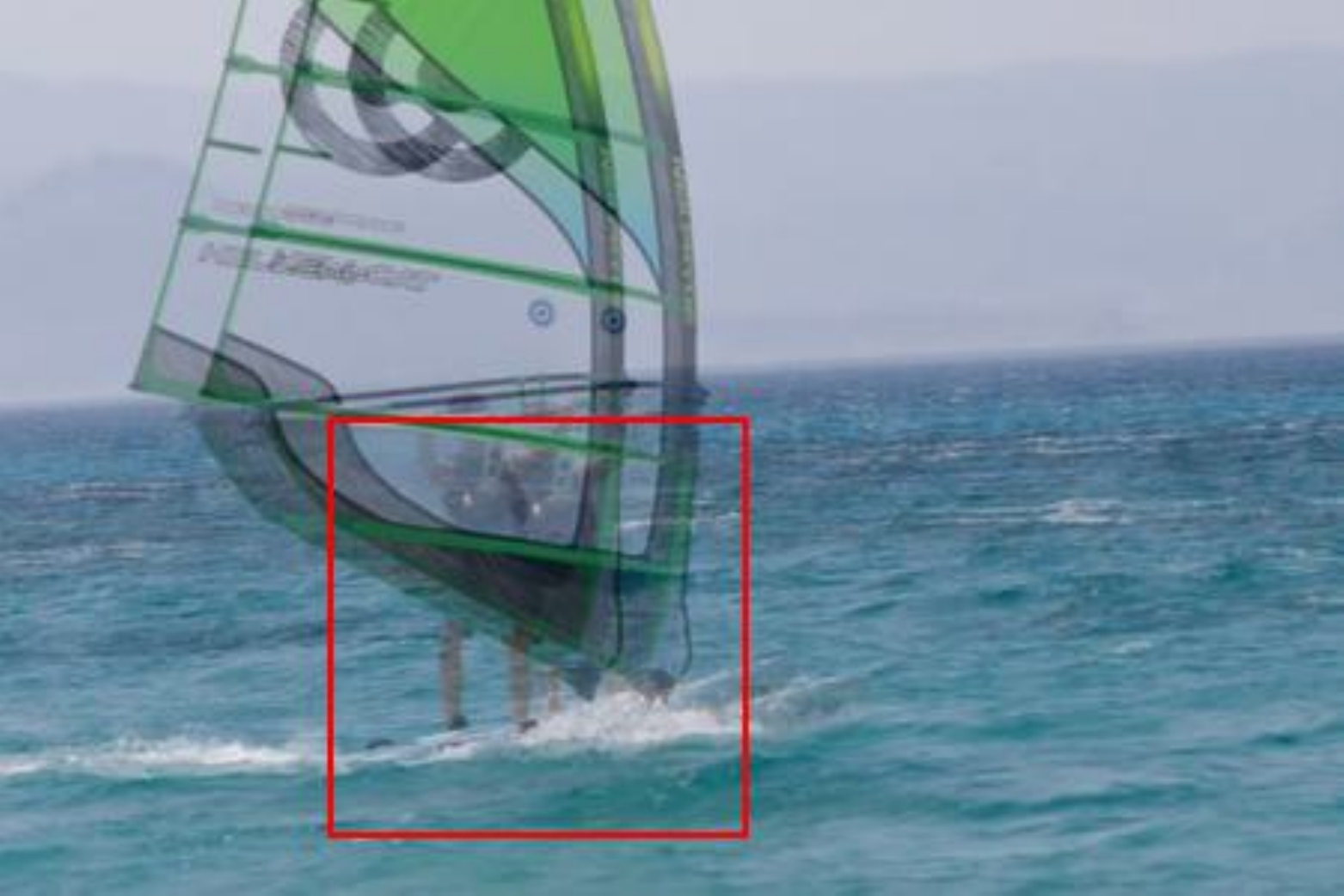}}\;\!\!
	\subfloat {\includegraphics[width=0.130\linewidth]{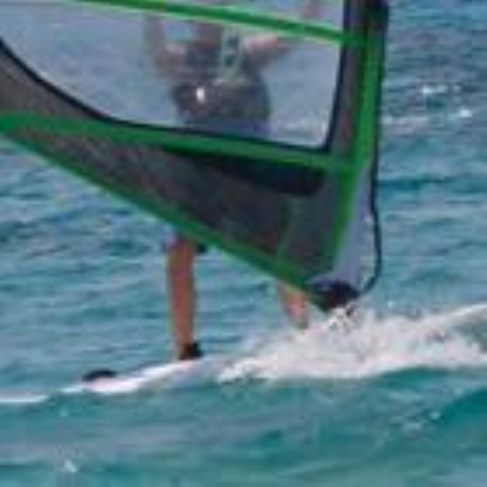}}\;\!\!
	\subfloat {\includegraphics[width=0.130\linewidth]{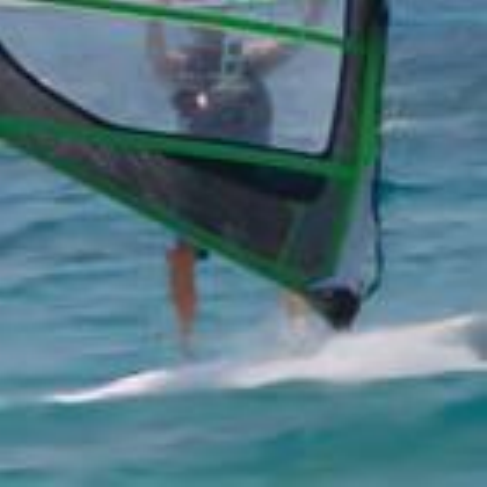}}\;\!\!
    \subfloat {\includegraphics[width=0.130\linewidth]{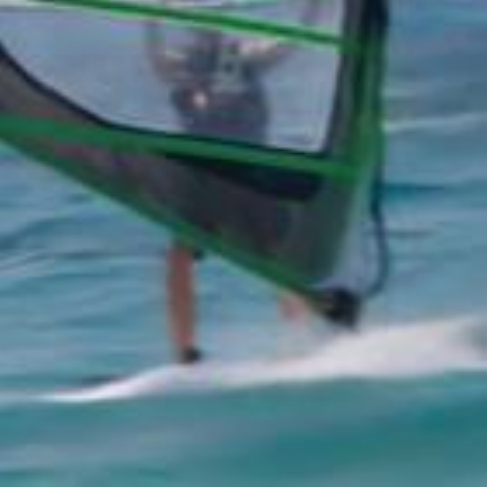}}\;\!\!
    \subfloat {\includegraphics[width=0.130\linewidth]{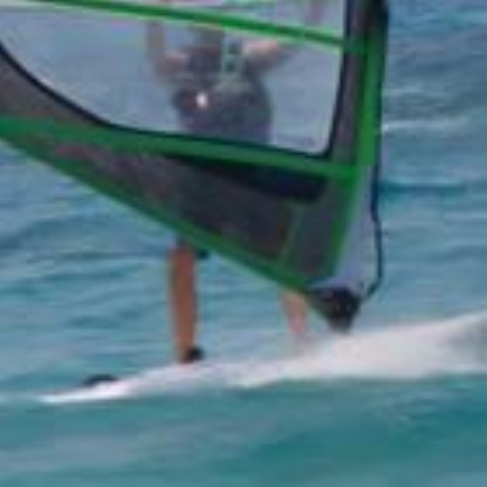}}\;\!\!
    \subfloat {\includegraphics[width=0.130\linewidth]{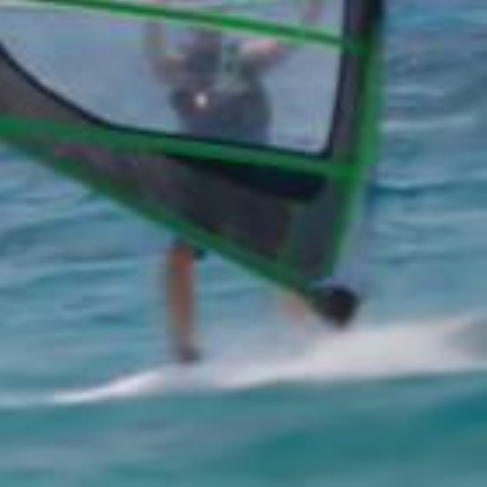}}\;\!\!
    \subfloat {\includegraphics[width=0.130\linewidth]{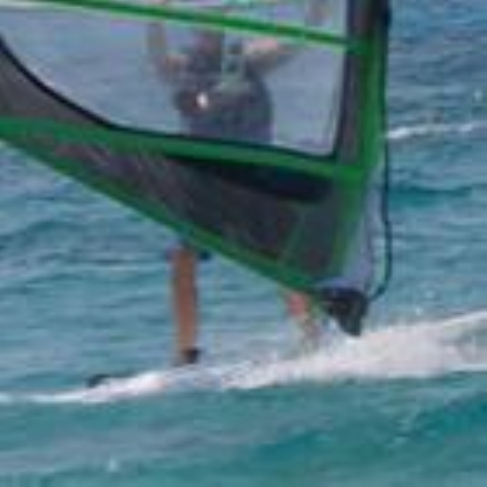}}\\
    
    \subfloat {\includegraphics[width=0.195\linewidth]{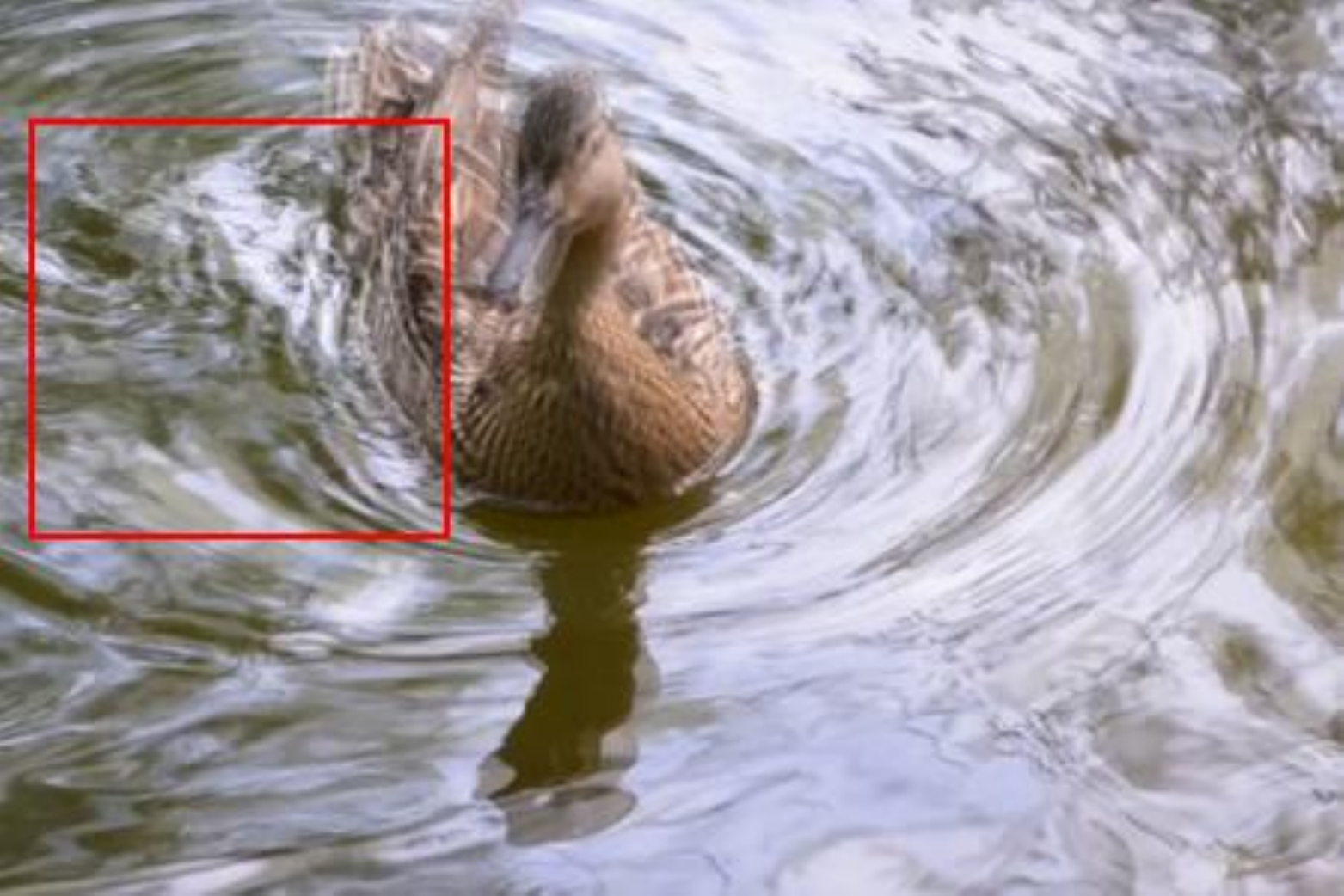}}\;\!\!
	\subfloat {\includegraphics[width=0.130\linewidth]{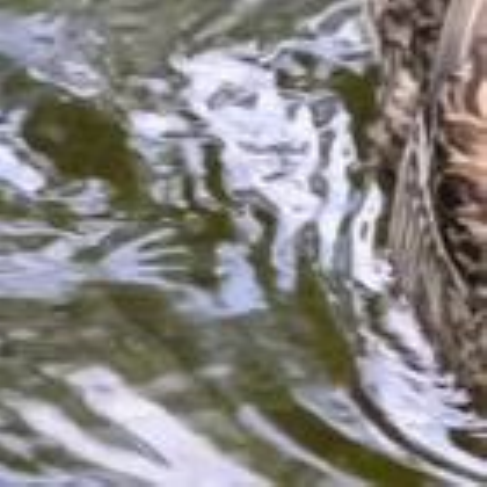}}\;\!\!
	\subfloat {\includegraphics[width=0.130\linewidth]{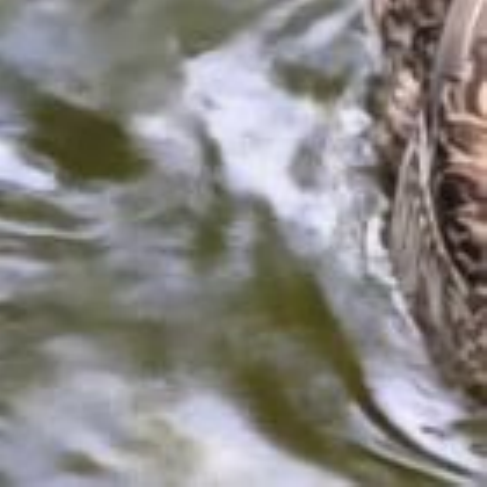}}\;\!\!
    \subfloat {\includegraphics[width=0.130\linewidth]{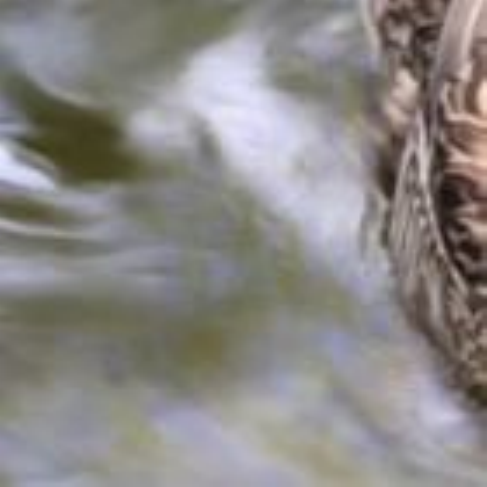}}\;\!\!
    \subfloat {\includegraphics[width=0.130\linewidth]{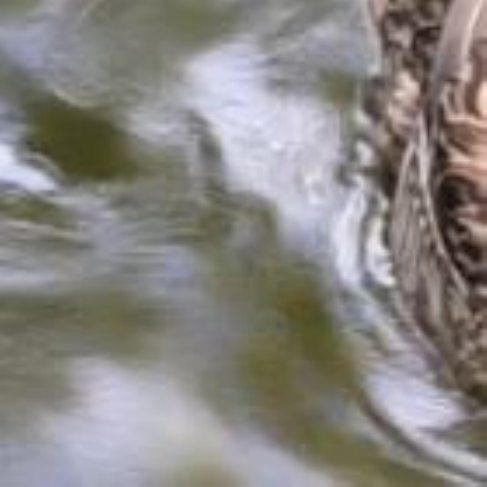}}\;\!\!
    \subfloat {\includegraphics[width=0.130\linewidth]{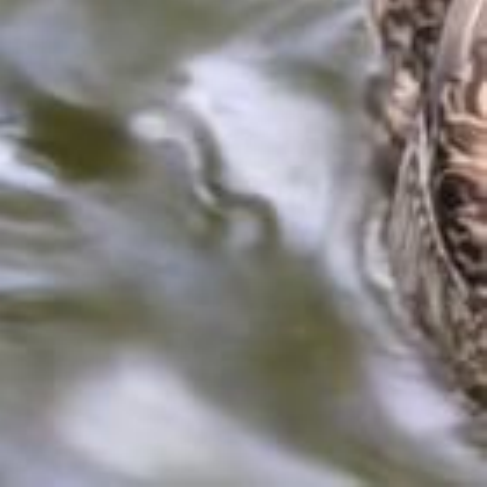}}\;\!\!
    \subfloat {\includegraphics[width=0.130\linewidth]{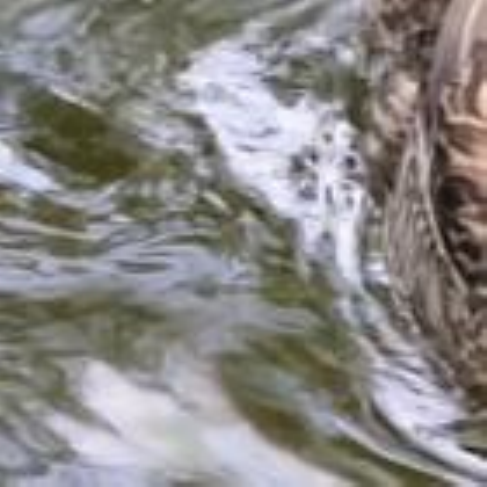}}\\

    \subfloat {\includegraphics[width=0.195\linewidth]{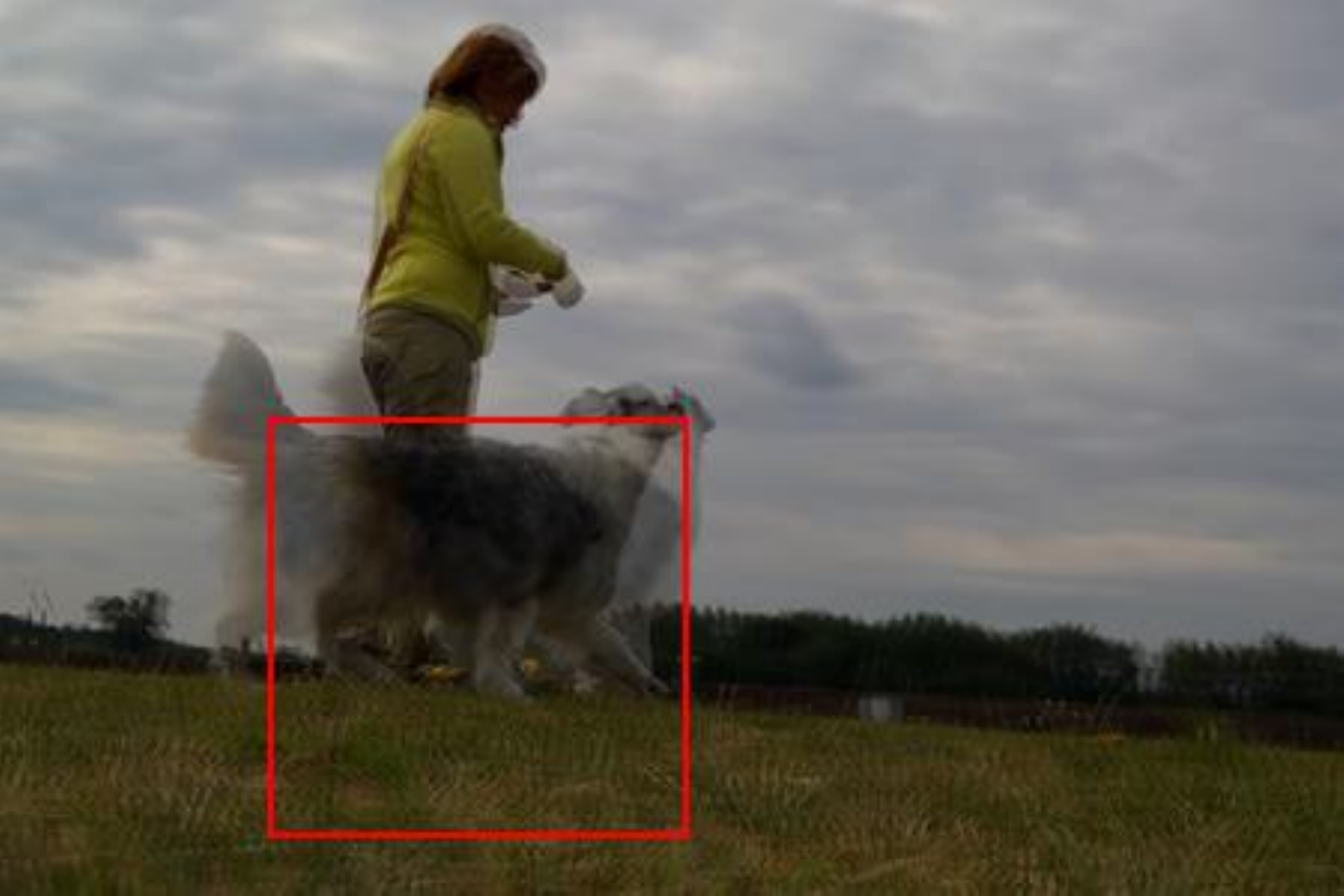}}\;\!\!
	\subfloat {\includegraphics[width=0.130\linewidth]{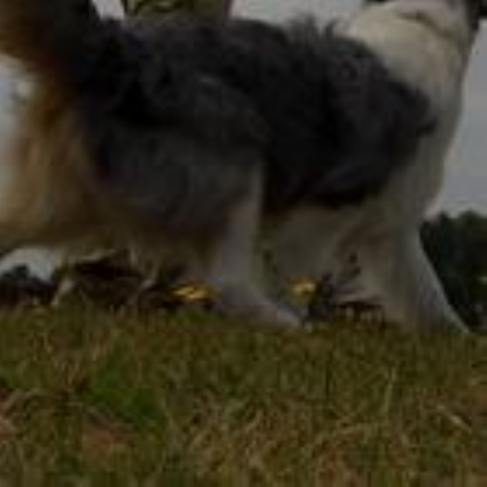}}\;\!\!
	\subfloat {\includegraphics[width=0.130\linewidth]{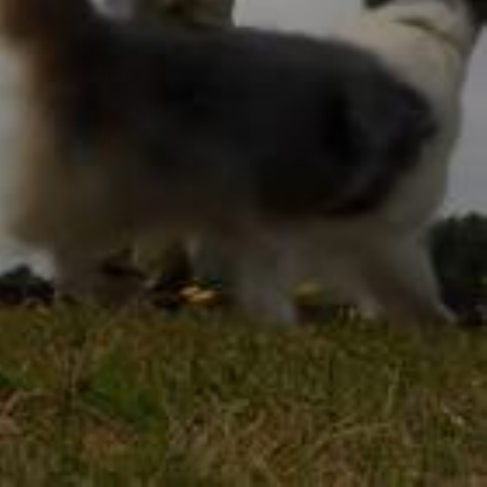}}\;\!\!
    \subfloat {\includegraphics[width=0.130\linewidth]{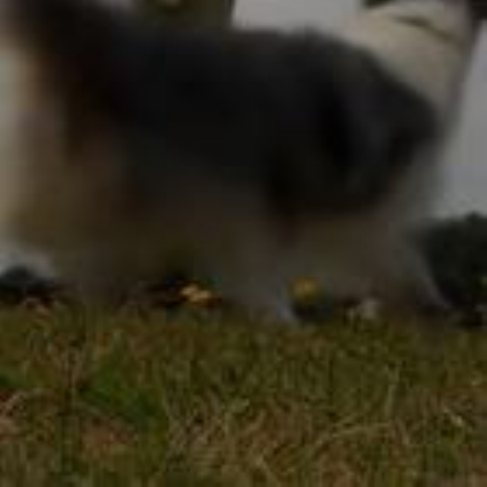}}\;\!\!
    \subfloat {\includegraphics[width=0.130\linewidth]{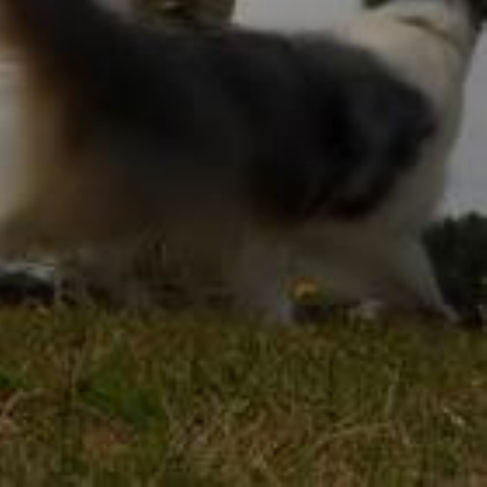}}\;\!\!
    \subfloat {\includegraphics[width=0.130\linewidth]{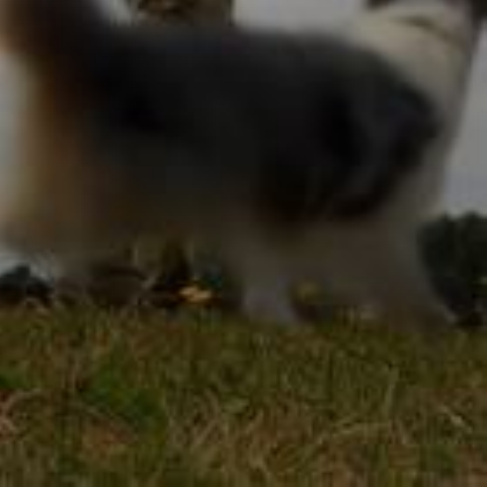}}\;\!\!
    \subfloat {\includegraphics[width=0.130\linewidth]{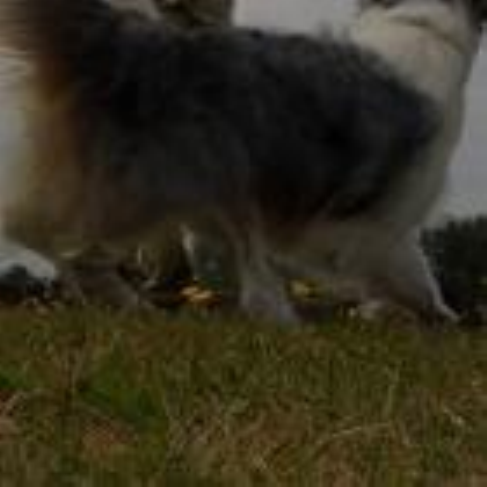}}\\

    \subfloat {\includegraphics[width=0.195\linewidth]{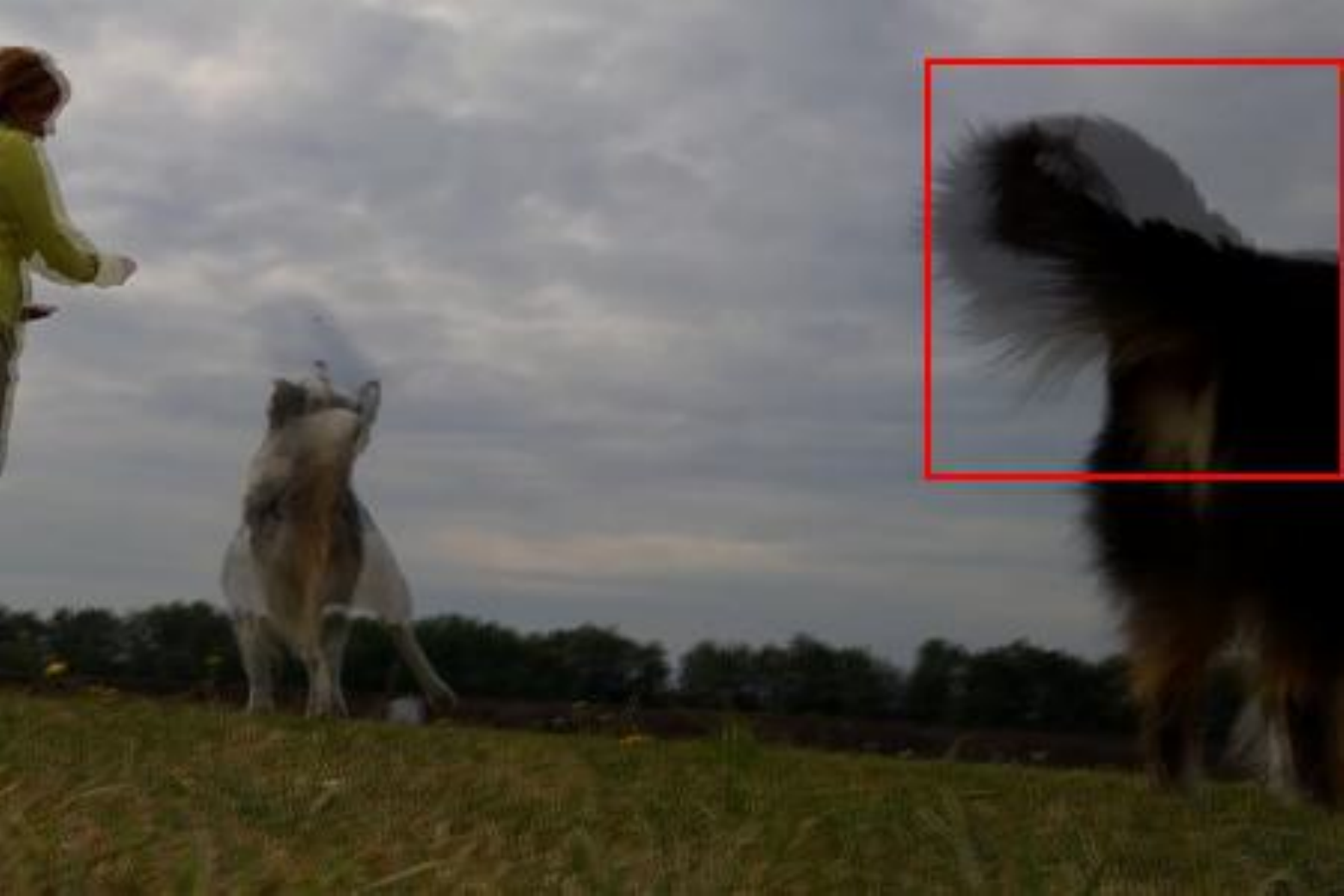}}\;\!\!
	\subfloat {\includegraphics[width=0.130\linewidth]{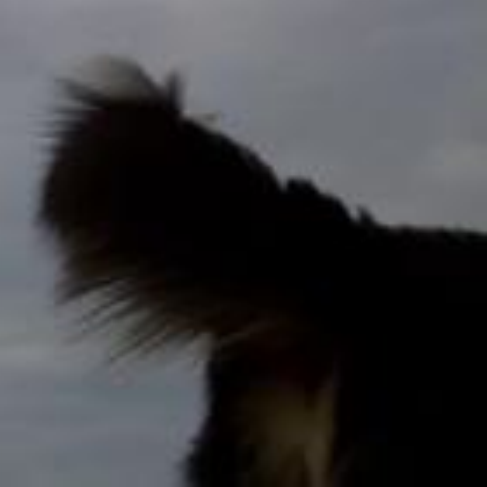}}\;\!\!
	\subfloat {\includegraphics[width=0.130\linewidth]{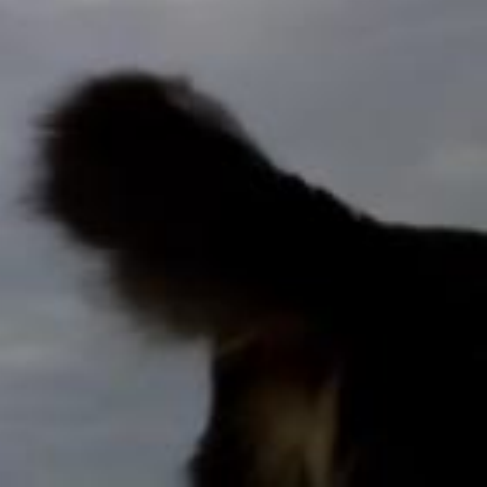}}\;\!\!
    \subfloat {\includegraphics[width=0.130\linewidth]{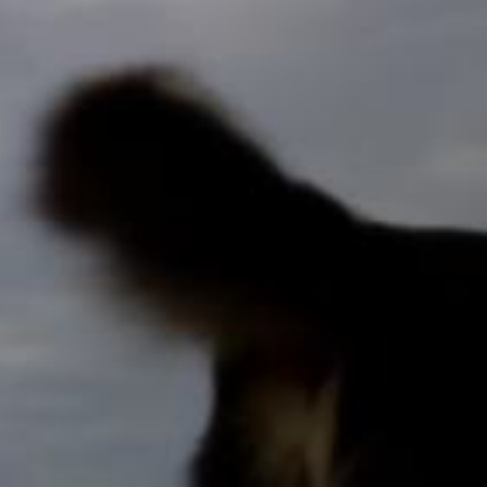}}\;\!\!
    \subfloat {\includegraphics[width=0.130\linewidth]{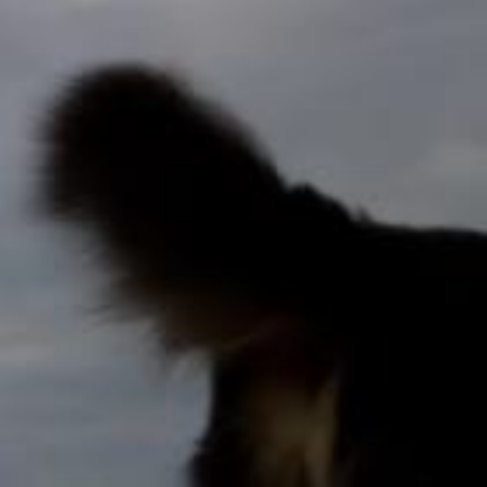}}\;\!\!
    \subfloat {\includegraphics[width=0.130\linewidth]{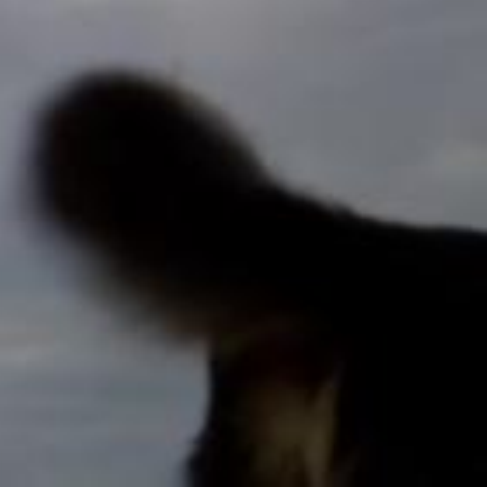}}\;\!\!
    \subfloat {\includegraphics[width=0.130\linewidth]{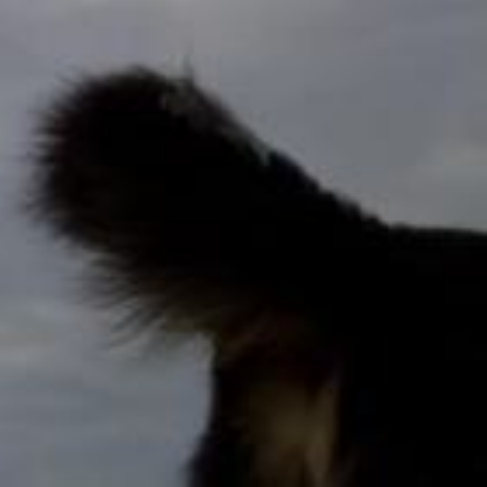}}\\

    \subfloat {\includegraphics[width=0.195\linewidth]{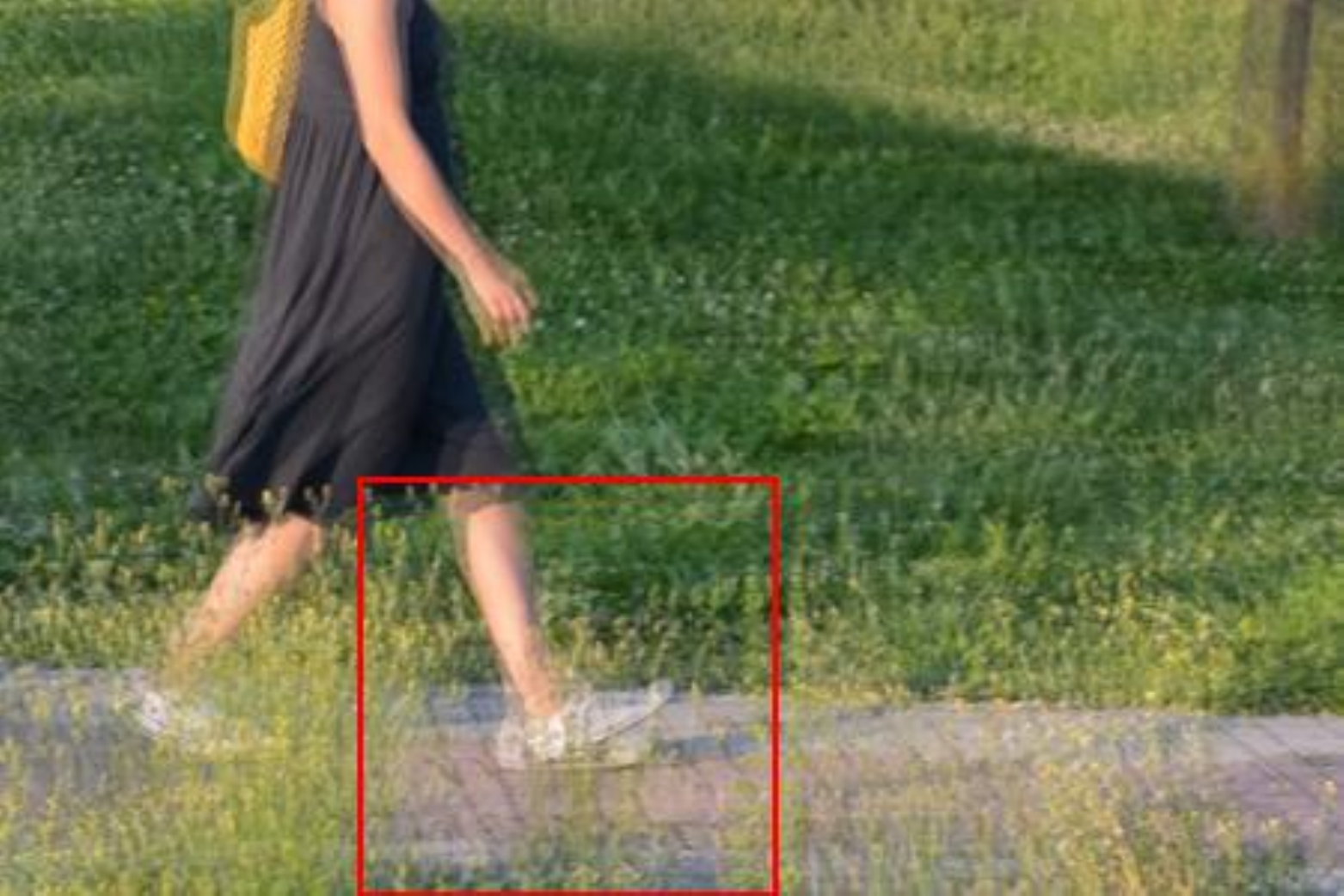}}\;\!\!
	\subfloat {\includegraphics[width=0.130\linewidth]{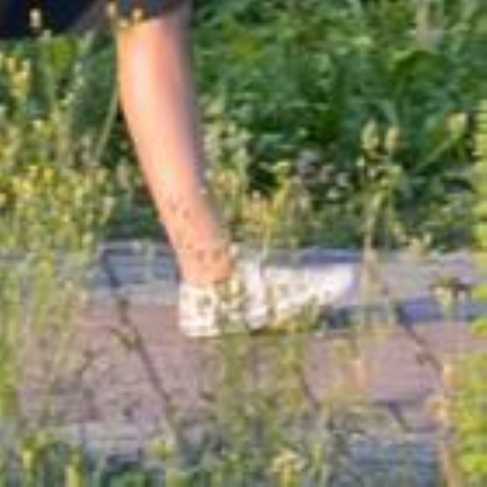}}\;\!\!
	\subfloat {\includegraphics[width=0.130\linewidth]{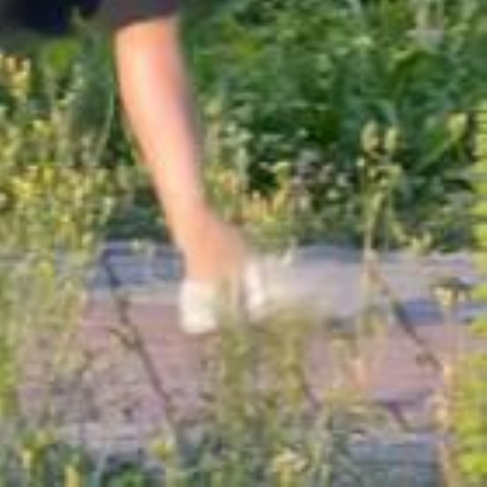}}\;\!\!
    \subfloat {\includegraphics[width=0.130\linewidth]{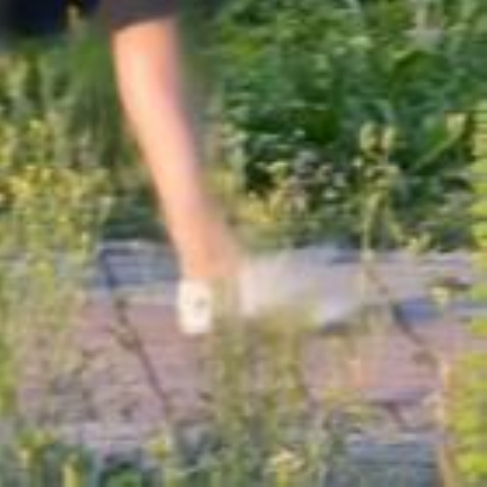}}\;\!\!
    \subfloat {\includegraphics[width=0.130\linewidth]{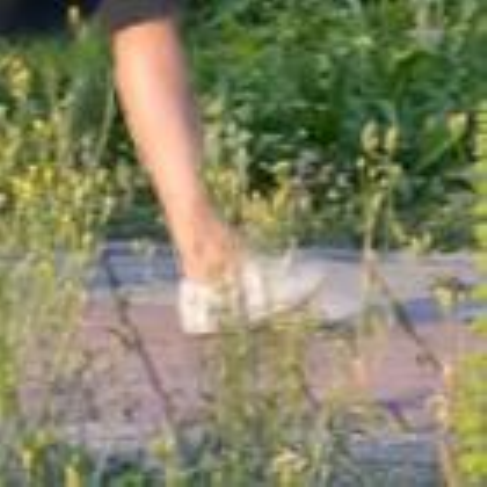}}\;\!\!
    \subfloat {\includegraphics[width=0.130\linewidth]{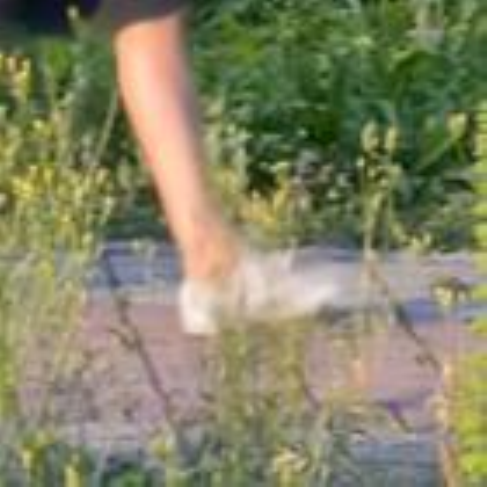}}\;\!\!
    \subfloat {\includegraphics[width=0.130\linewidth]{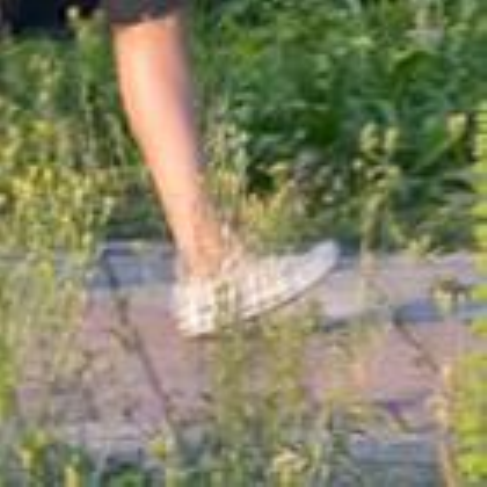}}\\

    \subfloat {\includegraphics[width=0.195\linewidth]{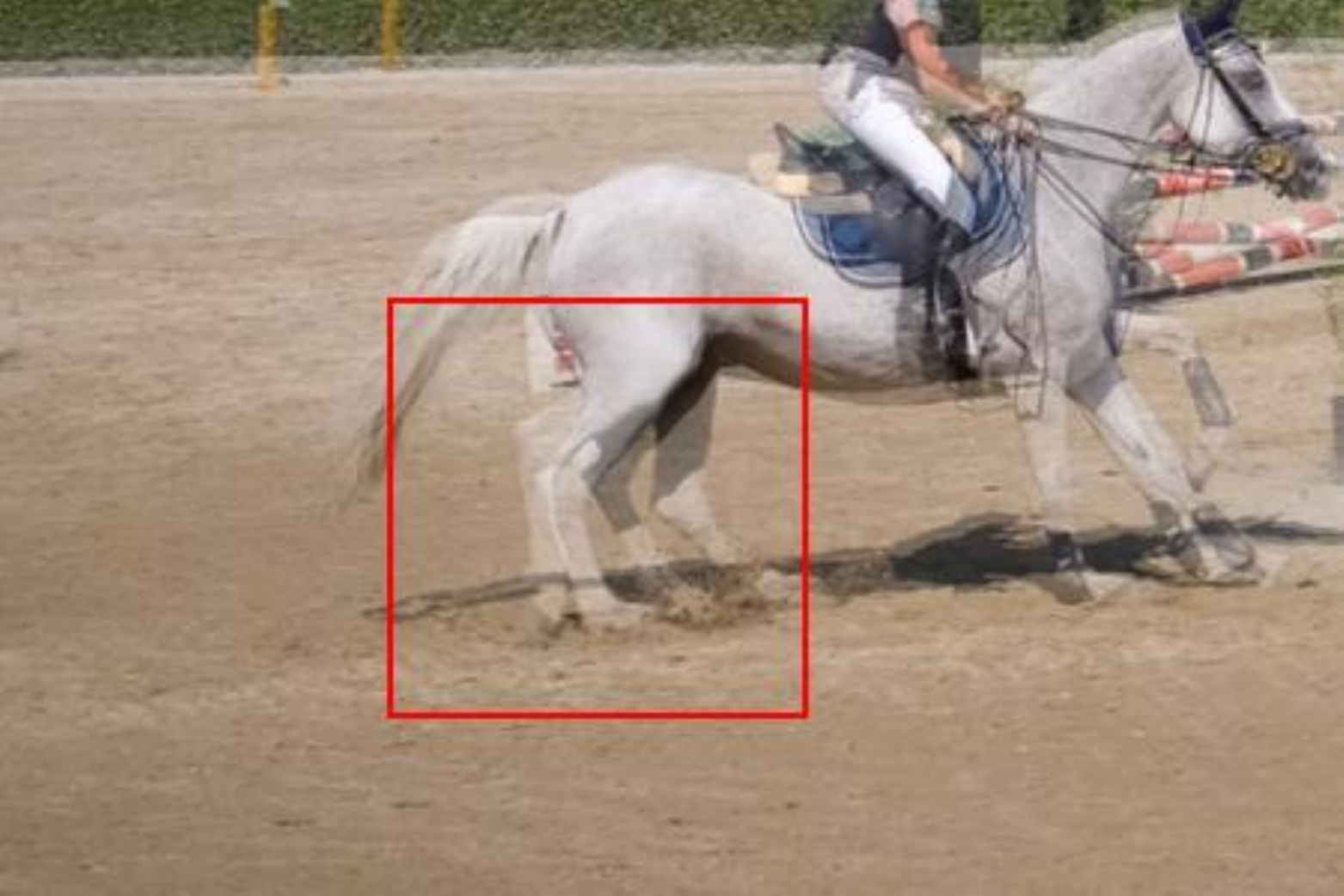}}\;\!\!
	\subfloat {\includegraphics[width=0.130\linewidth]{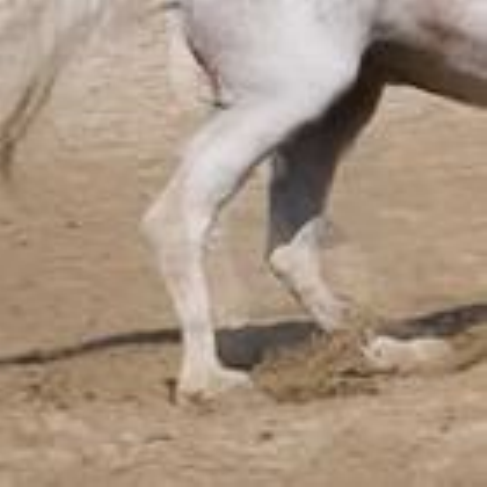}}\;\!\!
	\subfloat {\includegraphics[width=0.130\linewidth]{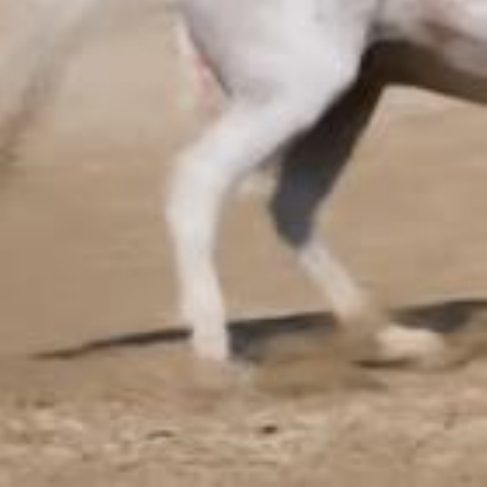}}\;\!\!
    \subfloat {\includegraphics[width=0.130\linewidth]{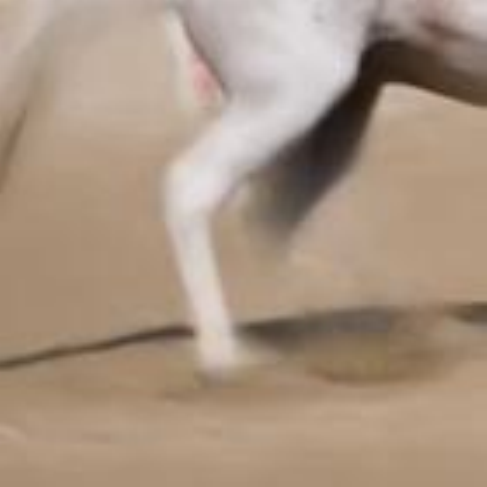}}\;\!\!
    \subfloat {\includegraphics[width=0.130\linewidth]{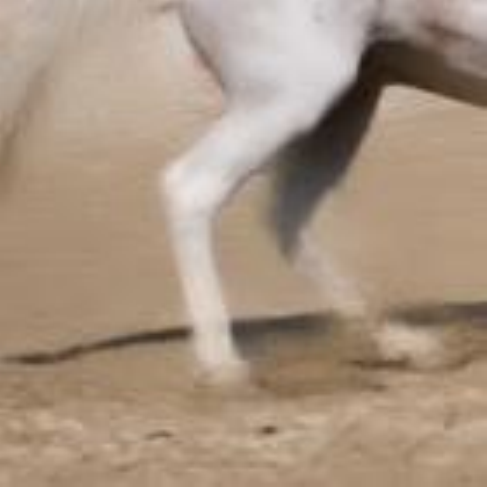}}\;\!\!
    \subfloat {\includegraphics[width=0.130\linewidth]{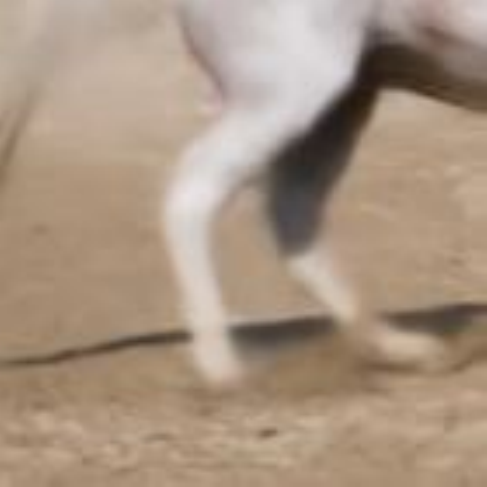}}\;\!\!
    \subfloat {\includegraphics[width=0.130\linewidth]{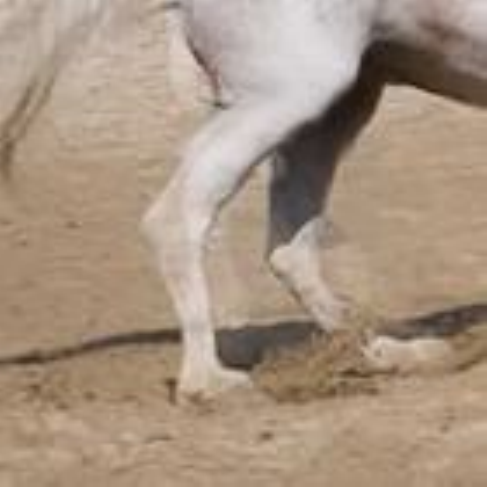}}\\

    \setcounter{subfigure}{0}
    \subfloat[Overlaid inputs] {\includegraphics[width=0.195\linewidth]{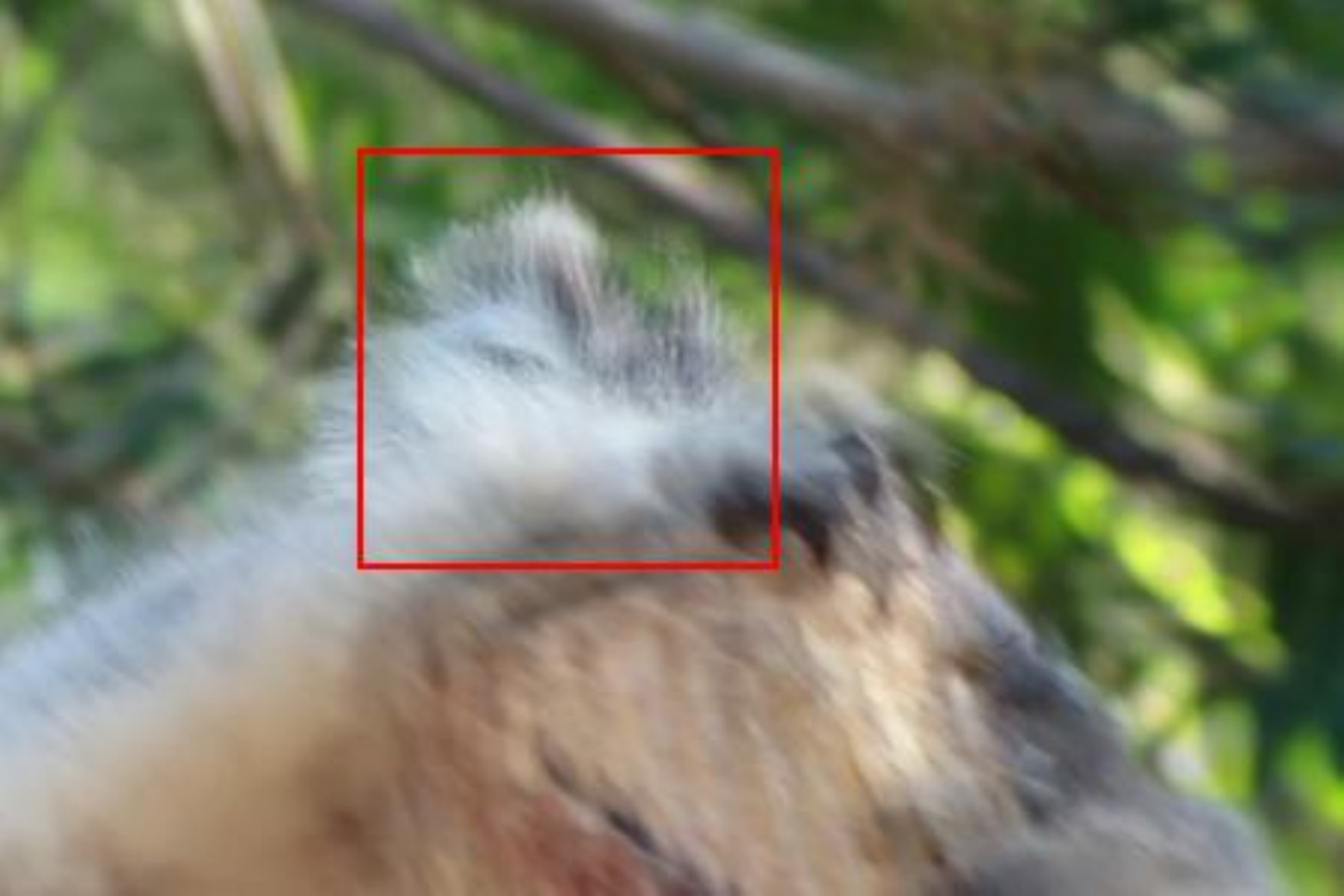}}\;\!\!
	\subfloat[GT] {\includegraphics[width=0.130\linewidth]{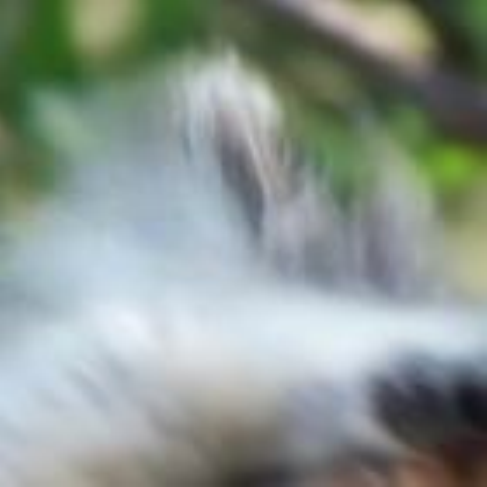}}\;\!\!
	\subfloat[BMBC] {\includegraphics[width=0.130\linewidth]{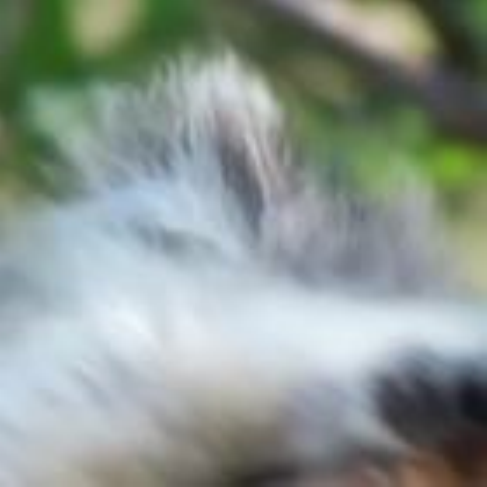}}\;\!\!
    \subfloat[VFIformer] {\includegraphics[width=0.130\linewidth]{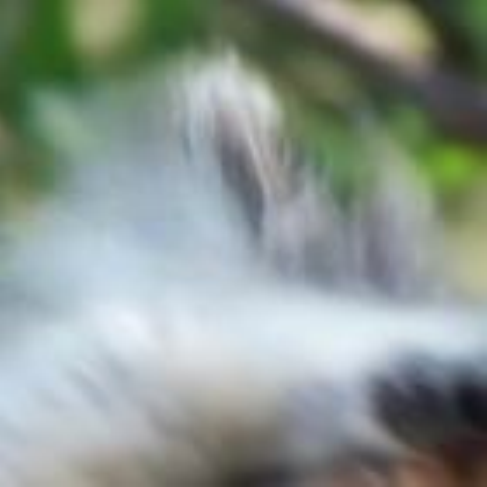}}\;\!\!
    \subfloat[IFRNet] {\includegraphics[width=0.130\linewidth]{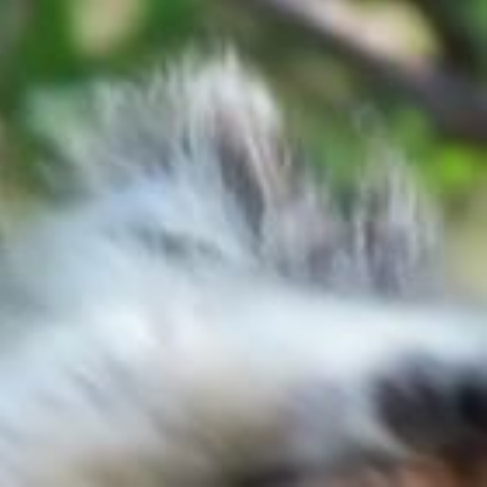}}\;\!\!
    \subfloat[ST-MFNet] {\includegraphics[width=0.130\linewidth]{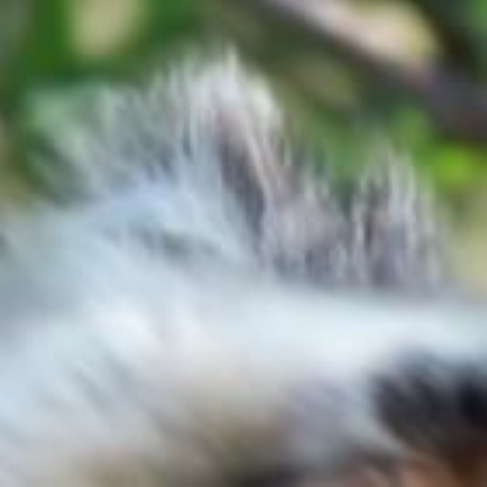}}\;\!\!
    \subfloat[LDMVFI (ours)] {\includegraphics[width=0.130\linewidth]{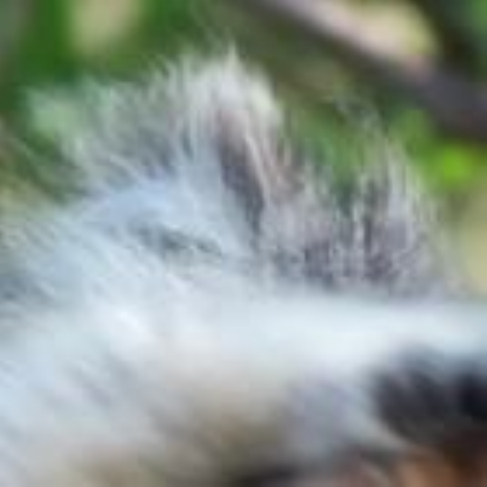}}\\
    \vspace{1em}
    \caption{Visual interpolation examples.}
	\label{fig:qualitative}
\end{figure}

\begin{figure}[H]
    \centering
    \subfloat {\includegraphics[width=0.195\linewidth]{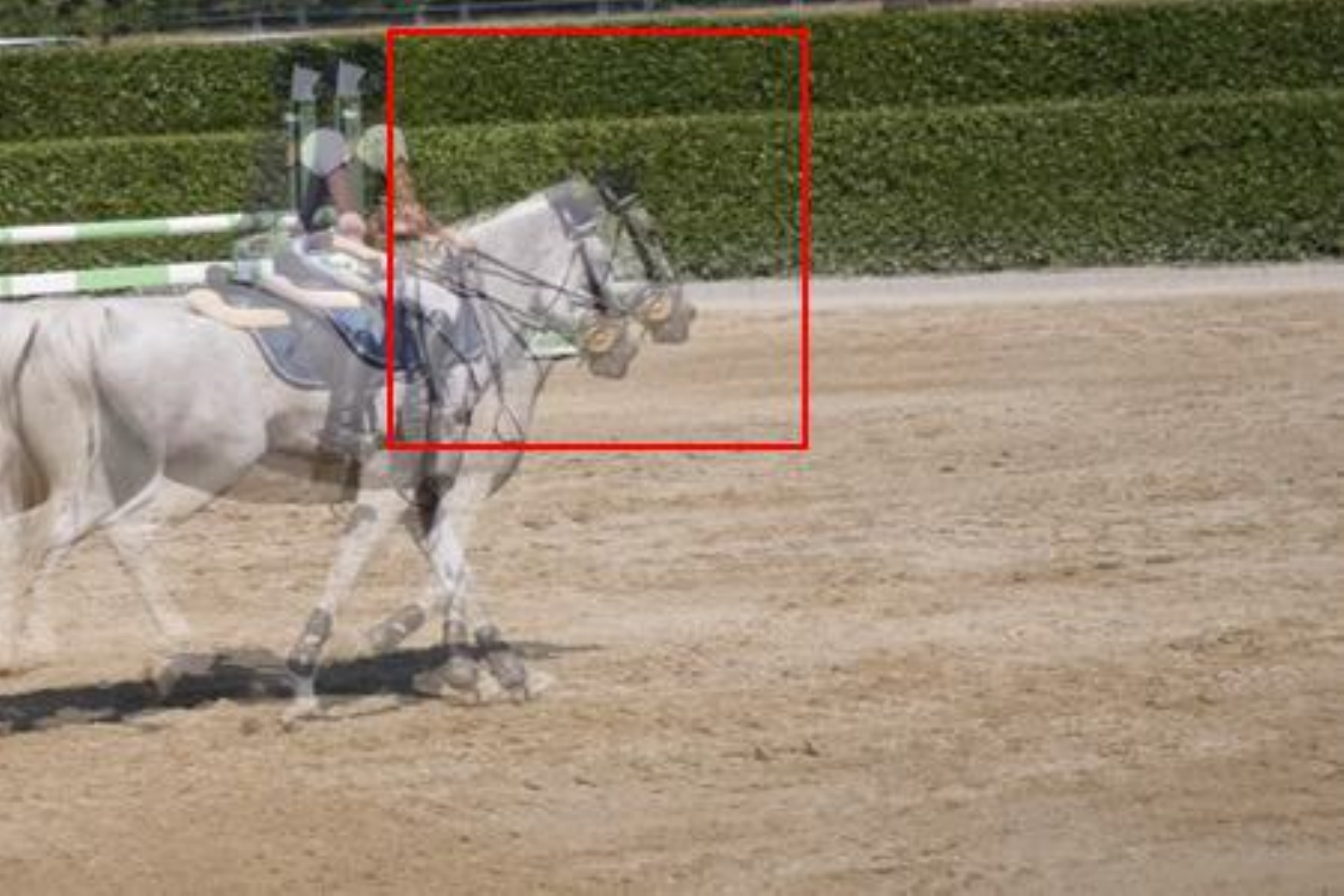}}\;\!\!
	\subfloat {\includegraphics[width=0.130\linewidth]{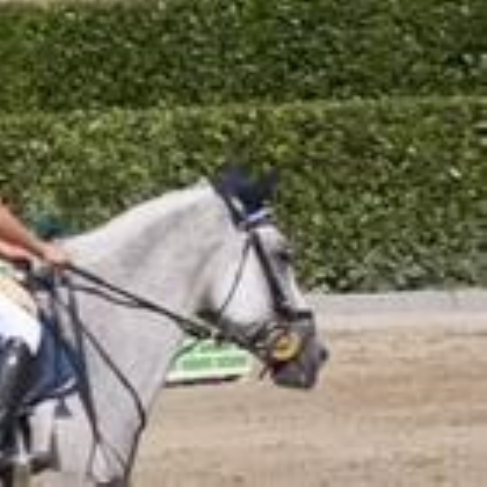}}\;\!\!
	\subfloat {\includegraphics[width=0.130\linewidth]{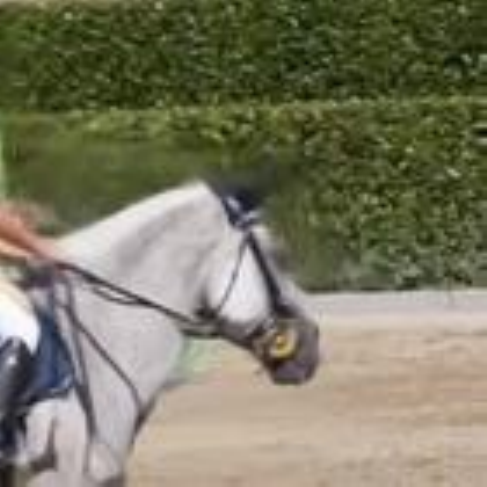}}\;\!\!
    \subfloat {\includegraphics[width=0.130\linewidth]{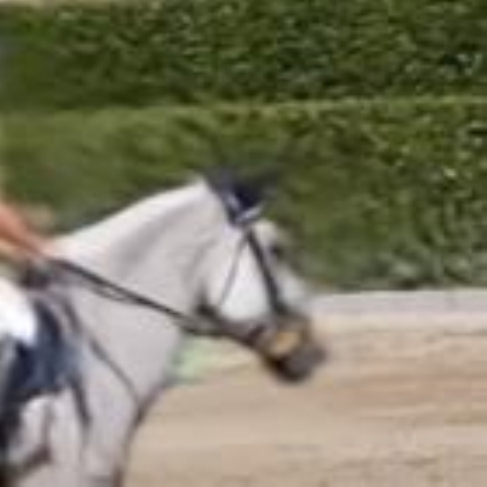}}\;\!\!
    \subfloat {\includegraphics[width=0.130\linewidth]{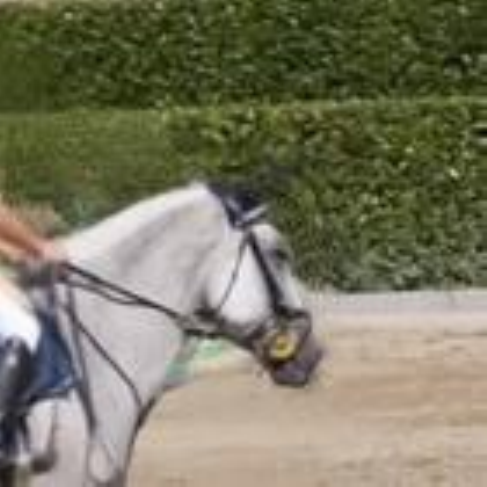}}\;\!\!
    \subfloat {\includegraphics[width=0.130\linewidth]{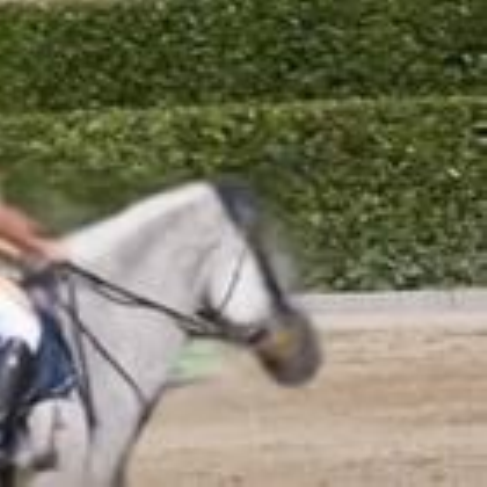}}\;\!\!
    \subfloat {\includegraphics[width=0.130\linewidth]{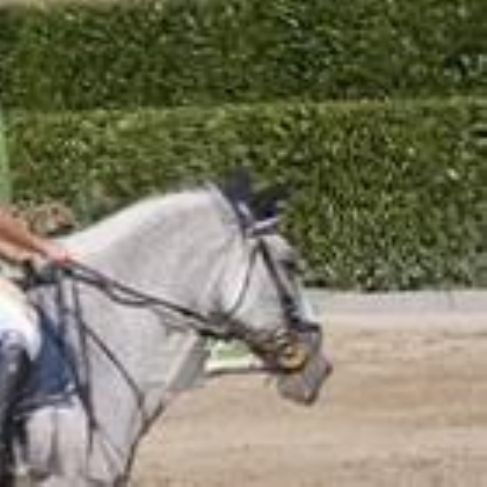}}\\

    \subfloat {\includegraphics[width=0.195\linewidth]{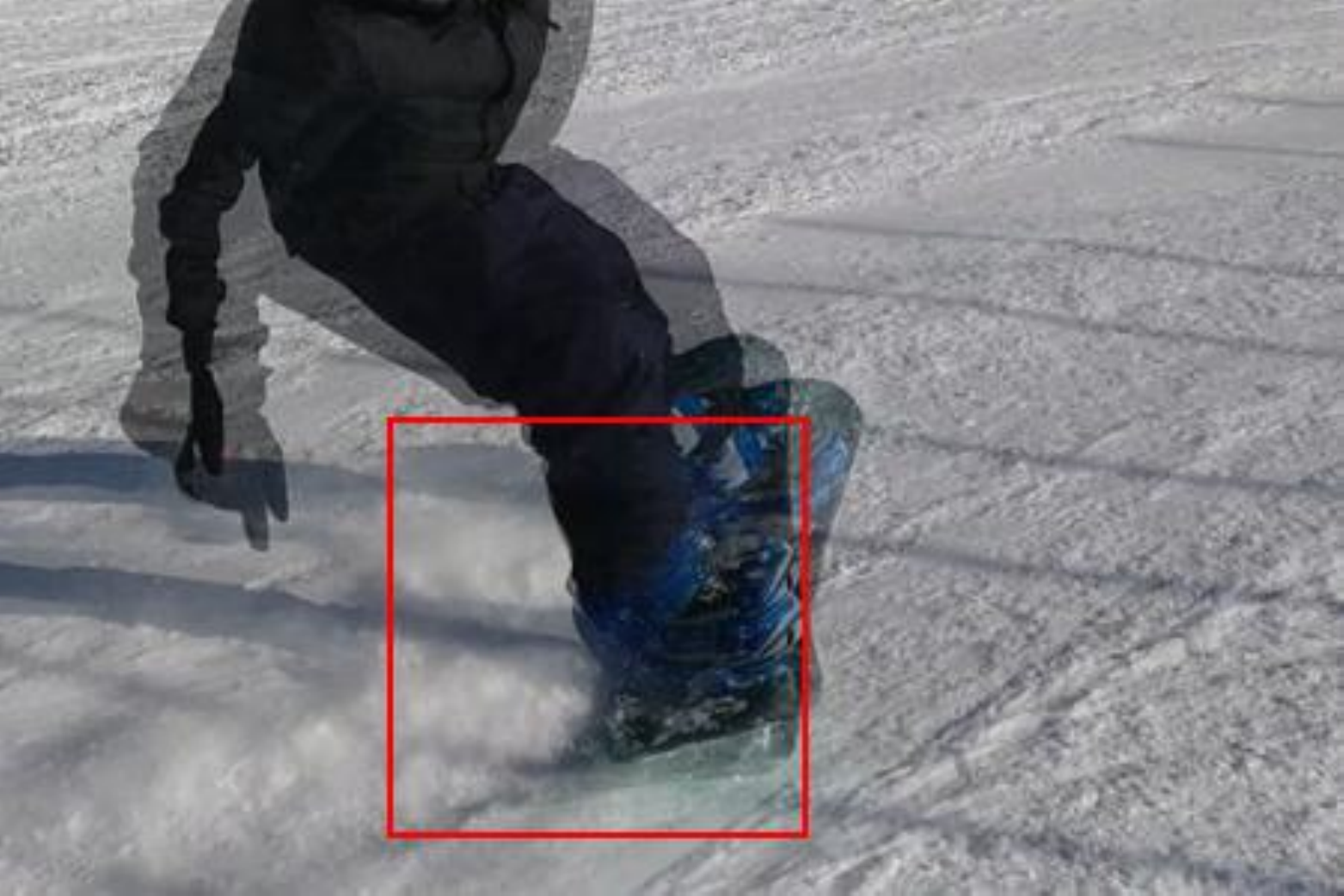}}\;\!\!
	\subfloat {\includegraphics[width=0.130\linewidth]{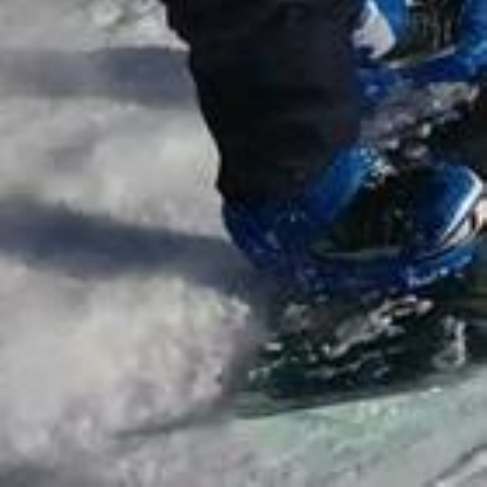}}\;\!\!
	\subfloat {\includegraphics[width=0.130\linewidth]{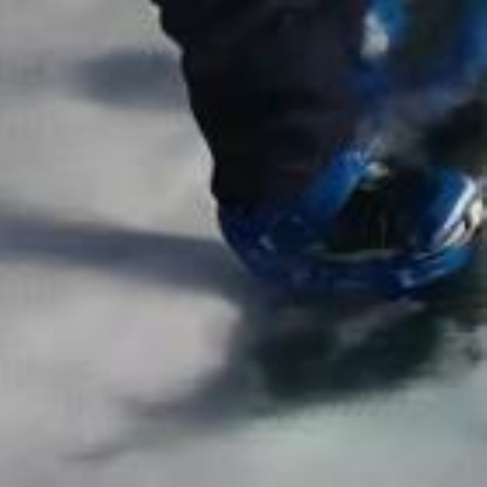}}\;\!\!
    \subfloat {\includegraphics[width=0.130\linewidth]{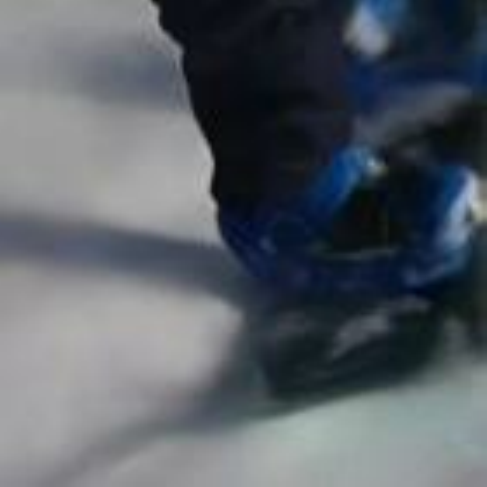}}\;\!\!
    \subfloat {\includegraphics[width=0.130\linewidth]{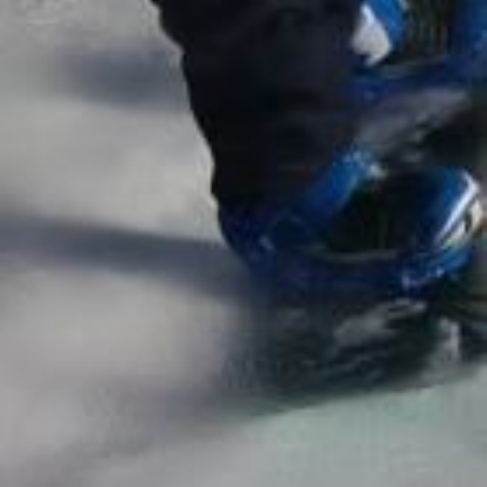}}\;\!\!
    \subfloat {\includegraphics[width=0.130\linewidth]{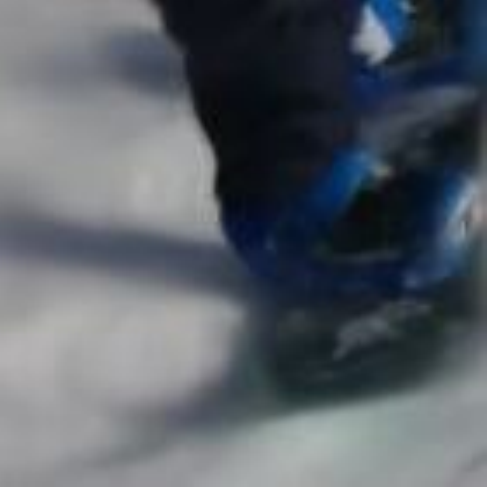}}\;\!\!
    \subfloat {\includegraphics[width=0.130\linewidth]{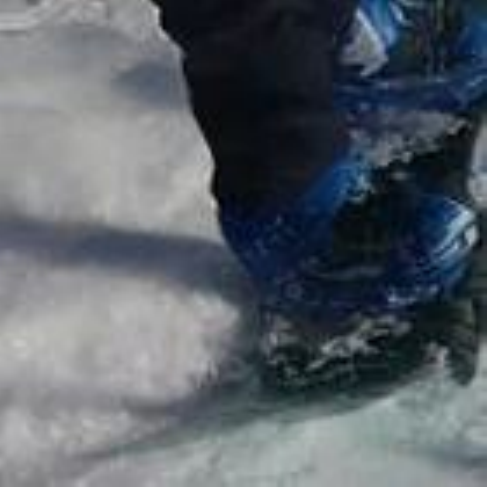}}\\

    \subfloat {\includegraphics[width=0.195\linewidth]{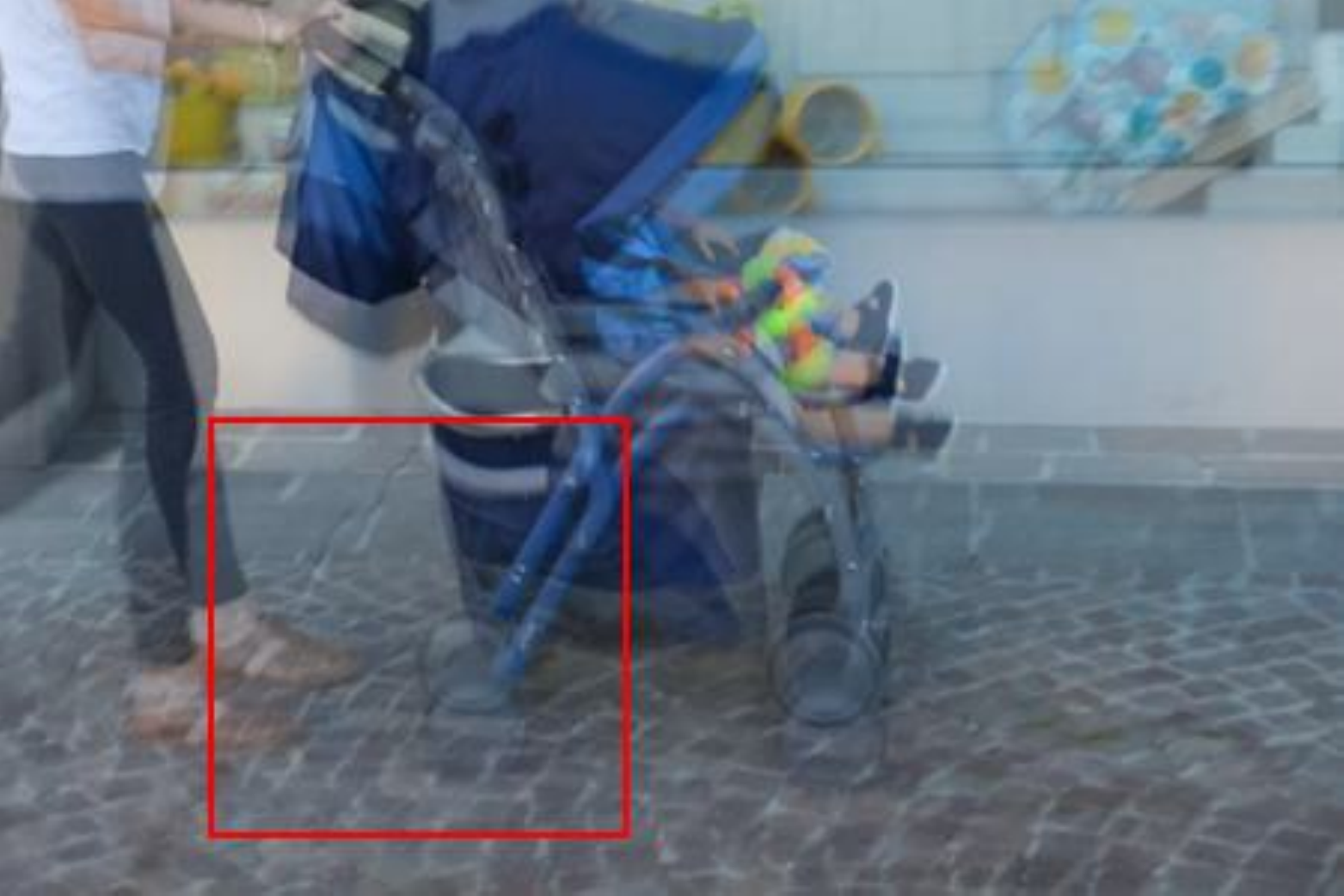}}\;\!\!
	\subfloat {\includegraphics[width=0.130\linewidth]{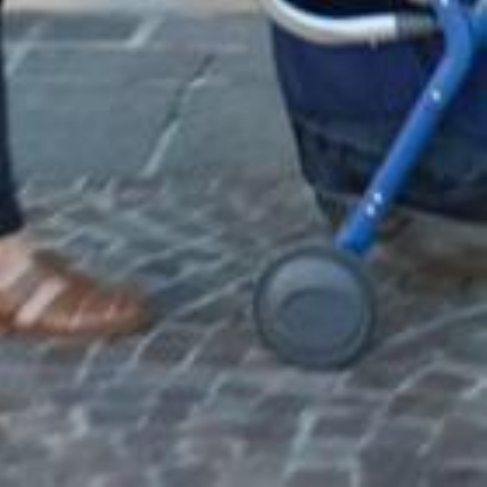}}\;\!\!
	\subfloat {\includegraphics[width=0.130\linewidth]{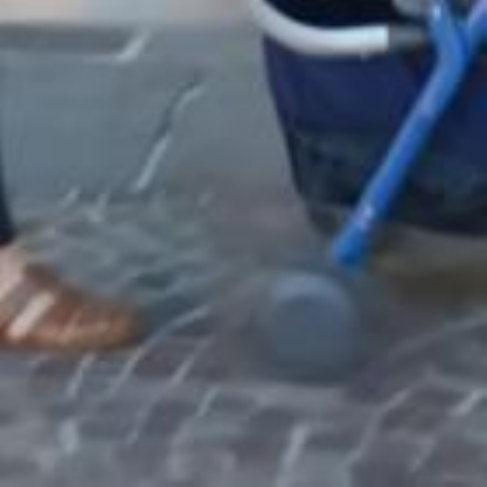}}\;\!\!
    \subfloat {\includegraphics[width=0.130\linewidth]{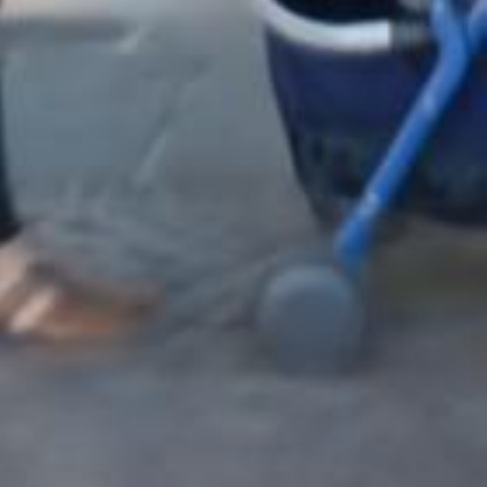}}\;\!\!
    \subfloat {\includegraphics[width=0.130\linewidth]{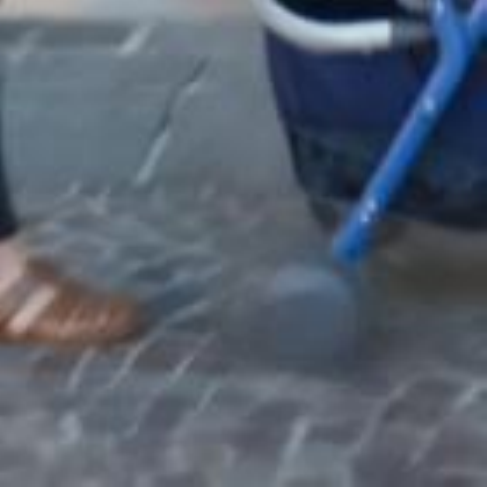}}\;\!\!
    \subfloat {\includegraphics[width=0.130\linewidth]{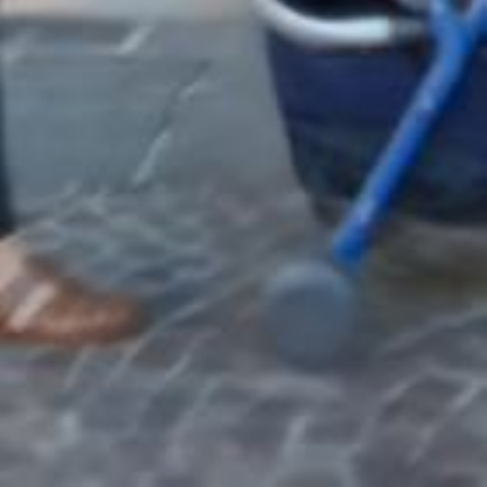}}\;\!\!
    \subfloat {\includegraphics[width=0.130\linewidth]{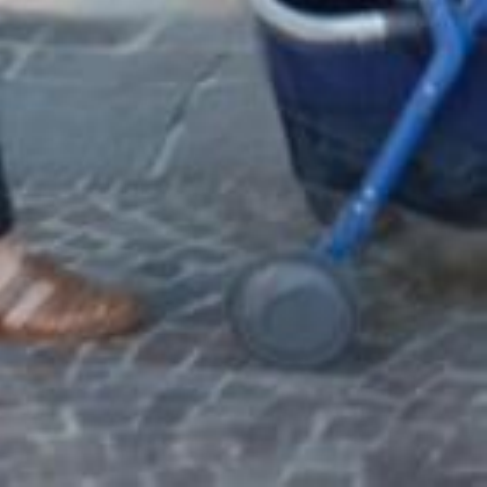}}\\

    \setcounter{subfigure}{0}
    \subfloat[Overlaid inputs] {\includegraphics[width=0.195\linewidth]{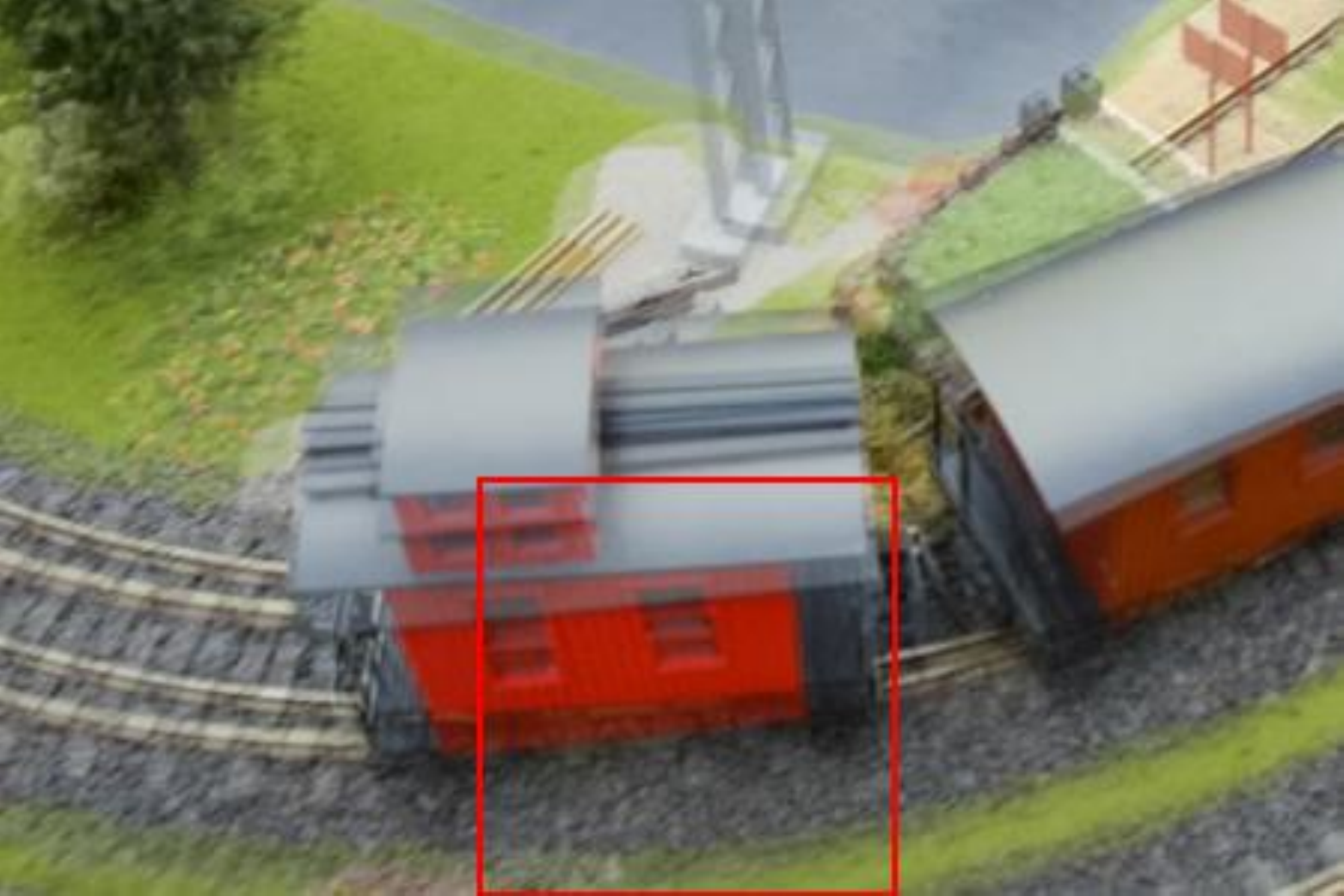}}\;\!\!
	\subfloat[GT] {\includegraphics[width=0.130\linewidth]{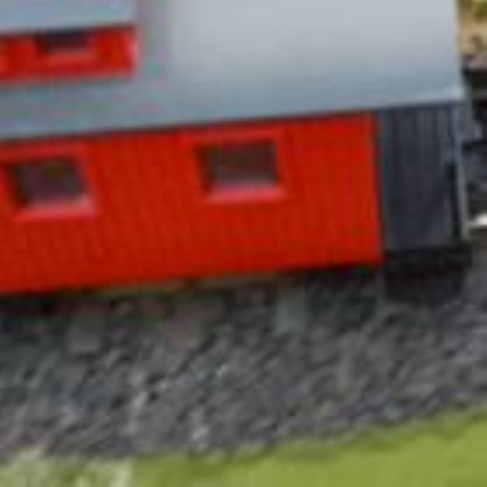}}\;\!\!
	\subfloat[BMBC] {\includegraphics[width=0.130\linewidth]{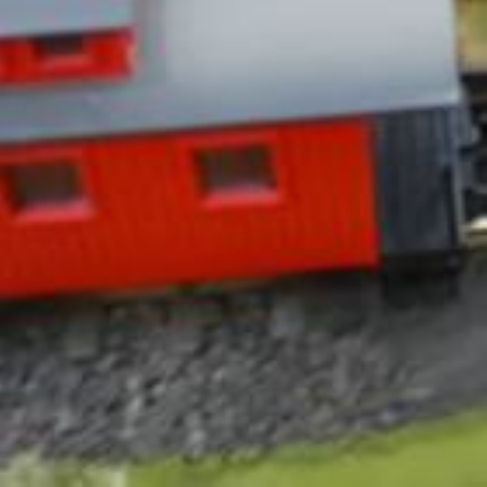}}\;\!\!
    \subfloat[VFIformer] {\includegraphics[width=0.130\linewidth]{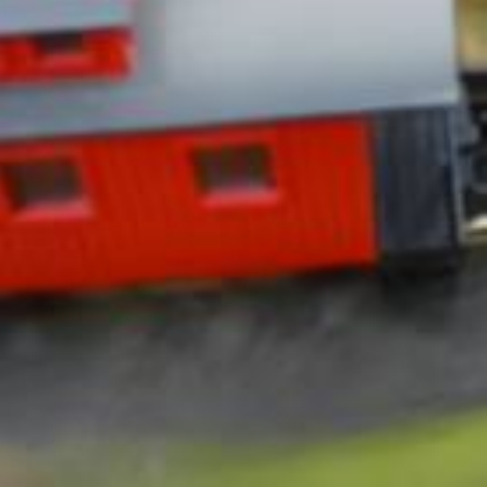}}\;\!\!
    \subfloat[IFRNet] {\includegraphics[width=0.130\linewidth]{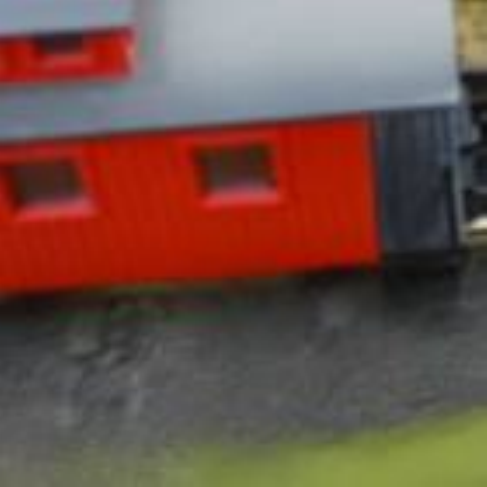}}\;\!\!
    \subfloat[ST-MFNet] {\includegraphics[width=0.130\linewidth]{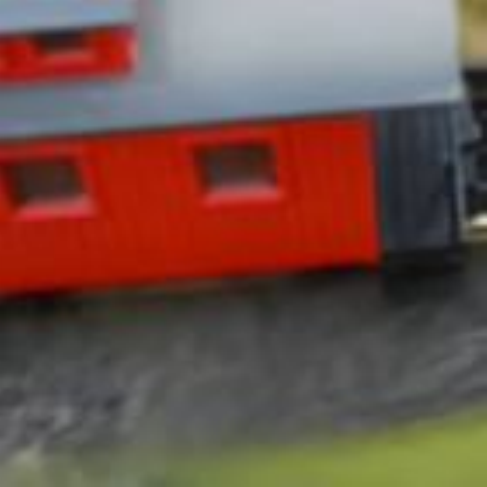}}\;\!\!
    \subfloat[LDMVFI (ours)] {\includegraphics[width=0.130\linewidth]{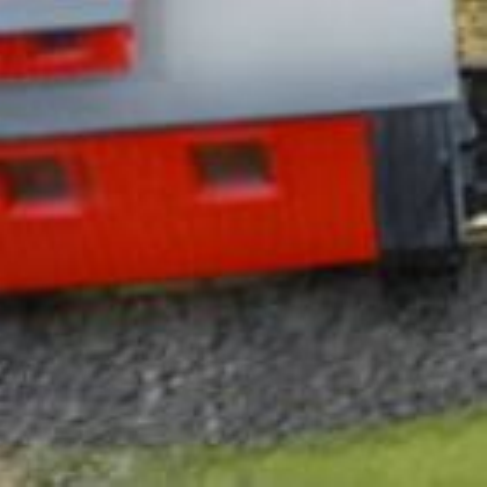}}\\
    \vspace{1em}
    \caption{Visual interpolation examples.}
	\label{fig:qualitative2}
\end{figure}

\end{document}